\documentclass{lmcs}
\pdfoutput=1

\usepackage{lastpage}
\lmcsdoi{19}{1}{11}
\lmcsheading{}{\pageref{LastPage}}{}{}%
{Aug.~28,~2020}{Feb.~09,~2023}{}

\usepackage[utf8]{inputenc}

\usepackage{etex}

\usepackage{subcaption}
\usepackage{hyperref}
\usepackage{xspace}
\usepackage{tikz}
\usetikzlibrary{arrows,positioning,arrows.meta}
\usepackage{amssymb,amsmath,amsthm}
\usepackage{fancybox}
\usepackage{booktabs}

\def\eg{{\em e.g.}}

\def\ie{{\em i.e.}}

\usepackage{ourmacros}


\begin{document}
%
%
\title{Countdown games, and simulation on\texorpdfstring{\\}{ }(succinct) one-counter nets}

\titlecomment{This paper has arisen by a thorough rewriting of
a preliminary version
in the proceedings of the Reachability Problems
conference 2018, and by adding the topics elaborated in
Sections~\ref{sec:structsimulOCN} and~\ref{sec:doubleexpperiod}.}

\thanks{P.~Jan\v{c}ar and P.~Osi\v{c}ka
acknowledge the support by the project 18-11193S of the Grant Agency of the
Czech Republic. Z.~Sawa is partially supported by Grant
of SGS No.~SP2022/123, V\v{S}B Technical University of Ostrava, Czech Republic.}

\author[P.~Jan\v{c}ar]{Petr Jan\v{c}ar\lmcsorcid{0000-0002-8738-9850}}[a]
\author[P.~Osi\v{c}ka]{Petr Osi\v{c}ka\lmcsorcid{0000-0002-7495-7724}}[a]
\author[Z.~Sawa]{Zden\v{e}k Sawa\lmcsorcid{0000-0003-1268-3681}}[b]

\address{Dept of Comp. Sci., Faculty of Science, Palack\'y Univ. Olomouc, Czechia}
\email{petr.jancar@upol.cz, petr.osicka@upol.cz}

\address{Dept of Comp. Sci., FEI, Techn.~Univ.~Ostrava, Czechia}
\email{zdenek.sawa@vsb.cz}

%
%

\begin{abstract}
  We answer an open complexity question by Hofman, Lasota, Mayr,
  Totzke (LMCS 2016) for simulation preorder on the class of succinct
  one-counter nets (\ie, one-counter automata with no zero tests where
  counter increments and decrements are integers written in binary);
  the problem was known to be PSPACE-hard and in EXPSPACE.  We show
  that all relations between bisimulation equivalence and simulation
  preorder are EXPSPACE-hard for these nets; simulation preorder is
  thus EXPSPACE-complete.  The result is proven by a reduction from
  reachability games whose EXPSPACE-completeness in the case of
  succinct one-counter nets was shown by Hunter (RP 2015), by using
  other results.  We also provide a direct self-contained
  EXPSPACE-completeness proof for a special case of such reachability
  games, namely for a modification of countdown games that were shown
  EXPTIME-complete by Jurdzinski, Sproston, Laroussinie (LMCS 2008);
  in our modification the initial counter value is not given but is
  freely chosen by the first player.

  We also present an alternative proof for the upper bound by Hofman
  et al. In particular, we give a new simplified proof of the belt
  theorem that yields a~simple graphic presentation of simulation
  preorder on (non-succinct) one-counter nets and leads to a
  polynomial-space algorithm (which is trivially extended to an
  exponential-space algorithm for succinct one-counter nets).
\end{abstract}

\maketitle

%
%

\section{Introduction}
One-counter automata (OCA), \ie, finite automata
equipped with a~nonnegative counter,
are studied as one of the simplest models
of infinite-state systems. They can be viewed as a~special case of
Minsky counter machines, or as a~special case of pushdown automata.
In general, OCA can test the value of
the counter for zero, \ie,~some transitions could be enabled only if the value
of the counter is zero. One-counter nets (OCN) are a~``monotonic'' subclass of OCA where every
transition enabled for zero is also enabled for nonzero values. As
usual, we can consider
 deterministic, nondeterministic, and/or alternating versions of OCA
and/or OCN.
The basic versions are \emph{unary}, where the
counter can be incremented and decremented by one in one step,
while in the \emph{succinct} versions
the possible changes can be arbitrary integers
(but fixed for a~given transition);
as usual, the changes are assumed to be
written in binary in a~description of a~given automaton.
(Remark. In some papers the term ``unary'' is used differently,
for machines with a single-letter input alphabet.)

Problems that have been studied on OCA and OCN include reachability,
equivalence, model checking, and also different kinds of
games played on these automata.
One of the earliest results
showed decidability of (language)
equivalence for deterministic OCA~\cite{DBLP:journals/jcss/ValiantP75}.
The open polynomiality
question in~\cite{DBLP:journals/jcss/ValiantP75}
was positively answered in~\cite{DBLP:conf/stoc/BohmGJ13}, by showing
that shortest distinguishing words of non-equivalent deterministic OCA have
polynomial lengths. (We can refer, e.g.,
to~\cite{DBLP:journals/lmcs/ChistikovCHPW19} for precise bounds in
some related cases.)

Later other behavioural equivalences (besides language equivalence)
were studied. Most relevant for us is the research started by
Abdulla and \v{C}er\={a}ns who showed in~\cite{DBLP:conf/concur/AbdullaC98} that
simulation preorder on one-counter nets is decidable. An alternative proof
of this fact was given in~\cite{DBLP:conf/sofsem/JancarMS99};
it was also noted that simulation
equivalence is undecidable for OCA.
A relation to bisimulation problems
was shown in~\cite{DBLP:conf/stacs/JancarKM00}.
Ku\v{c}era showed some lower bounds
in~\cite{DBLP:conf/icalp/Kucera00};
Mayr~\cite{DBLP:conf/icalp/Mayr03} showed the undecidability of weak
bisimulation equivalence on OCN.

Simulation preorder
on one-counter nets turned out \PSPACE-complete:
the lower bound was shown by
Srba~\cite{DBLP:journals/corr/abs-0901-2068},
and the upper bound
by Hofman, Lasota, Mayr, and
Totzke~\cite{DBLP:journals/corr/HofmanLMT16}.
It was also shown in~\cite{DBLP:journals/corr/HofmanLMT16}
that deciding weak simulation on OCN can be reduced to deciding strong simulation
on OCN, and thus also solved in polynomial space.
(Strong) bisimulation equivalence on OCA is also known to be
\PSPACE-complete~\cite{DBLP:journals/jcss/BohmGJ14}
(which also holds for a probabilistic version of
OCA~\cite{DBLP:journals/lmcs/ForejtJKW18}).
We note that \PSPACE-membership of problems for the unary case easily
yields \EXPSPACE-membership for the succinct (binary) case.

Succinct (and
parametric)
OCA were considered, e.g., in~\cite{DBLP:conf/concur/HaaseKOW09}, where
reachability on succinct OCA was shown to be \NP-complete.
Games studied on OCA include, e.g., parity games
on one-counter processes (with test for
zero)~\cite{DBLP:conf/fossacs/Serre06},
and are closely related to counter reachability games
(e.g.~\cite{DBLP:journals/fuin/Reichert16}).
Model checking problems on OCA were studied for many types of
logics, \eg,~ LTL~\cite{DBLP:conf/fossacs/DemriLS08},
branching time logics~\cite{DBLP:conf/stacs/GollerL10}, or
first-order logics~\cite{DBLP:conf/lics/GollerMT09}.
\DP-lower bounds for some model-checking (and also equivalence checking)
problems were shown in~\cite{DBLP:journals/iandc/JancarKMS04}.
A recent study~\cite{AlmagorBokerHofmanTotzke:2020}
deals with parameterized universality problems for one-counter
nets.

An involved result by G\"oller, Haase, Ouaknine,
Worrell~\cite{DBLP:conf/icalp/GollerHOW10} shows that model
checking a~fixed CTL formula on succinct one-counter automata
is \EXPSPACE-hard.
The proof is nontrivial, using two involved
results from complexity theory.
The technique of this proof was
referred to by Hunter~\cite{DBLP:conf/rp/Hunter15},
to derive \EXPSPACE-hardness of reachability games
on succinct OCN.

\paragraph{Our contribution}
In this paper we close a~complexity gap for the simulation problem on
succinct OCN that
was mentioned in~\cite{DBLP:journals/corr/HofmanLMT16}, noting that
there was a \PSPACE lower bound
and an \EXPSPACE upper bound for the problem.
We show \EXPSPACE-hardness (and thus \EXPSPACE-completeness)
of the problem, using a defender-choice technique
(cf.,~e.g.,~\cite{DBLP:journals/jacm/JancarS08})
to reduce reachability games to any relation between
simulation preorder and bisimulation equivalence.
Further contributions are explained below in more detail.

As already mentioned, the \EXPSPACE-hardness of reachability games on
succinct OCNs was shown
in~\cite{DBLP:conf/rp/Hunter15}
by using~\cite{DBLP:conf/icalp/GollerHOW10}.
Here we present a direct proof
of \EXPSPACE-hardness (and completeness) even for a special case
of reachability games,
 which we call the ``existential countdown
games''. It is a mild relaxation
of the countdown games from~\cite{DBLP:journals/lmcs/JurdzinskiSL08}
(or their variant from~\cite{DBLP:journals/ipl/Kiefer13}),
which is an interesting \EXPTIME-complete problem.
We thus provide a  simple \EXPSPACE-hardness proof (in fact, by a
master reduction via a natural intermediate problem dealing with
ultimately periodic words)
that is
independent
of~\cite{DBLP:conf/rp/Hunter15} (and of the
involved technique from~\cite{DBLP:conf/icalp/GollerHOW10}
used by~\cite{DBLP:conf/rp/Hunter15}).

We now give an informal sketch
of the results for countdown games.
\begin{figure}[h]
  \begin{center}
  \begin{tikzpicture}[scale=0.8,every node/.style={scale=1}]

  \begin{scope}
  \node (dummy) {};
  \node[stav,left of=dummy,xshift=-0.4cm] (p) {$p_1$};
  \node[stav,right of=dummy,xshift=0.4cm] (q) {$p_2$};

  \draw[->]
  (p) edge[pil] node[above] {$-1$} (q)
  (p) edge[pil,loop left] node[left] {$-24$} (p)
  (q) edge[pil,loop right] node[right] {$-25$} (q)
  (q) edge[pil,loop below] node[below] {$-3$} (q);

  \node[above of=p,yshift=-0.2cm] (e) {$E$};
  \node[above of=q,yshift=-0.2cm] (e) {$A$};
  \end{scope}

  \begin{scope}[shift={(7,-3)}]
    \draw[->] (0,-0.5) -- (0,7);

    \node at (1,-0.5) {$p_1$};
    \node at (2,-0.5) {$p_2$};

    \foreach \y in {0,0.25,...,1.75}
    \drawpoint{(1,\y)}{black};

    \foreach \y in {0.25,1,1.75}
    \drawpoint{(1,\y)}{white};

    \foreach \y in {0,0.25,...,1.75}
    \drawpoint{(2,\y)}{black};

    \foreach \y in {0,0.75,1.5}
    \drawpoint{(2,\y)}{white};

    \marker{0}{$0$}
    \marker{1.75}{$7$}

    \draw[dotted] (1.5,2) -- (1.5,2.5);

    \foreach \y in {2.75,3,...,4} {
      \drawpoint{(1,\y)}{black};
      \drawpoint{(2,\y)}{black};
    }
    
    \foreach \y in {3,3.75}
    \drawpoint{(1,\y)}{white};

    \foreach \y in {2.75}
    \drawpoint{(2,\y)}{white};

    \marker{2.75}{$24$}
    \marker{4}{$29$}

    \draw[dotted] (1.5,4.25) -- (1.5,4.75);

    \foreach \y in {5, 5.25,...,6.25} {
      \drawpoint{(1,\y)}{black};
      \drawpoint{(2,\y)}{black};
    }

    \foreach \y in {5.25,6}
    \drawpoint{(1,\y)}{white};

    \marker{5}{$48$}
    \marker{6.25}{$53$}
    
  \end{scope}

\end{tikzpicture}
  \end{center}
	\caption{A simple countdown game (with $p_{win}=p_2$), and its
	solution.}
  \label{fig:excountdowngame}
\end{figure}
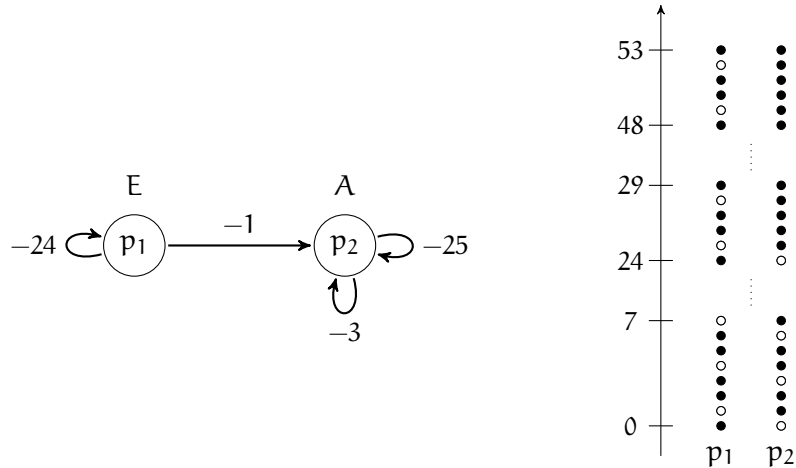
The left-hand part of Figure~\ref{fig:excountdowngame} shows an
example of a very simple countdown
game. It is, in fact, a special finite automaton, with Eve's states, in our case
just $p_1$, and Adam's states,
in our case just $p_2$. The game assumes a nonnegative counter whose
value is modified by transitions; in the \emph{countdown}
games the counter is only \emph{decreased} by transitions. E.g., in the configuration
$p_1(185)$, i.e.\ in the situation where the current state is $p_1$ and the current counter value is
$185$, Eve can choose the transition $p_1\gt{-24}p_1$, which changes the
current configuration to $p_1(185-24)$, i.e.\ to $p_1(161)$, or the transition $p_1\gt{-1}p_2$, which changes the
current configuration to $p_2(184)$. In $p_2(184)$ Adam has two choices,
either going to $p_2(159)$ or to $p_2(181)$.
Since the counter cannot become negative, in $p_2(17)$ Adam has
one choice only, necessarily reaching $p_2(2)$ by several steps, where the respective
play finishes.
Eve's goal is that the play finishes
in $p_{win}(0)$ for a distinguished state $p_{win}$; in our example we
put $p_{win}=p_2$.

It is obvious that an initial configuration $p_2(n)$ is
winning for Eve iff $n \in\{0,3,6,9,12,15,\allowbreak 18,21,24\}$ as partly
depicted in the right-hand part of
Figure~\ref{fig:excountdowngame} by the white points.
(The black points thus correspond to the configurations where
Adam has a winning strategy; e.g., $p_2(7)$ is winning for Adam.)
An initial configurations $p_1(n)$ is winning for Eve
(i.e., Eve has a strategy guaranteeing reaching
$p_2(0)$), iff  $n\bmod 3=1$.

Our concrete example is simple, but also
in the general case, with states $p_1,p_2,\dots,p_k$ (distributed
between Eve and Adam),
it is straightforward to stepwise fill the respective
``black-white-points table''
in the bottom-up fashion that we now describe.
The ``points'' $p_1(0), p_2(0),\dots, p_k(0)$ are black except of
$p_i(0)$ where $p_i=p_{win}$ since $p_{win}(0)$ is white.
If we have filled the rows $0,1,\dots,\ell$ (hence each $p_i(j)$ where
$i\in\{1,2,\dots,k\}$ and $j\in\{0,1,\dots,\ell\}$ has been determined to
be black or white), it is trivial to fill the row $\ell{+}1$:
$p_i(\ell{+}1)$ becomes white iff $p_i$ belongs to Eve and there is a
move from $p_i(\ell{+}1)$ to an already established white point (in the
rows $0,1,\dots,\ell$), or
 $p_i$ belongs to Adam, there is a move from $p_i(\ell{+}1)$, and each
such move
leads to an already established white point.

It is thus clear that the question if $p(n)$ is Eve's win is in \EXPTIME:
we fill the ``rows'' for $0,1,2,\dots,n$, using exponential time and
space in the input size, since $n$ and the countdown values are given
in binary or in decimal notation.
We also note that if $m$  is the maximum countdown value (which is
$25$ in Figure~\ref{fig:excountdowngame}),
the row $\ell\geq m$ is determined by the rows $\ell{-}1,
\ell{-}2,\dots,\ell{-}m$.
It is thus clear that deciding the ``existential version'' of the
countdown games, i.e.\ the question, given a state $p$, if there is
$n$ such that $p(n)$ is winning for Eve, can be decided in exponential
space. (When filling the table in the bottom-up fashion, we can always keep just last $m$ rows
in memory.) We can also observe that
the black-white table is thus (ultimately) periodic,
with an at most double-exponential period; the period is indeed
double-exponential in concrete cases, as we also show in this paper.

The lower bounds might look more surprising:
deciding if $p(n)$ is Eve's win is
\EXPTIME-complete~\cite{DBLP:journals/lmcs/JurdzinskiSL08}, while
deciding if there is $n$ such that  $p(n)$ is Eve's win is
\EXPSPACE-complete, as we show in this paper.
In fact, these lower bounds (\EXPTIME-hardness and \EXPSPACE-hardness)
can be also relatively easily established,
when looking at the ``bottom-up'' computation-table
of an exponential-space Turing machine in a convenient way, as is
depicted in and discussed around Figure~\ref{fig:conftable}.

\begin{figure}[h]
  \begin{center}
  \begin{tikzpicture}[scale=0.8,every node/.style={scale=1}]

  \begin{scope}[shift={(2.5,0)},scale={(2)}]
  \node (dummy) {};
  \node[stav,left of=dummy,xshift=-0.4cm] (p) {$p_1$};
  \node[stav,right of=dummy,xshift=0.4cm] (q) {$p_2$};

  \draw[->]
  (p) edge[pil] node[above] {$a, 5$} (q)
  (p) edge[pil,loop left] node[left] {$a,-1$} (p)
  (q) edge[pil,loop right] node[right] {$b,-2$} (q)
  (q) edge[pil,loop above] node[above] {$a,-2$} (q);

  \end{scope}

\end{tikzpicture}
  \end{center}
	\caption{An example of a succinct one-counter net.}
\label{fig:exsocn}
\end{figure}
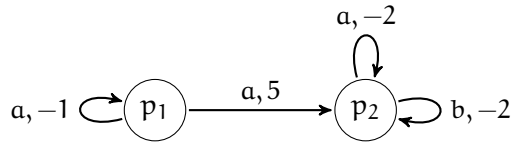

Regarding the simulation problem, we give an example of a succinct one-counter
net in Figure~\ref{fig:exsocn}; it is also a finite automaton equipped
with a nonnegative counter. Now there is no Eve or Adam, and the
counter changes associated with transitions can be also nonnegative; moreover, the
transitions have action-labels ($a,b$ in our example).
On the set of all configurations $p(n)$, the \emph{simulation preorder}
 $\simul$
is the maximal relation such that
for each pair $p(m)\simul q(n)$ and each move
$p(m)\gtl{a}p'(m')$ there is a move $q(n)\gtl{a}q'(n')$ (with the same
label $a$) such that
$p'(m')\simul q'(n')$.
In our example we can note that
$p_2(3)\notsimul p_1(58)$, since the move
$p_2(3)\gt{b}p_2(1)$ cannot be answered by any $b$-transition from
$p_1(58)$; on the other hand, we have $p_2(1)\simul p_1(58)$, since
 no transition is enabled in $p_2(1)$. Slightly more subtle is to note that
 $p_1(0)\simul p_2(6)$: the move $p_1(0)\gt{a}p_2(5)$ is answered by
 $p_2(6)\gt{a}p_2(4)$, and we check that $p_2(5)\simul p_2(4)$.

We can also think in terms of games here.
In the simulation game a position is not just one configuration, but a
pair $(p(m),q(n))$ of configurations. The first player, called
Attacker (or Spoiler), chooses a move $p(m)\gtl{a}p'(m')$ (Attacker
loses if there is no such move), and the
other player, called Defender (or Duplicator), answers by some
 $q(n)\gtl{a}q'(n')$ (with the same label $a$); Defender loses if
 there is no such answer. The play then continues with the next round,
with the current pair  $(p'(m'),q'(n'))$.
 An infinite play is deemed to be winning for
 Defender. It is standard to observe that $p(m)\notsimul q(n)$ iff
 Attacker has a winning strategy from $(p(m),q(n))$
 (and $p(m)\simul q(n)$ iff
 Defender has a winning strategy from $(p(m),q(n))$).

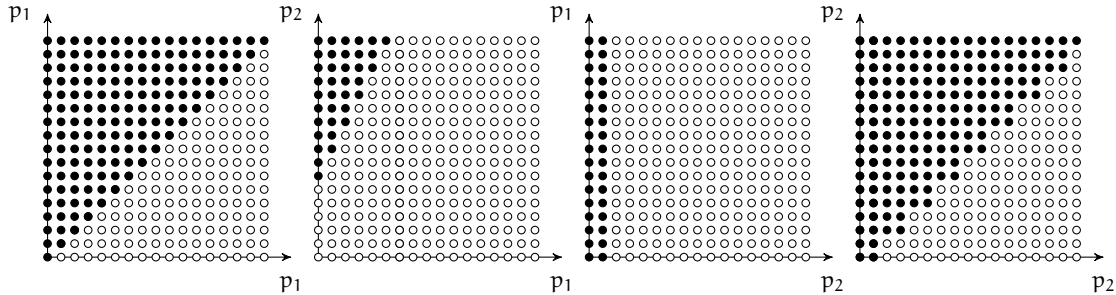
\begin{figure}[h]
  \begin{center}

\begin{tikzpicture}[scale=0.8,every node/.style={scale=1}]
    \begin{scope}[shift={(3, -7)},scale=(0.9)]
    \drawaxis{4.5}{4.5}
    \node[scale=0.8] at (4.5, -0.5) {$p_2$};
    \node[scale=0.8] at (-0.5, 4.5) {$p_1$};

    \foreach \x in {0,0.25}
    {
      \foreach \y in {0,0.25,...,4}
      \drawpoint{(\x,\y)}{black};
    }

    \foreach \x in {0.5,0.75,...,4}
    {
      \foreach \y in {0,0.25,...,4}
      \drawpoint{(\x,\y)}{white};
    }
    
  \end{scope}

  \begin{scope}[shift={(-1.5, -7)},scale=(0.9)]
    \drawaxis{4.5}{4.5}
    \node[scale=0.8] at (4.5, -0.5) {$p_1$};
    \node[scale=0.8] at (-0.5, 4.5) {$p_2$};

    \foreach \y in {0,0.25,...,1.25}
    \drawpoint{(0,\y)}{white};
    
    \foreach \y in {1.5,1.75,...,4}
    \drawpoint{(0,\y)}{black};

    \foreach \y in {0,0.25,...,1.75}
    \drawpoint{(0.25,\y)}{white};
    
    \foreach \y in {2,2.25,...,4}
    \drawpoint{(0.25,\y)}{black};

    \foreach \y in {0,0.25,...,2.25}
    \drawpoint{(0.5,\y)}{white};
    
    \foreach \y in {2.5,2.75,...,4}
    \drawpoint{(0.5,\y)}{black};

    \foreach \y in {0,0.25,...,2.75}
    \drawpoint{(0.75,\y)}{white};
    
    \foreach \y in {3,3.25,...,4}
    \drawpoint{(0.75,\y)}{black};

    \foreach \y in {0,0.25,...,3.25}
    \drawpoint{(1,\y)}{white};
    
    \foreach \y in {3.5,3.75,...,4}
    \drawpoint{(1,\y)}{black};

    \foreach \y in {0,0.25,...,3.75}
    \drawpoint{(1.25,\y)}{white};

    \drawpoint{(1.25,4)}{black};

    \foreach \y in {0,0.25,...,4}
    \drawpoint{(1.5,\y)}{white};

    \foreach \x in {1.5,1.75,...,4}
    {
      \foreach \y in {0,0.25,...,4}
      \drawpoint{(\x, \y)}{white};
    }
    
  \end{scope}

    \begin{scope}[shift={(7.5, -7)}, scale=(0.9)]
    \drawaxis{4.5}{4.5}
    \node[scale=0.8] at (4.5, -0.5) {$p_2$};
    \node[scale=0.8] at (-0.5, 4.5) {$p_2$};

    \foreach \x in {0,0.25}
    {
      \foreach \y in {0,0.25,...,4}
      \drawpoint{(\x, \y)}{black};
    }

    \foreach \x in {0,0.25} 
    \drawvertsqn{0,0.25}{4}{\x}{black};

    \foreach \x in {0.5,0.75}
    {
      \drawvertsqn{0.5,0.75}{4}{\x}{black};
      \drawvertsqn{0,0.25}{0.25}{\x}{white};
    }

    \foreach \x in {1,1.25}
    {
      \drawvertsqn{1,1.25}{4}{\x}{black};
      \drawvertsqn{0,0.25}{0.75}{\x}{white};
    }
     
    \foreach \x in {1.5,1.75}
    {
      \drawvertsqn{1.5,1.75}{4}{\x}{black};
      \drawvertsqn{0,0.25}{1.25}{\x}{white};
    }

    \foreach \x in {2,2.25}
    {
      \drawvertsqn{2,2.25}{4}{\x}{black};
      \drawvertsqn{0,0.25}{1.75}{\x}{white};
    }

    \foreach \x in {2.5,2.75}
    {
      \drawvertsqn{2.5,2.75}{4}{\x}{black};
      \drawvertsqn{0,0.25}{2.25}{\x}{white};
    }

    \foreach \x in {3,3.25}
    {
      \drawvertsqn{3,3.25}{4}{\x}{black};
      \drawvertsqn{0,0.25}{2.75}{\x}{white};
    }

    \foreach \x in {3.5,3.75}
    {
      \drawvertsqn{3.5,3.75}{4}{\x}{black};
      \drawvertsqn{0,0.25}{3.25}{\x}{white};
    }

    \drawpoint{(4,4)}{black};
    \drawvertsqn{0,0.25}{3.75}{4}{white};
  
  \end{scope}

  \begin{scope}[shift={(-6,-7)},scale=(0.9)]
    \drawaxis{4.5}{4.5}
    \node[scale=0.8] at (4.5, -0.5) {$p_1$};
    \node[scale=0.8] at (-0.5, 4.5) {$p_1$};

    \drawvertsqn{0,0.25}{4}{0}{black};
    \foreach \x in {0.25,0.5,...,3.75}
    {
      \drawvertsqn{0,0.25}{\x}{\x}{white};
      \drawvertsqn{4,3.75}{\x}{\x}{black};
    }
    \drawvertsqn{0,0.25}{3.75}{4}{white};
    \drawpoint{(4,4)}{black};
    
  \end{scope}  
 
\end{tikzpicture}
  \end{center}
	\caption{A depiction of the simulation preorder $\simul$
	related to
	the net from Figure~{\ref{fig:exsocn}}.}
\label{fig:solvedsimul}
\end{figure}

Given a succinct one-counter net,
we can represent the relation $\simul$ by
the respective ``black-white colourings'' $C_{\tu{p,q}}$ of the integer points in the
first quadrant of the plane, for each ordered pair of states
$\tu{p,q}$; white points correspond to Attacker's
wins, black points correspond to Defender's
wins.
In our example we have four ordered pairs of states
($(p_1,p_1)$,$(p_1,p_2)$,$(p_2,p_1)$,$(p_2,p_2)$), and the respective four
colourings are depicted in Figure~\ref{fig:solvedsimul}.
E.g., $p_2(1)\simul p_1(n)$ and $p_2(2)\notsimul p_1(n)$
for all $n$, $p_1(0)\notsimul p_2(5)$ and   $p_1(0)\simul p_2(6)$,
etc.

We can easily observe the general monotonicity: if
 $p(m)\simul q(n)$ then $p(m')\simul q(n')$ for all $m'\leq m$ and
 $n'\geq n$.
Our example also suggests that the black-white frontier in each
colouring
is contained in a ``linear belt'', with a rational (or infinite) slope
and a certain width. Such a belt (two parallel lines)
in $C_{\tu{p_1,p_2}}$ in
Figure~\ref{fig:solvedsimul} can be described as follows:
the frontier points, which we can define as
the rightmost black points in each row, have
coordinates $(0,6), (0,7), (1,8), (1,9), (2,10), (2,11), \cdots$ and
are contained in a belt with the slope $\frac{2}{1}$ and the vertical
width $1$.
(A more general form is depicted in Figure~\ref{fig:belttheoremclaim}.)

Contemplating a bit,
it is not difficult to get an intuition captured by a
``belt theorem'', claiming that also in any general case of (succinct)
one-counter nets the frontier in each plane is contained in a linear
belt; moreover, it is intuitively obvious that each frontier is
periodic, from some row onwards.
(In this paper, we also show that the periods can be, and are at most,
double-exponential.)
By the previous figures and discussions, one can
also easily get an intuition that deciding the countdown games,
even in their ($\EXPSPACE$-complete) existential form, could be reduced
to deciding the simulation preorder on (succinct)
one-counter nets; finding a solution like the one depicted in
Figure~\ref{fig:excountdowngame} seems intuitively simpler than
finding a solution like the one   depicted in
Figure~\ref{fig:solvedsimul}.
This is confirmed in this paper, by a
 reduction based on a ``defender-choice technique'' that in particular
 enables to mimic Adam's choices in a countdown game by Defender's
 choices in the corresponding simulation game.

The above mentioned belt theorem
is important for the upper
bounds, namely for showing that deciding simulation preorder is in \PSPACE
for one-counter nets where the counter changes are presented in unary,
and in \EXPSPACE
for succinct one-counter nets.
Proving the belt theorem had
turned out surprisingly difficult; we also contribute to this topic
here, as we explain below.
The (non)succinctness of the one-counter
nets plays no important role in this discussion, so we restrict ourselves to the
cases in which the counter can change by at most $1$ in one move.
The \emph{qualitative form} of the belt theorem (claiming just the existence of
belts, not caring about their slopes, widths, and positions)
followed from~\cite{DBLP:conf/concur/AbdullaC98} where an involved
mechanism of two-player games was used; this qualitative result was shown
in~\cite{DBLP:conf/sofsem/JancarMS99} by another technique, based
rather on ``geometric'' ideas. The \emph{quantitative form}, stating that the
linear belts, assumed to start in the origin $(0,0)$,
have the slopes and widths that can be presented as
(fractions of) numbers with polynomially bounded values, was shown
in~\cite{DBLP:journals/corr/HofmanLMT16}, by enhancing the technique
of games from~\cite{DBLP:conf/concur/AbdullaC98}; this is the crux
of the \PSPACE-membership of simulation preorder for
OCN~\cite{DBLP:journals/corr/HofmanLMT16} (which also yields the
\EXPSPACE-membership for succinct OCN).

In this paper we give a new self-contained proof for both the qualitative and quantitative
versions of the belt theorem.
One important new ingredient is a simple observation that we call
a \emph{black-white vector travel}.
We describe it here, using Figure~\ref{fig:neighbourvector},
but the description can be safely skipped if the
reader finds it too technical; it is later captured by
Proposition~\ref{prop:neighbour} and Corollary~\ref{cor:neighbour}.
\begin{figure}[h]
  \centering
  \begin{subfigure}[t]{0.6\textwidth}
  \centering
  \begin{tikzpicture}[scale=0.8,every node/.style={scale=1}]
  \draw[->] (0,0) -- (0,6);
  \draw[->] (0,0) -- (6,0);

  \drawvec{(3,3)}{black}{(4.5,4.5)}{white}{solid}
  \draw[dotted] (3, 3) -- (3,0);
  \draw[dotted] (3, 3) -- (0, 3);

  \drawvec{(2.5,3.5)}{black}{(4,5)}{white}{solid}
  \draw[dotted] (2.5,3.5) -- (2.5,0);
  \draw[dotted] (2.5,3.5) -- (0,3.5);

  \draw[<->] (2.5,-0.25) -- (3,-0.25);
  \node[anchor=north, scale=0.8] at (2.75,-0.25) {$1$};

  \draw[<->] (-0.25,3) -- (-0.25,3.5);
  \node[anchor=east, scale=0.8] at (-0.25,3.25) {$1$};

  \node[anchor=south east] at (3.25,4.15) {$v_1$};
  \node[anchor=north] at (3.95,3.75) {$v_0$};
\end{tikzpicture}
	\caption{A neighbour smaller-rank black-white vector.}
  \label{fig:neighbourvector1}
  \end{subfigure}%
  \begin{subfigure}[t]{0.4\textwidth}
    \centering
  \begin{tikzpicture}[scale=0.8,every node/.style={scale=1}]
  \draw[->] (0,0) -- (0,6);
  \draw[->] (0,0) -- (6,0);

  \drawvec{(2.5,3.5)}{black}{(4,5)}{white}{solid}
  \drawvec{(3,3)}{black}{(4.5,4.5)}{white}{solid}
  \drawvec{(0,0.5)}{black}{(1.5,2)}{white}{solid}

  \draw[->,dotted] (3,3) -- (2.55, 3.45); 

  \draw[->, dotted]
  (2.5,3.5) .. controls (2,4) and (1,4) .. (0.04,0.55);
  
  \node[anchor=south east] at (3.25,4.15) {$v_1$};
  \node[anchor=north] at (3.95,3.75) {$v_0$};
  \node[anchor=north] at (0.95,1.25) {$v_k$};

  \node[anchor=north, scale=0.8] at (2.75,-0.25) {\phantom{$1$}};
\end{tikzpicture}
  \caption{A vector travel.}
  \label{fig:neighbourvector2}
\end{subfigure}
\caption{Black-white vector travel.}
\label{fig:neighbourvector}
\end{figure}
For any vector $v_0$ with a positive slope and with a black start and
a white end in some colouring $C_{\tu{p_0,q_0}}$,
 like the vector $v_0$ in Figure~\ref{fig:neighbourvector},
there are vectors $v_1,v_2,\dots,v_k$ of the same size and slope as
$v_0$, each $v_i$ being a black-white vector in some colouring $C_{\tu{p_i,q_i}}$,
such that
\begin{itemize}
	\item
		for all $i\in\{1,2,\dots,k\}$,
$v_i$ is a neighbour of $v_{i-1}$, i.e.,
the $x$-coordinates of the starts of $v_i$ and $v_{i-1}$ differ by at most $1$, and
the same constraint holds for their $y$-coordinates; and
\item
the start of $v_k$ is on the vertical axis.
\end{itemize}
This can be easily verified: in the pair of configurations
corresponding to the white end of $v_0$ Attacker has an optimal
 transition $p_0\gtl{a,z}p_1$,
for which each Defender's response $q_0\gtl{a,z'}q$
establishes a ``white'' pair with a smaller rank, i.e.,
closer
to Attacker's final win. Attacker can perform the transition
 $p_0\gtl{a,z}p_1$
also
in the pair of configurations
corresponding to the black start of $v_0$, if this start is not on the
vertical axis (which would entail that Attacker's counter is zero);
there is at least one Defender's response
$q_0\gtl{a,z_1}q_1$ establishing another ``black'' pair.
Hence by changing the $x$-coordinate of $v_0$ by $z$ and its
$y$-coordinate by $z_1$ we get the vector $v_1$ that is black-white in the
colouring $C_{\tu{p_1,q_1}}$. Etc.

This black-white vector travel, together with other enhancements,
allowed us to substantially simplify
the proof of the qualitative belt theorem
from~\cite{DBLP:conf/sofsem/JancarMS99}. Moreover,
the quantitative version can be now derived from the qualitative
version by a few simple observations.

\paragraph{Organization of the paper}
Section~\ref{sec:basicdefinitions} gives the basic definitions.
In Section~\ref{sec:excountgames} we show that the ``existential'' countdown
games are \EXPSPACE-complete (which also yields
an alternative proof for the known \EXPTIME-completeness of countdown
games).
Section~\ref{sec:sketchreducereachsimul} describes the reductions
from reachability games to (bi)simulation relations, in a general
framework and then in the framework of succinct OCN.
Section~\ref{sec:structsimulOCN} contains new proofs for both the
qualitative and quantitative versions of the belt theorem (leading to
the
\PSPACE membership for OCN).
Section~\ref{sec:doubleexpperiod} shows that the period of the belts
in the succinct case can be double-exponential.
We finish with some additional remarks in Section~\ref{sec:addrem}.

%
%

\section{Basic Definitions}\label{sec:basicdefinitions}

By $\Zset$, $\Nat$,
$\Nat_{>0}$
we denote
the sets of integers, of nonnegative integers, and of positive integers, respectively.
We use $[i,j]$, where $i,j\in\Zset$,
for denoting the set $\{i,i{+}1,\dots,j\}$.

\paragraph{Labelled transition systems and (bi)simulations}
A \emph{labelled transition system}, an \emph{LTS} for short, is a
tuple
\[\calL = (S, Act, (\gt{a})_{a\in Act})\]
  where $S$ is the set of \emph{states}, $Act$ is the set of
  \emph{actions},
  and $\mathop{\gt{a}}\subseteq S \times S$ is the set
of  \emph{$a$-transitions} (transitions labelled with $a$), for each
  $a\in Act$.
  We write $s\gta t$ instead of
  $(s,t) \in \mathop{\gta}$.
By $s \gt{a}$ we denote that $a$ is \emph{enabled in} $s$,
\ie, $s \gt{a} t$ for some $t$.

Given $\calL = (S, Act, (\gt{a})_{a\in Act})$,
a relation $R \subseteq S \times S$ is a \emph{simulation}
  if for every $(s,s')\in R$ and every
  $s \gta t$ there is $s' \gta t'$ such that $(t,t')\in R$;
  if, moreover, for every $(s,s')\in R$ and every
  $s' \gta t'$ there is $s \gta t$ such that $(t,t')\in R$, then $R$
is a \emph{bisimulation}.

The union of all simulations (on $S$) is the maximal
  simulation, denoted $\simul$; it is a preorder, called
  \emph{simulation preorder}.
 The union of all bisimulations is the maximal
 bisimulation, denoted $\bisim$; it is an equivalence, called
 \emph{bisimulation equivalence} (or \emph{bisimilarity}).
 We obviously have
 $\bisim\subseteq\simul$.

We can write
 $s_1\simul s_2$ or $s_1\bisim s_2$ also for states $s_1$,
 $s_2$ from different  LTSs $\calL_1$, $\calL_2$, in which case
 the LTS arising by the disjoint union of $\calL_1$ and $\calL_2$ is
 (implicitly) referred to.

It is useful to think in terms of two-player turn-based games,
played
by Attacker and Defender (or Spoiler and Duplicator).
A round
of the \emph{simulation game} from a (current) pair $(s,s')$
proceeds as follows:
Attacker chooses
a transition $s\gt{a}t$, and Defender responds with some
$s'\gt{a}t'$ (for the action $a$ chosen by Attacker); the play then
continues with another round, now from the current pair  $(t,t')$.
If a player has no legal move in a round, then the other player wins;
infinite plays are deemed to be Defender's wins.
It is standard that $s\notsimul s'$ iff Attacker has a winning
strategy from $(s,s')$.

The \emph{bisimulation game} is analogous, but
in any round starting from  $(s,s')$
Attacker can choose to
play  $s\gt{a}t$ or $s'\gt{a}t'$, and Defender has to respond
with some $s'\gt{a}t'$ or $s\gt{a}t$, respectively.
Here we have $s\notbisim s'$ iff Attacker has a winning
strategy from $(s,s')$.

\paragraph{Stratified simulation, and ranks of pairs of states}
Given
$\calL = (S, Act, (\gt{a})_{a\in Act})$,
we use (transfinite) induction to define the relations $\simul_{\lambda}$
where $\lambda$ ranges over the class $Ord$ of ordinals.
We put $\simul_0=S\times S$. For $\lambda>0$ we have
$s\simul_{\lambda}s'$ if for each transition $s\gt{a}t$
and each $\lambda'<\lambda$ there is a
transition $s'\gt{a}t'$ where $t\simul_{\lambda'}t'$.

We note that $\simul_{\lambda'}\supseteq \simul_{\lambda}$ when
$\lambda'\leq \lambda$, and that $\simul=\bigcap_{\lambda\in
Ord}\simul_{\lambda}$. For each pair $(s,s')\not\in\simul$, we define
its \emph{rank} $\rank(s,s')$ as the least ordinal $\lambda$
such that $s\notsimul_{\lambda}s'$.
We note in particular that $\rank(s,s')=1$ iff $s$ enables an action $a$ (i.e.,
$s\gt{a}$) that is not enabled in $s'$ (i.e., $s'\not\gt{a}$).

\medskip

\begin{rem}
We use such a general definition for the purpose of the general
reduction presented in Section~\ref{sec:sketchreducereachsimul}.
Otherwise   we
consider just (special cases of) LTSs that
are \emph{image-finite} (i.e., in which the sets
$\{t\mid s\gt{a}t\}$ are finite for all $s\in S$, $a\in Act$); in such
systems
we have $\simul=\bigcap_{i\in\Nat}\simul_i$ and
$\rank(s,s')\in\Nat$ for each $(s,s')\not\in \simul$.
\\
We could define the analogous concepts for bisimulation equivalence as
well.
\end{rem}

\paragraph{One-counter nets (OCNs and SOCNs), and their
associated LTSs}
A \emph{labelled one-counter net},
or just a \emph{one-counter net} or even
just an \emph{OCN} for short,
is a triple
\[ \calN=(Q,Act,\delta), \]
where $Q$ is the finite set of \emph{control states}, $Act$ the finite
set of \emph{actions},
and $\delta \subseteq Q \times Act\times \{{-}1,0,{+}1\} \times Q$ is
	the
	set of (\emph{labelled transition}) \emph{rules}.
By allowing $\delta$  to be
a finite subset of $Q \times Act\times \Zset \times Q$,
and presenting $z\in\Zset$ in the rules $(q,a,z,q')$
in binary, we
get a \emph{succinct one-counter net}, or a \emph{SOCN} for short.
A rule $(q,a,z,q')$ is usually presented as
$q\gtl{a,z}q'$.

Each OCN or SOCN $\calN=(Q,Act,\delta)$  has the \emph{associated LTS}
\begin{equation}\label{eq:ltsfornet}
\calL_\calN=(Q\times
\Nat,Act,(\gt{a})_{a\in Act})
\end{equation}
where $(q,m)\gtl{a}(q',n)$ iff $q\gtl{a,n-m}q'$ is a rule in $\delta$.
We often write a state $(q,m)$, which is also called a \emph{configuration},
in the form $q(m)$, and we view  $m$ as a value of a nonnegative counter.
A rule  $q\gtl{a,z}q'$ thus induces transitions $q(m)\gt{a}q'(m{+}z)$ for
all $m\geq \max\{0,{-}z\}$.

We remark that \emph{one-counter automata} extend one-counter nets
by the ability to test zero, i.e., by transitions that are enabled only if the counter value is zero.

\paragraph{Reachability games (r-games), winning areas, ranks of states}
We are interested in reachability games played in LTSs
	associated with (succinct) one-counter nets, but we first define the
	respective notions generally.

By a \emph{reachability game}, or an \emph{r-game} for short, we mean
a tuple
\[\calG=(V,V_\exists, \gt{}, \calT),\]
where $V$ is the set of \emph{states} (or \emph{vertices}),
$V_\exists\subseteq V$ is the set of \emph{Eve's states},
$\mathop{\gt{}}\subseteq {V\times V}$ is the \emph{transition relation}
(or the set of \emph{transitions}), and $\calT\subseteq V$ is the set
of \emph{target states}.
By  \emph{Adam's states} we mean  the elements of
$V_\forall=V\smallsetminus V_\exists$.

\emph{Eve's winning area}
is $\winareaE=\bigcup_{\lambda\in Ord}W_\lambda$, for $Ord$ being
the class of ordinals, where the sets $W_\lambda\subseteq V$ are defined
inductively as follows.

We put $W_0=\calT$; for $\lambda>0$ we put
$W_{<\lambda}=\bigcup_{\lambda'<\lambda}W_{\lambda'}$, and we
stipulate:
\begin{enumerate}[a)]
	\item
if $s\not\in W_{<\lambda}$, $s\in V_\exists$, and
$s\gt{}\bar{s}$ for some $\bar{s}\in W_{<\lambda}$, then $s\in W_{\lambda}$;
	\item
if $s\not\in W_{<\lambda}$, $s\in V_\forall$,
and we have $\emptyset\neq \{\bar{s}\mid s\gt{}\bar{s}\}\subseteq
W_{<\lambda}$,
then $s\in W_{\lambda}$.
\end{enumerate}
(If (a) applies, then $\lambda$ is surely a successor ordinal.)

For each $s\in\winareaE$, by $\rank(s)$ we denote (the unique)
$\lambda$ such that
$s\in W_\lambda$. A transition $s\gt{}\bar{s}$ is
\emph{rank-reducing} if $\rank({s})>\rank(\bar{s})$.
We note that for any $s\in\winareaE$ with $\rank(s)>0$ we have:
if $s\in V_{\exists}$, then there is at least one rank-reducing
transition $s\gt{}\bar{s}$ (in fact, $\rank(s)=\rank(\bar{s}){+}1$ in
this case); if $s\in V_{\forall}$, then there is at least one
transition $s\gt{}\bar{s}$ and all such transitions are rank-reducing.
This entails that $\winareaE$ is the set of states from which Eve has a
strategy that guarantees reaching (some state in) $\calT$ when Eve
is choosing a next transition in Eve's states
and Adam is choosing a next transition in Adam's states.

\medskip

\begin{rem}
We are primarily interested in the games that have  (at most)
countably many states and are finitely branching (the sets
$\{\bar{s}\mid s\gt{}\bar{s}\}$ are finite for all $s$). In such cases
we have $\rank(s)\in\Nat$ for each $s\in\winareaE$.
We have again introduced the general definition for the purpose of the
reduction in Section~\ref{sec:sketchreducereachsimul}.
\end{rem}

\paragraph{Reachability games on succinct one-counter nets}
We now define specific r-games,
presented by SOCNs
 with
partitioned
control-state sets; these succinct one-counter nets are
\emph{unlabelled}, which means that the set of actions can be always deemed to be
a singleton.

By a \emph{succinct one-counter net reachability game}, a \emph{socn-r-game} for short, we
mean a tuple
\[ \calN=(Q,Q_{\exists},\delta,p_{win}) \]
where $Q$ is the finite set of \emph{(control) states},
$Q_{\exists}\subseteq Q$ is the set of \emph{Eve's (control) states},
$p_{win}\in Q$ is the \emph{target (control) state}, and
$\delta\subseteq Q\times\Zset\times Q$ is the finite set of
\emph{(transition) rules}.
We often present a rule $(q,z,q')\in\delta$ as
$q\gt{z}q'$.
By \emph{Adam's (control) states} we
mean the elements of $Q_{\forall}=Q\smallsetminus Q_{\exists}$.
A socn-r-game $\calN=(Q,Q_{\exists},\delta,p_{win})$ has the
\emph{associated r-game}
\begin{equation}\label{eq:gamefornet}
\calG_\calN=(Q\times
\Nat,Q_{\exists}\times\Nat,\gt{},\{(\Pwin,0)\})
\end{equation}
where $(q,m)\gt{}(q',n)$ iff $q\gtl{n-m}q'$ is a rule (in $\delta$).
We often write $q(m)$ instead of $(q,m)$ for states of $\calG_\calN$.

We define the problem \SOCNRG (to decide succinct
one-counter net r-games) as follows:

\begin{Problem}
\ProblemName \SOCNRG
\Instance a socn-r-game $\calN$
(with integers $z$ in rules $q\gt{z}q'$ written in binary), \\
\hspace*{2.2em} and a control state $p_0$.
\Question is $p_0(0)\in\winareaE$ in the game $\calG_\calN$\,?
\end{Problem}

\medskip

\begin{rem}
We have defined the target states (in $\calG_\calN$) to be the
singleton set $\{\Pwin(0)\}$.
There are other natural variants
(e.g., one in~\cite{DBLP:conf/rp/Hunter15} defines the target set
 $\{p(0)\mid p\neq p_0\}$)
that can be easily shown to be essentially equivalent.
\end{rem}

\medskip

The \EXPSPACE-hardness of \SOCNRG
was announced in~\cite{DBLP:conf/rp/Hunter15}, where
an idea of a proof
is sketched, also using a reference
to an involved result~\cite{DBLP:conf/icalp/GollerHOW10}
(which is further discussed in Section~\ref{sec:addrem}).
In Section~\ref{sec:excountgames} we give a direct self-contained proof that does not rely
on~\cite{DBLP:conf/rp/Hunter15}
or involved techniques from~\cite{DBLP:conf/icalp/GollerHOW10}, and that
even shows
that \SOCNRG is \EXPSPACE-hard already in the special
case that slightly generalizes the countdown games
from~\cite{DBLP:journals/lmcs/JurdzinskiSL08}.
(The \EXPSPACE-membership follows from~\cite{DBLP:conf/rp/Hunter15},
but we add a short proof to be self-contained.)

\paragraph{Countdown games}
We define a~\emph{countdown game}
as a~socn-r-game $\calN=(Q,Q_{\exists},\delta,p_{win})$, where in every
rule $q \gt{z} q'$ in~$\delta$ we have~$z < 0$.
The problem \CG is defined as follows:

\begin{Problem}
	\ProblemName \CG (countdown games)
\Instance a~countdown game $\calN$ (with integers in
   rules written in binary), \\
   \hspace*{2.4em}and an initial configuration
   $p_0(n_0)$ where $n_0\in\Nat$ ($n_0$ in binary).
\Question is $p_0(n_0) \in\winareaE$\,?
\end{Problem}

The problem \CG (in an equivalent form) was shown \EXPTIME-complete
in~\cite{DBLP:journals/lmcs/JurdzinskiSL08}.
Here we define an existential version, \ie the problem \ECG:

\begin{Problem}
\ProblemName \ECG (existential countdown games)
\Instance a~countdown game $\calN$ and a control state $p_0$.
\Question is there some $n\in\Nat$ such that $p_0(n) \in\winareaE$\,?
\end{Problem}

We note that \ECG can be viewed as a~subproblem of \SOCNRG:
given an instance of \ECG, it suffices to add a~fresh
Eve's state $p'_0$  and rules $p'_0 \gt{1} p'_0$, $p'_0\gt{0}p_0$;
the question then is if $p'_0(0)\in\winareaE$.

%
%

\section{EXPSPACE-Completeness
of Existential Countdown Games}
\label{sec:excountgames}

In this section we prove the following new theorem.

\begin{thm}\label{th:ecgp_expspcomplete}
\ECG (existential countdown game) is \EXPSPACE-complete.
\end{thm}
Our \EXPSPACE-hardness proof of ECG
is, in fact, a particular instance of a simple general method.
A slight modification also yields a proof of
\EXPTIME-hardness of countdown games that is an alternative to the
proof in~\cite{DBLP:journals/lmcs/JurdzinskiSL08}
and in~\cite{DBLP:journals/ipl/Kiefer13}.

\begin{thm}\label{th:cg_exptimecomplete}
	\cite{DBLP:journals/lmcs/JurdzinskiSL08}
\CG is \EXPTIME-complete.
\end{thm}

In the rest of this section we present
the proofs of Theorems~\ref{th:ecgp_expspcomplete}
and~\ref{th:cg_exptimecomplete} together, since they
only differ by a small detail.
We start with the upper bounds since these are obvious.

\subsection{ECG is in EXPSPACE (and CG in
EXPTIME)}\label{subsec:ECGinEXPS}

Let us consider an \ECG-instance
$\calN=(Q,Q_{\exists},\delta,p_{win})$, $p_0$.
By $\textsc{m}$ we denote
the maximum value by which the counter can be decremented in one
step (i.e., $\textsc{m}=\max\,\{\,|z|\,;$ there is some $q \gt{z} q'$
in~$\delta \,\}$); the value $\textsc{m}$ is at most exponential
in the size of the instance $\calN, p_0$.

We can stepwise construct
 $W(0), W(1), W(2), \dots$ where $W(j)=\{q\in Q\mid q(j)\in
 \winareaE\}$;
the \ECG-instance $\calN, p_0$ is positive iff $p_0\in W(j)$ for some
$j$.
We have $W(0)=\{p_{win}\}$, and
for determining $W(n)$ ($n\geq 1$)
it suffices to know the segment
$W(n{-}\textsc{m'})$, $W(n{-}\textsc{m'}{+}1)$, $\dots, W(n{-}1)$
where $\textsc{m'}=\min\{\textsc{m},n\}$. Hence, during the
construction of $W(j)$, $j=0,1,2,\dots$,
it suffices to remember just the segment
$W(j{-}\textsc{m'})$, $W(j{-}\textsc{m'}{+}1)$, $\dots, W(j{-}1)$
(where $\textsc{m'}\leq \textsc{m}$). Obviously, if for some $j,j'$, where $\textsc{m}\leq j < j'$, the
segments
$W(j{-}\textsc{m}),\,\dots, W(j{-}1)$ and
$W(j'{-}\textsc{m}),\,\dots, W(j'{-}1)$ are the same
(\ie,~if $W(j{-}k) = W(j'{-}k)$ for each $k=1,2,\ldots,\textsc{m}$),
then also $W(j{+}k) = W(j'{+}k)$ for each $k\in\Nat$, so the sequence
repeats itself with a~period~$j'-j$.
By the pigeonhole principle, this surely happens for some
$j,j'$ such that $j<j'\leq \textsc{m}+2^{|Q|\cdot\textsc{m}}$
because there are at most $2^{|Q|\cdot\textsc{m}}$ segments of
length~$\textsc{m}$.
This means that in the algorithm that checks whether $p_0\in W(j)$ for
some $j\in\Nat$, the computation can be stopped after
constructing $W(\textsc{m}+2^{|Q|\cdot\textsc{m}})$.
The size of a~binary counter serving to count till the
(at most double-exponential) value $\textsc{m}+2^{|Q|\cdot\textsc{m}}$
is at most exponential.
Therefore \ECG belongs to \EXPSPACE.

It is also clear that \CG is in \EXPTIME, since for the instance
$\calN, p_0(n_0)$ we can simply construct
$W(0), W(1), W(2), \dots, W(n_0)$.

\subsection{ECG is EXPSPACE-hard (and CG EXPTIME-hard)}

In principle, we use a~``master'' reduction.
We fix an arbitrary language~$L$
in \EXPSPACE, in an alphabet $\Sigma$ (hence $L\subseteq\Sigma^*$),
decided by a~(deterministic) Turing machine~$\calM$
in space $2^{p(n)}$ for a fixed polynomial~$p$.
For any word~$w\in\Sigma^*$, $\abs{w}=n$,
there is the respective computation of $\calM$ using
at most $m=2^{p(n)}$ tape cells, which is accepting iff $w\in L$.
Our aim is to show a construction of a~countdown game $\NMwm$,
with a specified control state~$p_0$, such that
there is $k\in\Nat$ for which $p_0(k)\in\winareaE$ if, and only if,
$\calM$ accepts~$w$. The construction of  $\NMwm$ will be polynomial,
in the size $n=\abs{w}$; this will establish that $\ECG$ is
$\EXPSPACE$-hard.
(In fact, this polynomial construction easily yields a
logspace-reduction, but this detail is unimportant at our level of
discussion.)

In fact, the same construction of $\NMwm$ will also show
$\EXPTIME$-hardness of \CG. In this case we assume that
$L$ is decided by a Turing machine~$\calM$
in time (and thus also space) $2^{p(n)}$, and we construct a concrete
exponential value $n_0$ guaranteeing that
$p_0(n_0)\in\winareaE$ iff
$\calM$ accepts~$w$.

\begin{figure}[h!]
  \begin{center}
  {\tikzpicture[x=+4143sp, y=+4143sp]
\clip(4440,-6522) rectangle (8922,-965);
\tikzset{inner sep=+0pt, outer sep=+0pt}
\pgfsetlinewidth{30000sp}
\pgfsetstrokecolor{black}
\draw (4770,-1170)--(8910,-1170);
\draw (4770,-1350)--(8910,-1350);
\draw (8730,-990)--(8730,-6390);
\draw (8550,-990)--(8550,-6390);
\draw (8370,-990)--(8370,-6390);
\draw (5130,-990)--(5130,-6390);
\draw (5490,-990)--(5490,-6390);
\draw (4950,-990)--(4950,-6390);
\draw (4770,-6030)--(8910,-6030);
\draw (4770,-6210)--(8910,-6210);
\draw (4770,-5850)--(8910,-5850);
\draw (4770,-5670)--(8910,-5670);
\draw (6120,-5625)--(6120,-6390);
\draw (6300,-5625)--(6300,-6390);
\draw (6480,-5625)--(6480,-6390);
\draw (6660,-5625)--(6660,-6390);
\draw (6840,-5625)--(6840,-6390);
\draw (7200,-4230)--(7200,-4410)--(7740,-4410)--(7740,-4230)--(7200,-4230);
\draw (7380,-4410)--(7380,-4050)--(7560,-4050)--(7560,-4410);
\draw (4770,-990)--(4770,-6390)--(8910,-6390)--(8910,-990)--(4770,-990);
\draw (5310,-990)--(5310,-6390);
\draw (8550,-2250)--(8910,-2250);
\draw (8550,-2070)--(8910,-2070);
\draw (8730,-1890)--(8910,-1890);
\draw (4770,-1890)--(4950,-1890);
\draw (4770,-2070)--(4950,-2070);
\draw (4770,-3060)--(4770,-3240)--(5130,-3240)--(5130,-3060)--(4770,-3060);
\draw (4770,-2880)--(4950,-2880);
\draw (8730,-3420)--(8910,-3420);
\draw (8730,-3240)--(8910,-3240);
\pgfsetfillcolor{black}
\pgftext[base,at=\pgfqpointxy{7650}{-4365}] {\fontsize{8}{9.6}\usefont{T1}{phv}{m}{n}$\beta_3$}
\pgftext[base,at=\pgfqpointxy{5040}{-3180}] {\fontsize{8}{9.6}\usefont{T1}{phv}{m}{n}$\beta_3'$}
\pgftext[base,at=\pgfqpointxy{4857}{-2010}] {\fontsize{8}{9.6}\usefont{T1}{phv}{m}{n}$\beta_3''$}
\pgftext[base,at=\pgfqpointxy{7470}{-4365}] {\fontsize{8}{9.6}\usefont{T1}{phv}{m}{n}$\beta_2$}
\pgftext[base,at=\pgfqpointxy{4860}{-3186}] {\fontsize{8}{9.6}\usefont{T1}{phv}{m}{n}$\beta_2'$}
\pgftext[base,at=\pgfqpointxy{8817}{-2190}] {\fontsize{8}{9.6}\usefont{T1}{phv}{m}{n}$\beta_2''$}
\pgftext[base,at=\pgfqpointxy{4860}{-1305}] {\fontsize{8}{9.6}\usefont{T1}{phv}{m}{n}$\blank$}
\pgftext[base,at=\pgfqpointxy{8640}{-6345}] {\fontsize{8}{9.6}\usefont{T1}{phv}{m}{n}$\blank$}
\pgftext[base,at=\pgfqpointxy{8460}{-6345}] {\fontsize{8}{9.6}\usefont{T1}{phv}{m}{n}$\blank$}
\pgftext[base,at=\pgfqpointxy{7470}{-4185}] {\fontsize{8}{9.6}\usefont{T1}{phv}{m}{n}$\beta$}
\pgftext[base,at=\pgfqpointxy{7290}{-4365}] {\fontsize{8}{9.6}\usefont{T1}{phv}{m}{n}$\beta_1$}
\pgftext[base,at=\pgfqpointxy{6390}{-6345}] {\fontsize{8}{9.6}\usefont{T1}{phv}{m}{n}$\blank$}
\pgftext[base,at=\pgfqpointxy{6570}{-6345}] {\fontsize{8}{9.6}\usefont{T1}{phv}{m}{n}$\blank$}
\pgftext[base,at=\pgfqpointxy{6750}{-6345}] {\fontsize{8}{9.6}\usefont{T1}{phv}{m}{n}$\blank$}
\pgftext[base,at=\pgfqpointxy{6207}{-6324}] {\fontsize{8}{9.6}\usefont{T1}{phv}{m}{n}$a_n$}
\pgftext[base,at=\pgfqpointxy{8820}{-6345}] {\fontsize{8}{9.6}\usefont{T1}{phv}{m}{n}$\blank$}
\pgftext[base,at=\pgfqpointxy{5219}{-6330}] {\fontsize{8}{9.6}\usefont{T1}{phv}{m}{n}$a_2$}
\pgftext[base,at=\pgfqpointxy{5409}{-6327}] {\fontsize{8}{9.6}\usefont{T1}{phv}{m}{n}$a_3$}
\pgftext[base,at=\pgfqpointxy{4860}{-6345}] {\fontsize{8}{9.6}\usefont{T1}{phv}{m}{n}$\blank$}
\pgftext[base,at=\pgfqpointxy{4860}{-6165}] {\fontsize{8}{9.6}\usefont{T1}{phv}{m}{n}$\blank$}
\pgftext[base,at=\pgfqpointxy{4860}{-5985}] {\fontsize{8}{9.6}\usefont{T1}{phv}{m}{n}$\blank$}
\pgftext[base,at=\pgfqpointxy{4860}{-5805}] {\fontsize{8}{9.6}\usefont{T1}{phv}{m}{n}$\blank$}
\pgftext[base,at=\pgfqpointxy{8820}{-2025}] {\fontsize{8}{9.6}\usefont{T1}{phv}{m}{n}$\beta''$}
\pgftext[base,at=\pgfqpointxy{4860}{-3015}] {\fontsize{8}{9.6}\usefont{T1}{phv}{m}{n}$\beta'$}
\pgftext[base,at=\pgfqpointxy{8817}{-3363}] {\fontsize{8}{9.6}\usefont{T1}{phv}{m}{n}$\beta_1'$}
\pgftext[base,at=\pgfqpointxy{8637}{-2193}] {\fontsize{8}{9.6}\usefont{T1}{phv}{m}{n}$\beta_1''$}
\pgftext[base,at=\pgfqpointxy{5040}{-6360}] {\fontsize{6}{7.2}\usefont{T1}{phv}{m}{n}$a_1$}
\pgftext[base,at=\pgfqpointxy{5043}{-6274}] {\fontsize{5}{6}\usefont{T1}{phv}{m}{n}$q_0$}
\pgftext[base,at=\pgfqpointxy{4867}{-1052}] {\fontsize{4}{4.8}\usefont{T1}{phv}{m}{n}$q_{\tiny acc}$}
\pgftext[base,at=\pgfqpointxy{4857}{-1161}] {\fontsize{6}{7.2}\usefont{T1}{phv}{m}{n}$\blank$}
\pgftext[base,at=\pgfqpointxy{5220}{-6480}] {\fontsize{5}{6}\usefont{T1}{phv}{m}{n}2}
\pgftext[base,at=\pgfqpointxy{5400}{-6480}] {\fontsize{5}{6}\usefont{T1}{phv}{m}{n}3}
\pgftext[base,at=\pgfqpointxy{6210}{-6480}] {\fontsize{5}{6}\usefont{T1}{phv}{m}{n}$n$}
\pgftext[base,at=\pgfqpointxy{6390}{-6480}] {\fontsize{5}{6}\usefont{T1}{phv}{m}{n}$n{+}1$}
\pgftext[base,at=\pgfqpointxy{7470}{-6480}] {\fontsize{5}{6}\usefont{T1}{phv}{m}{n}$j$}
\pgftext[base,at=\pgfqpointxy{8820}{-6480}] {\fontsize{5}{6}\usefont{T1}{phv}{m}{n}$m{-}1$}
\pgftext[base,at=\pgfqpointxy{4860}{-6480}] {\fontsize{5}{6}\usefont{T1}{phv}{m}{n}0}
\pgftext[base,at=\pgfqpointxy{5040}{-6480}] {\fontsize{5}{6}\usefont{T1}{phv}{m}{n}1}
\pgftext[base,left,at=\pgfqpointxy{4455}{-1125}] {\fontsize{8}{9.6}\usefont{T1}{phv}{m}{n}$C^w_t$}
\pgftext[base,left,at=\pgfqpointxy{4455}{-1305}] {\fontsize{8}{9.6}\usefont{T1}{phv}{m}{n}$C^w_{t{-}1}$}
\pgftext[base,left,at=\pgfqpointxy{4455}{-6345}] {\fontsize{8}{9.6}\usefont{T1}{phv}{m}{n}$C^w_0$}
\pgftext[base,left,at=\pgfqpointxy{4455}{-6165}] {\fontsize{8}{9.6}\usefont{T1}{phv}{m}{n}$C^w_1$}
\pgftext[base,left,at=\pgfqpointxy{4455}{-5985}] {\fontsize{8}{9.6}\usefont{T1}{phv}{m}{n}$C^w_2$}
\pgftext[base,left,at=\pgfqpointxy{4455}{-5805}] {\fontsize{8}{9.6}\usefont{T1}{phv}{m}{n}$C^w_3$}
\pgftext[base,left,at=\pgfqpointxy{4455}{-4185}] {\fontsize{8}{9.6}\usefont{T1}{phv}{m}{n}$C^w_i$}
\pgftext[base,left,at=\pgfqpointxy{4455}{-4365}] {\fontsize{8}{9.6}\usefont{T1}{phv}{m}{n}$C^w_{i{-}1}$}
\pgfsetdash{{60000sp}{180000sp}}{60000sp}
\draw (7560,-990)--(7560,-6390);
\draw (7380,-990)--(7380,-6390);
\draw (4770,-4230)--(8910,-4230);
\draw (4770,-4050)--(8910,-4050);
\draw (4770,-4410)--(8910,-4410);
\draw (4770,-2250)--(8910,-2250);
\draw (4770,-2070)--(8910,-2070);
\draw (4770,-1890)--(8910,-1890);
\draw (4770,-3420)--(8910,-3420);
\draw (4770,-3240)--(8910,-3240);
\draw (4770,-3060)--(8910,-3060);
\draw (4770,-2880)--(8910,-2880);
\endtikzpicture}%
  \end{center}
  \caption{A~computation table of $\calM$ on a word
	$w=a_1a_2\ldots a_n$ (in the bottom-up fashion).}
  \label{fig:conftable}
\end{figure}

\medskip

\paragraph{Construction informally}
The construction of the countdown game $\NMwm$ elaborates an
idea that is already present in
\cite{DBLP:journals/jacm/ChandraKS81} (in Theorem 3.4) and that was
also used, e.g., in~\cite{DBLP:journals/ipl/JancarS07}.
We first present the construction informally.

Figure~\ref{fig:conftable} presents an accepting computation of $\calM$, on a
word
$w=a_1a_2\ldots a_n$; it starts in the initial control state $q_0$ with the
head scanning $a_1$.  The computation is a sequence of configurations
$C^w_0,C^w_1,\dots,C^w_t$, where $C^w_t$ is accepting (since the control
state is $\qacc$). We assume that $\calM$ never leaves cells $0\ldots m{-}1$
of its tape during the computation. Hence each $C^w_i$ can be
presented
as a word of length~$m$ over
the alphabet~$\Delta=(Q\times\Gamma) \cup \Gamma$ where $Q$ and $\Gamma$ are the
set of control states and the tape alphabet of~$\calM$, respectively;
by $\blank\in\Gamma$ we denote the special blank tape symbol.
We refer to the (bottom-up) presentation of
$C^w_0,C^w_1,\dots,C^w_t$ depicted in Figure~\ref{fig:conftable} as to
a~\emph{computation table}.

Given $m$,
each number $k\in\Nat$ determines the cell $j$
in the ``row'' $i$ (i.e., in the potential $C^w_i$) where $i=k\div m$
($\div$ being integer division)
and $j=k\bmod m$; we refer by $\cell(k)$ to this cell $j$ in the row
$i$.

For $k>m$, if $\cell(k)$ is in the computation table ($k\leq t\cdot m$
in Figure~\ref{fig:conftable}), then
the symbol $\beta$ in $\cell(k)$ in the table is surely
 determined by the symbols $\beta_1,\beta_2,\beta_3$ in
the cells  $\cell(k{-}m{-}1)$,  $\cell(k{-}m)$, $\cell(k{-}m{+}1)$
(and by the transition function of the
respective Turing machine $\calM$); see
Figure~\ref{fig:conftable} for illustration (where also the cases
$\beta'$ and $\beta''$ on the ``border'' are depicted). The transition function of $\calM$ allows
us to define which \emph{triples} $(\beta_1,\beta_2,\beta_3)$ are
\emph{eligible for} $\beta$, i.e., those that can
be in the cells
$\cell(k{-}m{-}1)$,  $\cell(k{-}m)$, $\cell(k{-}m{+}1)$ when $\beta$
is in $\cell(k)$ (these triples are independent of $k$, assuming
$k>m$).

Let us now imagine a game between Eve and Adam
where Eve, given $w$, claims that $w$ is
accepted by $\calM$, in space $m$
(in our case $m=2^{p(|w|)}$ for a respective polynomial $p$).
Eve does not present a respective accepting
computation table
but she starts a play by producing a tape-symbol $x$ and
a~number $k_0\in\Nat$, i.e. sets a counter
to $k_0$,
claiming that $\cell(k_0)$ in the computation table
contains $(\qacc,x)$ (in Figure~\ref{fig:conftable} the correct values
are $k_0=t\cdot m$ and $x=\blank$).
Then the play proceeds as follows.
If Eve claims that $\beta$ is the symbol in $\cell(k)$ for the current
counter value $k$, while also claiming that $k>m$, then she
 decreases the counter by $m{-}2$ and produces a triple
$(\beta_1,\beta_2,\beta_3)$ that is eligible for $\beta$,
thus claiming that $\cell(k{-}m{-}1), \cell(k{-}m),\cell(k{-}m{+}1)$
contain $\beta_1,\beta_2,\beta_3$, respectively.
Adam then decreases the counter either by $3$, asking to verify the
claim that
$\cell(k{-}m{-}1)$ contains $\beta_1$, or by $2$, asking to verify
that $\cell(k{-}m)$ contains $\beta_2$, or by $1$, asking to verify
that  $\cell(k{-}m{+}1)$ contains $\beta_3$. Eve then again produces
an eligible triple for the current symbol, etc., until claiming that
the counter value is $k\leq m$ (which can be contradicted by Adam
when he is able to decrease the counter by $m{+}1$). The last phase
just implements checking if the symbol claimed for $\cell(k)$
corresponds to the initial configuration.

It is clear that Eve has a winning strategy in this game if
$w$~is accepted by $\calM$ (in space $m$). If $w$ is not accepted,
then Eve's first claim does not correspond to the real computation
table. Moreover, if a claim by Eve is incorrect, then at least
one claim for any respective
eligible triple is also incorrect (as can be easily
checked), hence Adam can be always asking to
verify incorrect claims, which is revealed when the consistency with
the initial
configuration is verified in the end. Hence Eve has
no winning strategy if $w$ is not accepted by $\calM$.

In the formal construction presented below
we proceed by introducing an intermediate auxiliary problem (in
two versions) that
allows us to avoid some technicalities in the construction
of countdown games $\NMwm$. Roughly speaking, instead of the
computation of $\calM$ on $w$ we consider the computation of $\calM_w$
on the empty word, which first writes $w$ on the tape and then
invokes $\calM$;
checking
the consistency with the (empty) initial configuration is then
technically easier
to handle in the constructed countdown game.

\paragraph{Construction formally}
Now we formalize the above idea, using the announced intermediate
problem.

By a~\emph{sequence description} we mean a~tuple $\calD=(\Delta, D, m)$,
where $\Delta$ is its~finite alphabet, always containing two special symbols
$\hash$ and $\blank$
(and other symbols),
$D: \Delta^3 \to \Delta$ is its~\emph{description function},
and $m\geq 3$ is its \emph{initial length}.
The sequence description $\calD=(\Delta,D,m)$ defines the infinite
sequence $\calSD$ in $\Delta^{\omega}$, i.e.
the function  $\calSD: \Nat \to \Delta$, that is defined
 inductively as follows:
\begin{itemize}
\item $\calSD(0)=\hash$,
\item $\calSD(1)=\calSD(2)=\cdots=\calSD(m)=\blank$,
\item for $i > m$ we have $\calSD(i)=D(\beta_1,\beta_2,\beta_3)$
	where $\beta_1=\calSD(i{-}m{-}1)$,
		$\beta_2=\calSD(i{-}m)$, and $\beta_3=\calSD(i{-}m{+}1)$.
\end{itemize}
The two versions of the announced intermediate problem
are defined as follows:
\begin{Problem}
\ProblemName \GenSeqProblem (sequence problem)
\Instance A~sequence description~$\calD=(\Delta, D, m)$, $n_0\in\Nat$,
   and $\beta_0\in\Delta$ (with $m$ and $n_0$ written in binary).
\Question Is $\calSD(n_0)=\beta_0$\,?
\end{Problem}

\begin{Problem}
\ProblemName \ExGenSeqProblem (existential sequence problem)
\Instance A~sequence description~$\calD=(\Delta, D, m)$, and
   $\beta_0\in\Delta$ (with $m$ written in binary).
\Question Is there $i\in\Nat$ such that $\calSD(i)=\beta_0$\,?
\end{Problem}

Our informal discussion around Figure~\ref{fig:conftable}
(including the remark on $\calM_w$ starting with the empty tape)
makes (almost) clear that
\begin{itemize}
	\item
\GenSeqProblem is \EXPSPACE-complete, and
\item
 \ExGenSeqProblem is \EXPTIME-complete.
\end{itemize}

\begin{rem}
In fact, we would get the same complexity results even if
we restricted $D$ in $\calD=(\Delta, D, m)$ to $D:\Delta^2\to \Delta$
and defined $\calSD(i)=D(\calSD(i{-}m{-}1,\calSD(i{-}m))$ for $i>m$.
(We would simulate $\calM$ by a Turing machine $\calM'$ that only
moves to the right in each step, while working on a circular tape of
length $m$.) But this technical enhancement would not simplify our construction of the
countdown games $\NMwm$, in fact.
\end{rem}

Formally it suffices for us to claim just the lower bounds:

\begin{prop}
\label{prop:EGSP_exspace_complete}
\ExGenSeqProblem is
\EXPSPACE-hard
	and
\GenSeqProblem is
\EXPTIME-hard.
\end{prop}
\begin{proof}
To show \EXPSPACE-hardness
	of \ExGenSeqProblem ,
we	assume an arbitrary
fixed language~$L\subseteq\Sigma^*$ in \EXPSPACE,
decided by a Turing machine~$\calM$ in space  $2^{p(n)}$ for a
	polynomial $p$.

Using a standard notation,
	let $\calM=(Q,\Sigma,\Gamma,\delta,q_0,\{\qacc,\qrej\})$
where $\Sigma\subseteq \Gamma$,
$\blank\in\Gamma\smallsetminus\Sigma$, $\hash\not\in\Gamma$,
and  $\delta: Q\times\Gamma\rightarrow
  Q\times\Gamma\times\{{-}1,0,{+}1\}$ is the transition function of
	$\calM$, satisfying $\delta(\qacc,x)=(\qacc,x,0)$ and
	$\delta(\qrej,x)=(\qrej,x,0)$ (hence an accepting or
	rejecting configuration is formally viewed as repeated forever).
Moreover, w.l.o.g. we assume that
the computation of $\calM$ on $w=a_1a_2\cdots a_n\in\Sigma^*$
starts with $w$ written in tape cells $1,2,\dots,n$ with the head
	scanning the cell~$1$ (the control state being $q_0$),
	the computation never leaves the cells $0,1,\dots,m{-}1$ for
 $m=2^{p(n)}$, never rewrites $\blank$ in the cell $0$,
	and the state
	$\qacc, \qrej$ can only be entered when the head is scanning
	the cell $0$ (as is also depicted in
	Figure~\ref{fig:conftable});
	we also assume that $p$ is such that $m=2^{p(n)}$ satisfies
	$m>n$
	and $m\geq 3$.

Given $w\in\Sigma^*$, we now aim to show a polynomial construction
	of $\calD_w=(\Delta_w,D_w,m)$, where
	$m=2^{p(|w|)}$, such that
	$w\in L$ iff there is $i\in\Nat$ such that
	$\calSDw(i)=(\qacc,\blank)$. (Hence $w$ is reduced to the
	$\ExGenSeqProblem$-instance $\calD_w$, $\beta_0$ where
	$\beta_0=(\qacc,\blank)$.)

	As already suggested in the previous discussion, for
	$w=a_1a_2\cdots a_n$
	we first construct a Turing machine
	$M_w$ that starts with the empty tape while scanning the cell $0$
	in its initial state $q'_0$, then by moving to the right it
	writes $w=a_1a_2\cdots a_n$ in
	the cells $1,2,\dots,n$, after which it moves the head to the cell $1$ and
	enters $q_0$, thus invoking the computation of $\calM$ on $w$.

When we consider the computation of $M_w$ on the empty word
as the sequence $S=C_0C_1C_2\dots$ of configurations of length $m$,
	where $m=2^{p(|w|)}$, and we view the symbol $(q'_0,\blank)$ as
	$\hash$, then we observe that $S(0)=\hash$,
	$S(1)=S(2)=\cdots=S(m)=\blank$, and for $i>m$ the symbol
	$S(i)$ is determined by $S(i{-}m{-}1)$, $S(i{-}m)$,
$S(i{-}m{+}1)$ and the transition function of $\calM_w$, independently
	of the actual value of $i$. The symbols in
	$\Delta_w$ and the function $D_w$, guaranteeing
	 $\calSDw=S$, are thus obvious.

A polynomial reduction from $L$ to $\ExGenSeqProblem$ is therefore clear,
yielding $\EXPSPACE$-hardness of $\ExGenSeqProblem$.

In the case of $\GenSeqProblem$, we assume that $\calM$ deciding $L$
works in time (and thus also space) $2^{p(n)}$; to the
$\ExGenSeqProblem$-instance $\calSDw$, $\beta_0=(\qacc,\blank)$
constructed to $w$ as  above we
simply add $n_0=m^2$ (for $m=2^{p(|w|)}$), to get a $\GenSeqProblem$-instance.
Here it is clear that $w\in L$ iff
$\calSDw(n_0)=\beta_0$.
This yields $\EXPTIME$-hardness of $\GenSeqProblem$.
\end{proof}

We now show polynomial (in fact, logspace) reductions from \ExGenSeqProblem to \ECG and
from \GenSeqProblem to \CG. Again, we present both reductions together.

Given a~sequence description $\calD=(\Delta,D,m)$ (where $\hash,
\blank\in\Delta$), we construct the
countdown game
\[
	\ND = (\overline{Q},\overline{Q}_{\exists}, \delta, \Pwin)
\]
where
\begin{itemize}
\item
    $\overline{Q}_{\exists}=\{\Pwin, \Pbad,p_1\} \;\cup\;
                \{s_\beta\mid \beta\in\Delta\}$,
\item
    $\overline{Q}_{\forall}=\{p_2\} \;\cup\;
         \{\,t_{(\beta_1,\beta_2,\beta_3)} \,\mid\,
                \beta_1,\beta_2,\beta_3\in\Delta\}$
\end{itemize}
(recall that $\overline{Q}=\overline{Q}_{\exists}\cup
\overline{Q}_{\forall}$),
and the set $\delta$ consists of the rules
in Figure~\ref{fig:rulesofND}
(for all $\beta,\beta_1,\beta_2,\beta_3\in\Delta$).
(Note that the rules for states $\EStateBlank$ and $\EStateHash$ also
 include the rules of the form~(5).)

\begin{figure}
  \begin{center}
  \begin{tabular}{ccr}
  \toprule
  States & Rules & \\
  \midrule
  $\Pwin$ ($\exists$) & --- & \\
  \midruleThin
  $\Pbad$ ($\exists$) & --- & \\
  \midruleThin
  $p_1$ ($\exists$) &
         $p_1 \gt{-1} p_1$  \qquad $p_1 \gt{-1} \Pwin$ & (1) \\
  \midruleThin
  $p_2$ ($\forall$) &
         $p_2 \gt{-1} p_1$  \qquad $p_2 \gt{-(m{+}2)} \Pbad$ & (2) \\
  \midruleThin
  $\EStateBlank$ ($\exists$) &
         $\EStateBlank \gt{-1} p_2$  & {(3)} \\
  \midruleThin
  $\EStateHash$ ($\exists$) &
         $\EStateHash \gt{-2} \Pwin$
         &  (4) \\
  \midruleThin
  $s_{\beta}$ ($\exists$) &
         $s_{\beta} \gt{-(m{-}2)} \AStateIII$\ \
         when\ \ $D(\beta_1,\beta_2,\beta_3)=\beta$ & (5) \\
  \midruleThin
       & $\AStateIII \gt{-3}s_{\beta_1}$  &     \\
  $\AStateIII$ ($\forall$) &
         $\AStateIII \gt{-2} s_{\beta_2}$ & (6) \\
       & $\AStateIII \gt{-1}s_{\beta_3}$  &     \\
  \bottomrule
  \end{tabular}
  \end{center}
  \caption{Rules of $\ND$.}\label{fig:rulesofND}
\end{figure}

The idea is that the configuration $s_\beta(k)$ of $\ND$
should ``claim'' that $\calSD(k)=\beta$ (and Eve should have a winning
strategy from  $s_\beta(k)$ iff this claim is correct). For technical
reasons we add $2$ to the counter, hence it is $s_\beta(k{+}2)$ that ``claims''
  $\calSD(k)=\beta$.

\begin{lem}
\label{lem:ecg_winarea}
For each $\beta\in\Delta$ we have $s_{\beta}(0) \not\in\winareaE$,
	$s_{\beta}(1) \not\in\winareaE$,
	and for each $k\in\Nat$
	we have $s_{\beta}(k{+}2)\in\winareaE$ iff
$\calSD(k)=\beta$.
\end{lem}
\begin{proof}
We start by noting the following
facts that are easy to check (recall that Eve wins iff the configuration
$\Pwin(0)$ is reached):
\begin{enumerate}[a)]
  \item $\Pwin(k)\in\winareaE$ iff $k=0$;\quad $\Pbad(k)\not\in\winareaE$ for all $k\in\Nat$;
  \item $p_1(k)\in\winareaE$ iff $k \geq 1$;\quad
        $p_2(k)\in\winareaE$ iff $k\in[2, m{+}1]$;
  \item for each $\beta\in\Delta$, $s_{\beta}(0) \not\in\winareaE$ and
$s_{\beta}(1) \not\in\winareaE$;
  \item $\EStateHash(2)\in\winareaE$ (recall that
	  $\calSD(0)=\hash$);

  \item $\EStateBlank(k{+}2) \in\winareaE$ for each $k\in[1, m]$
	\\
		(recall that
	$\calSD(1)=\calSD(2)=\cdots=\calSD(m)=\blank$).
\end{enumerate}
	The statement of the lemma for $s_\beta(0)$ and  $s_\beta(1)$
	follows from the fact~(c). To prove
that $s_{\beta}(k{+}2)\in\winareaE$ iff $\calSD(k)=\beta$,
	we proceed by induction on~$k$:
\begin{itemize}
\item \emph{Base case} $k\in[0,m]$: First we note that Eve cannot win
   in $s_{\beta}(k{+}2)$ by playing any rule of the form~(5) because
		either this
   cannot be played at all, or
		\[s_{\beta}(k{+}2) \gt{-(m{-}2)} \AStateIII(k')\]
   where $k'\leq 4$, and Adam either cannot continue in $\AStateIII(k')$, or he
   can play to $s_{\beta_{\ell}}(k'')$ for some $\ell\in\{1,2,3\}$ with $k''<2$,
   which is losing for Eve (fact~(c)).
		Now it easily follows from the facts~(d) and~(e) that Eve wins in
   $s_{\beta}(k{+}2)$ exactly in those cases where either $\beta=\hash$ and
   $k=0$ (fact~(d)), or $\beta=\blank$ and $k\in[1,m]$
   (fact~(e)).

\item \emph{Induction step} for $k>m$:~~Eve
	cannot win
   in~$s_{\beta}(k{+}2)$ by playing a~rule of the form~(3) or~(4),
 so she is forced to use a~rule from~(5), \ie,~to play a~transition of the form
\[s_{\beta}(k{+}2) \gt{-(m{-}2)} \AStateIII(k')\]
			where $D(\beta_1,\beta_2,\beta_3)=\beta$
		and $k'=k{-}m{+}4\geq 5$. By this, she chooses a~triple
   $(\beta_1,\beta_2,\beta_3)$ where $\beta_1$, $\beta_2$,
		$\beta_3$ are symbols supposedly occurring on
		positions $k{-}m{-}1$,
		$k{-}m$, and $k{-}m{+}1$ in $\calSD$.
   Now Adam can challenge some of the symbols $\beta_1$, $\beta_2$, $\beta_3$
   by choosing $\ell\in\{1,2,3\}$ and playing
		\[\AStateIII(k') \gt{-(4-\ell)}
		s_{\beta_{\ell}}(k'')\]
		where
		$k''= k{-}m{+}\ell\geq 2$.
   Either $\beta$ is correct (\ie,~$\calSD(k)=\beta$), and then Eve can choose correct
   $\beta_1$, $\beta_2$, $\beta_3$, or
   $\beta$ is incorrect (\ie,~$\calSD(k)\neq \beta$), and then at least one of
   $\beta_1$, $\beta_2$, $\beta_3$ must be also incorrect.
   Since, by the induction hypothesis, Adam can win in $\AStateIII(k')$  iff one of
   $\beta_1$, $\beta_2$, $\beta_3$ is incorrect (by choosing
   the corresponding move for this incorrect~$\beta_{\ell}$), Eve can
   win in~$s_{\beta}(k{+}2)$ iff $\beta$ is correct.
   From this we derive that
	$s_{\beta}(k{+}2)\in\winareaE$ iff $\calSD(k)=\beta$.
   \qedhere
\end{itemize}
\end{proof}

Hence,
for a given $\ExGenSeqProblem$-instance $\calD$, $\beta_0$
(where $\calD=(\Delta,D,m)$) there is $i\in\Nat$
such that $\calSD(i)=\beta_0$ iff there is $k\in\Nat$ such that
$s_{\beta_0}(k)\in\winareaE$ for $\ND$. Moreover,
 $\calSD(n_0)=\beta_0$ iff $s_{\beta_0}(n_0{+}2)\in\winareaE$.
Recalling Proposition~\ref{prop:EGSP_exspace_complete},
we have thus established the lower bounds in
Theorems~\ref{th:ecgp_expspcomplete} and~\ref{th:cg_exptimecomplete}.

%
%

\section{Reachability Game Reduces
to (Bi)simulation Game}
\label{sec:sketchreducereachsimul}

We show a reduction for general r-games,
and then apply it to the case of socn-r-games.
This yields a logspace reduction of \SOCNRG to behavioural
relations between bisimulation equivalence and simulation preorder.

Recalling the \EXPSPACE-hardness
of \SOCNRG (from~\cite{DBLP:conf/rp/Hunter15}, or
from the stronger statement of Theorem~\ref{th:ecgp_expspcomplete}),
the respective lemmas (Lemma~\ref{lem:reachreduced}
and~\ref{lem:logspreduc}) will yield the following theorem (which also
answers the respective open question from~\cite{DBLP:journals/corr/HofmanLMT16}):

\begin{thm}\label{th:equivhard}
	For succinct labelled one-counter net (SOCNs),
        deciding membership problem in any relation
        containing bisimulation equivalence and contained in simulation preorder
        (of the associated LTSs) is \EXPSPACE-hard.
\end{thm}

\subsection{Reduction in a General Framework}

We start with an informal introduction to the reduction, which is an
application of the technique
called ``Defender's forcing'' in~\cite{DBLP:journals/jacm/JancarS08}.

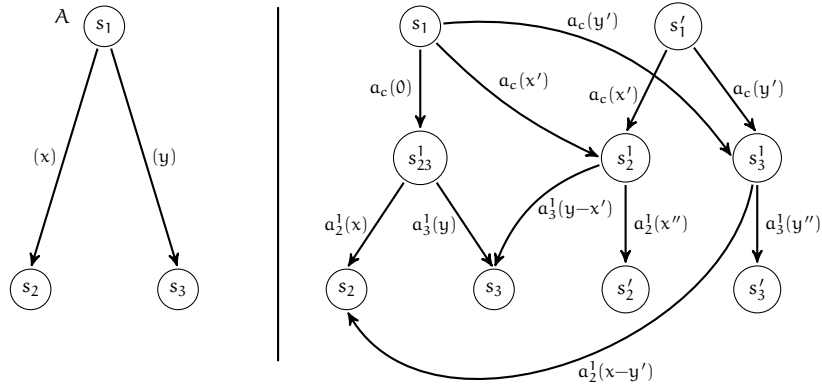
\begin{figure}[t]
  \centering
  \begin{tikzpicture}[scale=0.5,every node/.style={scale=0.7}]
    \node (dummy) {};
    \node[stav,left of=dummy,xshift=-0.4cm] (s2) {$s_2$};
    \node[stav,right of=dummy,xshift=0.4cm] (s3) {$s_3$};
   \node[stav,right of=s3,xshift=1.5cm] (s2prime) {$s'_2$};
\node[stav,right of=s2prime,xshift=1.5cm] (s3prime) {$s'_3$};

   \node[stav,above of=dummy,yshift=1.5cm] (s123) {$s^1_{23}$}
   edge[pil] node[left] {$a^1_2(x)$} (s2.north)
   edge[pil] node[left] {$a^1_3(y)$} (s3.north);

  \node[stav,above of=s2prime,yshift=1.5cm] (s12) {$s^1_{2}$}
  edge[pil] node[right] {$a^1_2(x'')$} (s2prime.north)
  edge[pil,bend right=25] node[right,xshift=-0.1cm,yshift=-0.1cm]
  {$a^1_3(y{-}x')$} (s3.north);

   \node[stav,above of=s3prime,yshift=1.5cm] (s13) {$s^1_{3}$}
   edge[pil, ] node[right] {$a^1_3(y'')$} (s3prime.north)
   edge[pil, out=-100, in=-60] node[right,yshift=-0.2cm]
   {$a^1_2(x{-}y')$} (s2.south);
 
  \node[stav,above of=s123,yshift=1.5cm] (s1) {$s_1$}
  edge[pil] node[left] {$a_c(0)$} (s123.north)
  edge[pil,bend right=10] node[right,xshift=-0.4cm,yshift=0.5cm]
  {$a_c(x')$} (s12.west)
    edge[pil,bend left=30] node[right,xshift=-0.9cm,yshift=0.5cm]
    {$a_c(y')$} (s13.west);

  \node[stav,above of=s12,yshift=1.5cm,xshift=1cm] (s1prime) {$s'_1$}
  edge[pil] node[left,xshift=-0cm,yshift=-0.1cm] {$a_c(x')$} (s12.north)
  edge[pil] node[right] {$a_c(y')$} (s13.north);

  \node[stav,left of=s2,xshift=-5cm] (s2orig) {$s_2$};

  \node[stav,left of=s3,xshift=-5cm] (s3orig) {$s_3$};
  \node[stav,left of=s1,xshift=-5cm] (s1orig) {$s_1$}
  edge[pil] node[left] {$(x)$} (s2orig.north)
  edge[pil] node[right] {$(y)$} (s3orig.north);

  \node [left of=s1orig,xshift=0.2cm,yshift=0.2cm] (dummy2) {$A$};

  \node [right of=s1orig,xshift=2.3cm,yshift=0.5cm] (dummy3) {};
  \node [below of=dummy3,yshift=-6.0cm] (dummy4) {};

  \draw [thick] (dummy3) -- (dummy4) ;

 \end{tikzpicture}
  \caption{In $(s_1,s'_1)$ it is, in fact, Defender
      who chooses $(s_2,s'_2)$ or
      $(s_3,s'_3)$ (when Attacker avoids pairs with equal states); to take
      the counter-changes into account correctly, we put
      $x'=\min{\{x,0\}}$, $x''=\max{\{x,0\}}$, and
      $y'=\min{\{y,0\}}$, $y''=\max{\{y,0\}}$
      (hence $x=x'{+}x''$ and $y=y'{+}y''$).
  }\label{fig:defendchoice}
\end{figure}

Any r-game $\calG=(V,V_\exists, \gt{}, \calT)$ gives rise to the LTS
$\calL=(V,Act,(\gt{a})_{a\in Act})$ where
$Act=\{a_{\langle s,\bar{s}\rangle}\mid
s\gt{}\bar{s}\}{\,\cup\,}\{a_{win}\}$,
$\gt{a_{\langle s,\bar{s}\rangle}}=\{(s,\bar{s})\}$, and
$\gt{a_{win}}=\{(s,s)\mid s\in\calT\}$; hence each
transition gets its unique action-label, and each target state gets a
loop labelled by Eve's winning action $a_{win}$.
Let $\calL'$ be a copy of $\calL$ with
the state set $V'=\{s'\mid s\in V\}$ but without the action $a_{win}$.
We thus have
$s\gt{a_{\langle s,\bar{s}\rangle}}\bar{s}$ in $\calL$ iff
$s'\gt{a_{\langle s,\bar{s}\rangle}}\bar{s}'$ in $\calL'$,
and  for $s\in\calT$ we have
$s\notsimul s'$
in the disjoint union of $\calL$ and $\calL'$
(since $a_{win}$ is enabled in
$s$ but not in $s'$).
If $V_\exists=V$ (in each state it is Eve who chooses the next
transition), then we easily observe that
\begin{equation}\label{eq:prelimfacts}
	\textnormal{$s\in\winareaE$ entails $s\notsimul s'$
	and  $s\not\in\winareaE$ entails $s\bisim s'$.}
\end{equation}
	A technical problem is how to achieve~(\ref{eq:prelimfacts}) in
	the case $V_\exists\neq V$; in this case also Adam's choices
	in the r-game have to be faithfully mimicked in the
	(bi)simulation game, where it is now Defender who should force
	the outcome of the relevant game rounds.
	This is accomplished by adding ``intermediate'' states and
	transitions, and the
	``choice action'' $a_c$, as
	depicted in Figure~\ref{fig:defendchoice} (and discussed later
	in detail); Attacker must let Defender to really choose since
	otherwise Defender wins by reaching a pair $(s,s)$ with the equal
	sides (where $s\bisim s$).

\begin{figure}[t]
  \centering
  \begin{tikzpicture}[scale=0.5,every node/.style={scale=0.7}]
    \node (dummy) {};
    \node[stav,left of=dummy,xshift=0cm] (s2) {$s_2$};
    \node[stav,right of=dummy,xshift=0cm] (s3) {$s_3$};
    \node[right of=dummy,xshift=4cm]  (dummy2) {};
    \node[stav,left of=dummy2] (s2prime) {$s'_2$};
    \node[stav,right of=dummy2] (s3prime) {$s'_3$};

  \node[stav,above of=dummy,yshift=1.5cm] (s1) {$s_1$}
  edge[pil] node[left] {$a^1_2(x)$} (s2.north)
  edge[pil] node[right] {$a^1_3(y)$} (s3.north);

  \node[stav,above of=dummy2,yshift=1.5cm] (s1prime) {$s'_1$}
  edge[pil] node[left] {$a^1_2(x)$} (s2prime.north)
  edge[pil] node[right] {$a^1_3(y)$} (s3prime.north);

  \node[stav,left of=s2,xshift=-5cm] (s2orig) {$s_2$};

  \node[stav,left of=s3,xshift=-5cm] (s3orig) {$s_3$};
  \node[stav,left of=s1,xshift=-5cm] (s1orig) {$s_1$}
  edge[pil] node[left] {$(x)$} (s2orig.north)
  edge[pil] node[right] {$(y)$} (s3orig.north);

  \node [left of=s1orig,xshift=0.2cm,yshift=0.2cm] (dummy2) {$E$};

  \node [right of=s1orig,xshift=2.3cm,yshift=0.5cm] (dummy3) {};
  \node [below of=dummy3,yshift=-2.8cm] (dummy4) {};

  \draw [thick] (dummy3) -- (dummy4) ;
 \end{tikzpicture}
  \caption{Eve's state $s_1$ in $\calG$ (left)
  is mimicked by the pair $(s_1,s'_1)$ in $\calL(\calG)$ (right);
  it is thus Attacker who chooses
  $(s_2,s'_2)$ or $(s_3,s'_3)$ as the next current pair.
}\label{fig:attchoice}
\end{figure}
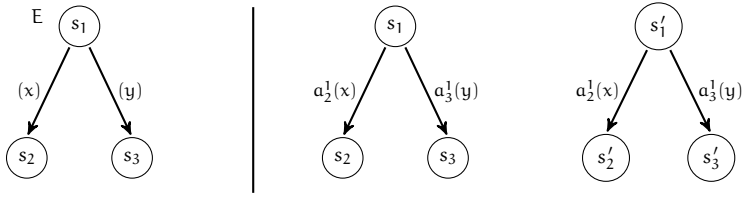

Now we formalize the above sketch.
We assume an r-game $\calG$, and we define a ``mimicking'' LTS
$\calL(\calG)$ (the enhanced union of the above LTSs $\calL$ and $\calL'$).
In illustrating Figures~\ref{fig:attchoice} and~\ref{fig:defendchoice} we now
ignore the bracketed parts of transition-labels;
hence, e.g., in Figure~\ref{fig:attchoice} we can see the
transition $s_1\gt{}s_2$ in $\calG$ on the left
and the (corresponding) transitions  $s_1\gt{a^1_2}s_2$
and $s'_1\gt{a^1_2}s'_2$
in $\calL(\calG)$ on the right.
Let $\calG=(V,V_\exists, \gt{}, \calT)$,
where $V_{\forall} = V\smallsetminus V_{\exists}$;
we define $\calL(\calG)=(S,Act,(\gt{a})_{a\in Act})$ as follows.
We put
\begin{center}
	$S=V\cup V'\cup
	\{\langle s,\bar{s}\rangle\mid s\in V_{\forall},
	s\gt{} \bar{s}\}\cup \{\langle s,X\rangle\mid
s\in V_{\forall}, X=\{\bar{s}\mid s\gt{}\bar{s}\}\neq\emptyset\}$
\end{center}
where $V'=\{s'\mid s\in V\}$ is a ``copy'' of $V$.
(In Figure~\ref{fig:defendchoice}
we write, e.g., $s^1_3$ instead of
$\langle s_1,s_3\rangle$,
and $s^1_{23}$ instead of
$\langle s_1,\{s_2,s_3\}\rangle$.)

We put
$Act=\{a_c,a_{win}\}\cup\{a_{\langle s,\bar{s}\rangle}\mid
s\gt{}\bar{s}\}$
and define $\gt{a}$ for $a\in Act$ as follows.
If $s\in V_{\exists}$ and $s\gt{} \bar{s}$, then
		$s\gtl{a_{\langle s,\bar{s}\rangle}} \bar{s}$ and
		$s'\gtl{a_{\langle s,\bar{s}\rangle}} \bar{s}'$
		(in Figure~\ref{fig:attchoice} we write, e.g., $a^1_3$ instead of
	$a_{\langle s_1,s_3\rangle}$).
If $s\in V_{\forall}$ and
 $X=\{\bar{s}\mid s\gt{}\bar{s}\}\neq\emptyset$,
 then:
\begin{enumerate}[a)]
	\item

 $s\gtl{a_c}\langle s,X\rangle$,
		and $s\gtl{a_c} \langle s,\bar{s}\rangle$,
		$s'\gtl{a_c} \langle s,\bar{s}\rangle$
		for all $\bar{s}\in X$
	(cf. Figure~\ref{fig:defendchoice}
	where $s=s_1$ and $X=\{s_2,s_3\}$;
        $a_c$ is a ``choice-action'');

\item
	for each $\bar{s}\in X$ we have
$\langle s,X\rangle\gtl{a_{\langle
s,\bar{s}\rangle}}\bar{s}$
and $\langle s,\bar{s}\rangle\gtl{a_{\langle
s,\bar{s}\rangle}}\bar{s}'$;
moreover, for each $\bar{\bar{s}}\in X\smallsetminus \{\bar{s}\}$
we have $\langle s,\bar{s}\rangle\gtl{a_{\langle
s,\bar{\bar{s}}\rangle}}\bar{\bar{s}}$
(e.g., in Figure~\ref{fig:defendchoice} we thus have
$s^1_2\gtl{a^1_2}s'_2$ and $s^1_2\gtl{a^1_3}s_3$).
\end{enumerate}

\noindent
For each $s\in\calT$ we have $s\gtl{a_{win}} s$
(for special $a_{win}$ that is not enabled in $s'$).

\begin{lem}\label{lem:reachreduced}
For an r-game $\calG=(V,V_\exists, \gt{}, \calT)$,
and the LTS $\calL(\calG)=(S,Act,(\gt{a})_{a\in Act})$, the following conditions
hold for every
$s\in V$ and every relation
	$\rho$ satisfying
	$\bisim{\subseteq}\mathop{\rho}{\subseteq}\simul$:
	\begin{enumerate}[a)]
		\item
if	$s\in\winareaE$ (in $\calG$), then $s\notsimul s'$ (in
$\calL(\calG)$) and thus $(s,s')\not\in \rho$;
\item
if $s\not\in\winareaE$, then $s\bisim s'$ and thus $(s,s')\in \rho$.
 \end{enumerate}
\end{lem}
\begin{proof}
a) For the sake of contradiction suppose that there is $s\in\winareaE$
such that $s\simul s'$; we consider such $s\in\winareaE$ with
the least rank.
We note that $\rank(s)>0$,
since $s\in\calT$ entails $s\notsimul s'$ due to the
	transition $s\gtl{a_{win}} s$.
If $s\in V_{\exists}$, then let $s\gt{} \bar{s}$ be a rank-reducing
transition. Attacker's move $s\gtl{a_{\langle s,\bar{s}\rangle}}
\bar{s}$, from the pair $(s,s')$,
must be responded with $s'\gtl{a_{\langle s,\bar{s}\rangle}} \bar{s}'$;
but we have $\bar{s}\notsimul \bar{s}'$ by the ``least-rank''
assumption,
which contradicts the assumption $s\simul s'$.
If $s\in V_{\forall}$, then $X=\{\bar{s}\mid s\gt{}\bar{s}\}$ is
nonempty (since $s\in\winareaE$) and
$\rank(\bar{s})<\rank(s)$ for all $\bar{s}\in X$.
For the pair  $(s,s')$
we now consider Attacker's move
$s\gtl{a_{c}} \langle s,X \rangle$. Defender can choose
$s'\gtl{a_{c}} \langle s,\bar{s} \rangle$ for any $\bar{s}\in X$
(recall that $\rank(\bar{s})<\rank(s)$).
In the current pair $(\langle s,X \rangle, \langle s,\bar{s} \rangle)$
Attacker can play $\langle s,X \rangle \gtl{a_{\langle s,\bar{s}
\rangle}} \bar{s}$, and this must be responded by
$\langle s,\bar{s} \rangle \gtl{a_{\langle s,\bar{s}
\rangle}} \bar{s}'$. But we again have
$\bar{s}\notsimul \bar{s}'$ by the ``least-rank'' assumption,
which contradicts $s\simul s'$.

b) It is easy to verify that the following set
is a bisimulation in $\calL(\calG)$:
\begin{center}
$I\cup\{(s,s')\mid s\in V\smallsetminus \winareaE\}
 \cup \{(\langle s,X \rangle, \langle s,\bar{s} \rangle)\mid
 s\in V_{\forall}\smallsetminus \winareaE, \bar{s}\in V\smallsetminus
 \winareaE\}$
\end{center}
where $I=\{(s,s)\mid s\in S\}$.
\end{proof}

We note that the transitions
$s_1\gt{a_c}s^1_2$ and $s_1\gt{a_c}s^1_3$
in Figure~\ref{fig:defendchoice}
could be omitted if we only wanted to show
that $s\in\winareaE$ iff $s\notsimul s'$.

\subsection{\texorpdfstring{\SOCNRG}{Socn-Rg} Reduces to Behavioural Relations on SOCNs}

We now note that the LTS $\calL(\calG_\calN)$
``mimicking'' the r-game $\calG_\calN$
associated with a socn-r-game $\calN$
 (recall~(\ref{eq:gamefornet})) can be presented as
 $\calL_{\calN'}$ for
a SOCN $\calN'$ (recall~(\ref{eq:ltsfornet}))
 that is efficiently constructible from $\calN$:

 \begin{lem}\label{lem:logspreduc}
There is a logspace algorithm that,
given
a socn-r-game $\calN$,
constructs a SOCN $\calN'$ such that
the LTSs $\calL(\calG_\calN)$ and $\calL_{\calN'}$ are
isomorphic.
\end{lem}
\begin{proof}
We again use Figures~\ref{fig:attchoice}
and~\ref{fig:defendchoice} for illustration;
now $s_i$ are viewed as control states
and the bracketed parts of edge-labels are counter-changes (in binary).

Given a socn-r-game $\calN=(Q,Q_{\exists}, \delta, \Pwin)$, we
first consider
the r-game
\[\calN^{csg}=(Q,Q_{\exists}, \gt{}, \{\Pwin\})\]
(``the control-state game of $\calN$'') arising from $\calN$ by
\emph{forgetting the counter-changes}; hence $q\gt{}\bar{q}$ iff
there is a rule $q\gt{z}\bar{q}$. In fact, we will assume that
there is at most one rule  $q\gt{z}\bar{q}$ in $\delta$ (of $\calN$)
for any pair $(q,\bar{q})\in Q\times Q$; this can be achieved by
harmless modifications.

We construct the (finite) LTS $\calL(\calN^{csg})$
(``mimicking'' $\calN$). Hence each $q\in Q$ has the copies $q,q'$ in
$\calL(\calN^{csg})$, and other states are added
(as also depicted in Figure~\ref{fig:defendchoice} where $s_i$ are
now in the role of control states); there are also the
respective labelled transitions in $\calL(\calN^{csg})$, with labels
$a_{\langle q,\bar{q}\rangle}$, $a_c$, $a_{win}$.

It remains to add the counter changes (integer increments and
decrements in binary), to create the required SOCN $\calN'$.
For $q\in Q_{\exists}$ this adding is simple, as
depicted in Figure~\ref{fig:attchoice}: if $q\gt{z}\bar{q}$ (in $\calN$),
then we
simply extend the label $a_{\langle q,\bar{q}\rangle}$
in $\calL(\calN^{csg})$
with $z$;
for $q\gtl{a_{\langle q,\bar{q}\rangle}}\bar{q}$ and
$q'\gtl{a_{\langle q,\bar{q}\rangle}}\bar{q}'$ in $\calL(\calN^{csg})$
we get
 $q\gtl{a_{\langle q,\bar{q}\rangle},z}\bar{q}$
 and  $q'\gtl{a_{\langle q,\bar{q}\rangle},z}\bar{q}'$ in $\calN'$.

 For $q\in Q_{\forall}$ (where $Q_{\forall}=Q\smallsetminus
 Q_{\exists}$) it is tempting to the same, i.e. to extend
the label $a_{\langle q,\bar{q}\rangle}$ with $z$ when
$q\gt{z}\bar{q}$, and extend $a_c$ with $0$.
 But this might allow cheating for Defender: she could thus
 mimic choosing a
 transition
 $q(k)\gt{x}\bar{q}(k{+}x)$ even if $k{+}x<0$. This is avoided by the
 modification that is demonstrated in Figure~\ref{fig:defendchoice}
 (by $x=x'{+}x''$, etc.); put simply: Defender must immediately prove
 that the transition she is choosing to mimic is indeed performable.
 Formally, if   $X=\{\bar{q}\mid q\gt{}\bar{q}\}\neq\emptyset$
 (in $\calL(\calN^{csg})$), then in $\calN'$
 we put $q\gtl{a_c,0}\langle q,X \rangle$ and
 $\langle q,X \rangle\gtl{a_{\langle q,\bar{q} \rangle},z}\bar{q}$
 for each $q\gt{z}\bar{q}$ (in $\calN$);
 for  each $q\gt{z}\bar{q}$ we also define $z'=\min\{z,0\}$,
  $z''=\max\{z,0\}$ and
 put $q'\gtl{a_c,z'}\langle q,\bar{q}\rangle$,
 $\langle q,\bar{q}\rangle\gtl{a_{\langle
 q,\bar{q}\rangle},z''}\bar{q}'$.
 Then for any pair $q\gtl{\bar{z}}\bar{q}$,
 $q\gtl{\bar{\bar{z}}}\bar{\bar{q}}$
 where $\bar{q}\neq \bar{\bar{q}}$ we put
$\langle q,\bar{q}\rangle\gtl{a_{\langle
q,\bar{\bar{q}}\rangle},\bar{\bar{z}}-\bar{z}'}\bar{\bar{q}}$.

Finally, $p_{win}\gtl{a_{win}}p_{win}$ in $\calL(\calN^{csg})$
is extended to $p_{win}\gtl{a_{win},0}p_{win}$ in $\calN'$.
\end{proof}

We have thus finished the proof of Theorem~\ref{th:equivhard}.

%
%

\section{Structure of simulation preorder
on one-counter nets}\label{sec:structsimulOCN}

In this section we give a new self-contained proof clarifying
the structure of simulation preorder $\simul$ on the LTS $\calL_{\calN}$
associated with a given one-counter net $\calN$;
this will also
yield a polynomial-space algorithm  generating a description
	of $\simul$ on $\calL_{\calN}$ that can be used to decide if $p(m)\simul q(n)$.

We first show
a natural graphic presentation
of the relation $\simul$, and in Section~\ref{sec:belttheorem} we
show
its linear-belt form; the result
is captured  by the \emph{belt theorem}.
The proof is inspired
by~\cite{DBLP:conf/sofsem/JancarMS99} but is substantially
different.
A main new ingredient is the notion of so called down-black and up-white
lines, and their limits in which we do not a priori exclude lines with
irrational slopes; this allows us to avoid many technicalities used in
previous proofs, while the presented ``geometric'' ideas should be
straightforward, and transparent due to the respective figures.

In Section~\ref{sec:quantitbelttheorem} we prove that
the slopes and the widths of belts are presented/bounded by small
integers.
Though this \emph{quantitative belt theorem} is in principle equivalent to the respective
theorem proved in~\cite{DBLP:journals/corr/HofmanLMT16}, our proof is
again conceptually different;
using the novel notions, the quantitative characteristics
are derived
from the (qualitative) belt theorem easily.

For completeness,
in Section~\ref{subsec:polspacealg} we briefly recall the idea
from the previous papers that shows how the achieved structural
results yield a polynomial-space algorithm generating a~description
of $\simul$ for a given one-counter net.

In the sequel we assume a fixed OCN $\calN=(Q,Act,\delta)$ if not said otherwise.
By
$\Real$, $\Real_{\geq 0}$, $\Real_{>0}$ we denote the sets of
reals, of nonnegative reals, and of positive reals, respectively.
We also use $\infty$ for an infinite amount (in particular for the
slope of a~vertical line); hence $\alpha<\infty$ for all
	$\alpha\in\Real$. By $\langle \gamma,\gamma'\rangle$, where
$\gamma\in\Real_{\geq 0}$ and $\gamma'\in \Real_{\geq
	0}\cup\{\infty\}$,
	we denote the set $\{\alpha\in\Real_{\geq 0}\cup\{\infty\}\mid
	\gamma\leq \alpha\leq\gamma'\}$; in particular,
 $\langle 0,\infty\rangle=\Real_{\geq 0}\cup\{\infty\}$.

\paragraph{Monotonic black-white presentation of the simulation
preorder \texorpdfstring{$\simul$}{<=}}
For each pair  $(p,q)\in Q\times Q$
we define the (black-white)
\emph{colouring} $C_{\tu{p,q}}$
 of the integer points in the first quadrant of the plane
$\Real\times\Real$; we put
 $C_{\tu{p,q}}: \nat \times \nat \to
\setof{\mathrm{black},\mathrm{white}}$ where
\begin{align*}
	C_{\tu{p,q}}(m, n) &=
		\begin{cases}
			\mathrm{black} & \mbox{if } p(m) \simul q(n), \\
			\mathrm{white} & \mbox{if } p(m) \notsimul q(n).
		\end{cases}
\end{align*}
We recall the definition of $\rank(s,s')$ in
the paragraph ``Stratified simulation, and ranks of pairs of states''
in Section~\ref{sec:basicdefinitions}; this yields
the definition of $\rank(p(m),q(n))$ for our fixed  OCN
$\calN=(Q,Act,\delta)$. Hence
for each $(p,q)\in Q\times Q$ we have
that each white point $(m,n)$ in
$C_{\tu{p,q}}$ has an associated finite rank, namely
$\rank(p(m),q(n))\in\Nat$.

The next proposition captures the trivial fact
that ``\emph{black is upwards- and leftwards-closed}'' and
 ``\emph{white is downwards- and rightwards-closed}''
(as is also depicted in Figure~\ref{fig:solvedsimul} in Introduction, which also
shows the ``linear-belt form'' of the colourings).

\begin{prop}[Black and white monotonicity]\label{prop:monotcolour}
\hfill\\
 If $C_{\tu{p,q}}(m,n) = \mathrm{black}$, then
$C_{\tu{p,q}}(m',n') = \mathrm{black}$ for all $m' \leq m$
 and $n'\geq n$.
Hence	if $C_{\tu{p,q}}(m,n) = \mathrm{white}$,
then			   $C_{\tu{p,q}}(m',n') = \mathrm{white}$ for all
$m'\geq m$ and  $n' \leq n$.
\end{prop}
\begin{proof}
Since OCNs are monotonic in the sense that $p(m)\gt{a}q(n)$ implies
$p(m{+}i)\gt{a}q(n{+}i)$ for all $i\in\Nat$, we will easily verify
that the relation
\[R=\big\{\big(p(m'),q(n')\big)\mid
p(m)\simul q(n) \textnormal{ for some } m\geq m'
\textnormal{ and } n\leq n' \big\}\] is a simulation relation.
	Since
	$\simul\subseteq R$, we will thus get $R=\simul$, and the proof
	will be finished.
To verify that $R$ is indeed a simulation,
we consider $\big(p(m'),q(n')\big)\in R$ and fix $m\geq m'$ and $n\leq n'$
so that $p(m)\simul q(n)$.
If $p(m')\gt{a}p'(m'{+}i)$ (for $i\in\{{-}1,0,{+}1\}$), then $p(m)\gt{a}p'(m{+}i)$. Since
	$p(m)\simul q(n)$, we have $q(n)\gt{a}q'(n{+}j)$ where
$j\in\{{-}1,0,{+}1\}$ and
$p'(m{+}i)\simul q'(n{+}j)$. Hence  $q(n')\gt{a}q'(n'{+}j)$,
and $\big(p'(m'{+}i),q'(n'{+}j)\big)\in R$.
\end{proof}

\subsection{Belt theorem}\label{sec:belttheorem}

\begin{figure}
	\begin{center}
\newcommand{\dhs}[4]{
  \foreach \x in {#1,...,#2}
   \drawpoint{(\x,#3)}{#4};
}

\begin{tikzpicture}[scale=0.7,every node/.style={scale=0.8}]

  
  \draw[->] (0,0) -- (9,0);
  \draw[->] (0,0) -- (0,10);

  \draw (1,-0.5) -- (5.75,9);
  \node[anchor=west] at (5.75,9) {$\ell_L$};
  
  \draw (4,-0.5) -- (8.75,9);
  \node[anchor=west] at (8.75,9) {$\ell_R$};

  \dhs{0,0.25}{1.25}{0.00}{black};
  \dhs{0,0.25}{1.25}{0.25}{black};
  \dhs{0,0.25}{1.50}{0.50}{black};
  \dhs{0,0.25}{1.50}{0.75}{black};
  \dhs{0,0.25}{1.75}{1.00}{black};
  \dhs{0,0.25}{1.75}{1.25}{black};
  \dhs{0,0.25}{2.00}{1.50}{black};
  \dhs{0,0.25}{2.00}{1.75}{black};
  \dhs{0,0.25}{2.25}{2.00}{black};
  \dhs{0,0.25}{2.25}{2.25}{black};
  \dhs{0,0.25}{2.50}{2.50}{black};
  \dhs{0,0.25}{2.50}{2.75}{black};
  \dhs{0,0.25}{2.75}{3.00}{black};
  \dhs{0,0.25}{2.75}{3.25}{black};
  \dhs{0,0.25}{3.00}{3.50}{black};
  \dhs{0,0.25}{3.00}{3.75}{black};
  \dhs{0,0.25}{3.25}{4.00}{black};
  \dhs{0,0.25}{3.25}{4.25}{black};
  \dhs{0,0.25}{3.50}{4.50}{black};
  \dhs{0,0.25}{3.50}{4.75}{black};
  \dhs{0,0.25}{3.75}{5.00}{black};
  \dhs{0,0.25}{3.75}{5.25}{black};
  \dhs{0,0.25}{4.00}{5.50}{black};
  \dhs{0,0.25}{4.00}{5.75}{black};
  \dhs{0,0.25}{4.25}{6.00}{black};
  \dhs{0,0.25}{4.25}{6.25}{black};
  \dhs{0,0.25}{4.50}{6.50}{black};
  \dhs{0,0.25}{4.50}{6.75}{black};
  \dhs{0,0.25}{4.75}{7.00}{black};
  \dhs{0,0.25}{4.75}{7.25}{black};
  \dhs{0,0.25}{5.00}{7.50}{black};
  \dhs{0,0.25}{5.00}{7.75}{black};
  \dhs{0,0.25}{5.25}{8.00}{black};
  \dhs{0,0.25}{5.25}{8.25}{black};

  \dhs{4.25,4.50}{8.5}{0.00}{white};

  \dhs{4.50,4.75}{8.5}{0.25}{white};
  \dhs{4.50,4.75}{8.5}{0.50}{white};
  \dhs{4.75,5.00}{8.5}{0.75}{white};
  \dhs{4.75,5.00}{8.5}{1.00}{white};
  \dhs{5.00,5.25}{8.5}{1.25}{white};
  \dhs{5.00,5.25}{8.5}{1.50}{white};
  \dhs{5.25,5.50}{8.5}{1.75}{white};
  \dhs{5.25,5.50}{8.5}{2.00}{white};
  \dhs{5.50,5.75}{8.5}{2.25}{white};
  \dhs{5.50,5.75}{8.5}{2.50}{white};
  \dhs{5.75,6.00}{8.5}{2.75}{white};
  \dhs{5.75,6.00}{8.5}{3.00}{white};
  \dhs{6.00,6.25}{8.5}{3.25}{white};
  \dhs{6.00,6.25}{8.5}{3.50}{white};
  \dhs{6.25,6.50}{8.5}{3.75}{white};
  \dhs{6.25,6.50}{8.5}{4.00}{white};
  \dhs{6.50,6.75}{8.5}{4.25}{white};
  \dhs{6.50,6.75}{8.5}{4.50}{white};
  \dhs{6.75,7.00}{8.5}{4.75}{white};
  \dhs{6.75,7.00}{8.5}{5.00}{white};
  \dhs{7.00,7.25}{8.5}{5.25}{white};
  \dhs{7.00,7.25}{8.5}{5.50}{white};
  \dhs{7.25,7.50}{8.5}{5.75}{white};
  \dhs{7.25,7.50}{8.5}{6.00}{white};
  \dhs{7.50,7.75}{8.5}{6.25}{white};
  \dhs{7.50,7.75}{8.5}{6.50}{white};
  \dhs{7.75,8.00}{8.5}{6.75}{white};
  \dhs{7.75,8.00}{8.5}{7.00}{white};
  \dhs{8.00,8.25}{8.5}{7.25}{white};
  \dhs{8.00,8.25}{8.5}{7.50}{white};
  \dhs{8.25,8.50}{8.5}{7.75}{white};
  \dhs{8.25,8.50}{8.5}{8.00}{white};
  \drawpoint{(8.5,8.25)}{white};

\end{tikzpicture}
	\end{center}
	\caption{Illustration of the belt theorem, for one colouring
	$C=C_{\tu{p,q}}$.}\label{fig:belttheoremclaim}
\end{figure}
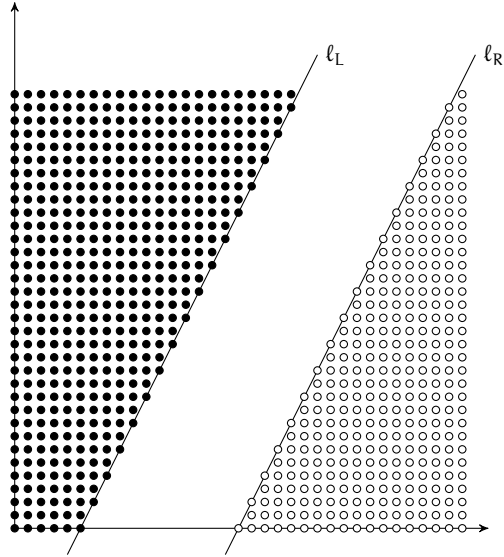

	Below we state the (qualitative) belt theorem, illustrated in
	Figure~\ref{fig:belttheoremclaim},
after
introducing the needed notions.
We can remark that the validity of the belt theorem can be easily intuitively
anticipated (by a quick thought about the problem) but it has turned out
surprisingly hard to be rigorously
proven. (It can be also intuitively anticipated that the colouring
inside the belt determined by the lines $\ell_L$, $\ell_R$ in
Figure~\ref{fig:belttheoremclaim} is ultimately periodic, but this is
only used in Section~\ref{subsec:polspacealg}.)

\begin{figure}
  \begin{tabular}{cp{0.95cm}c}
    \begin{tikzpicture}[scale=0.7,every node/.style={scale=0.8}]

  
  \draw[color=white] (0,0) -- (0,-0.5);
  
  \draw[->] (0,0) -- (9,0);
  \draw[->] (0,0) -- (0,9);
  
  \draw (-0.5, 0.5) -- (7,8);
  \node[anchor=west] at (7,8) {$\ell$};

  \draw (1,8) -- (8,1);
  \node[anchor=east] at (1,8) {$\ell'$};

  \node[align=left] at (2,5) {above $\ell$\\[2pt]below $\ell'$};
  \node[align=left] at (6,5) {below $\ell$\\[2pt]above $\ell'$};

\end{tikzpicture} && \begin{tikzpicture}[scale=0.7,every node/.style={scale=0.8}]

    \draw[->] (0,0) -- (9,0);
    \draw[->] (0,0) -- (0,9);

    \draw (4,-0.5) -- (4,8);
    \node[anchor=west] at (4,8) {$\ell$};

    \node at (2,4) {above, left of $\ell$};
    \node at (6,4) {below, right of $\ell$};

\end{tikzpicture}
  \end{tabular}
  \caption{Areas ``above $\ell$'', ``below $\ell$'', ``left
	of $\ell$'', and ``right of $\ell$''.}
  \label{fig:above-below}
\end{figure}
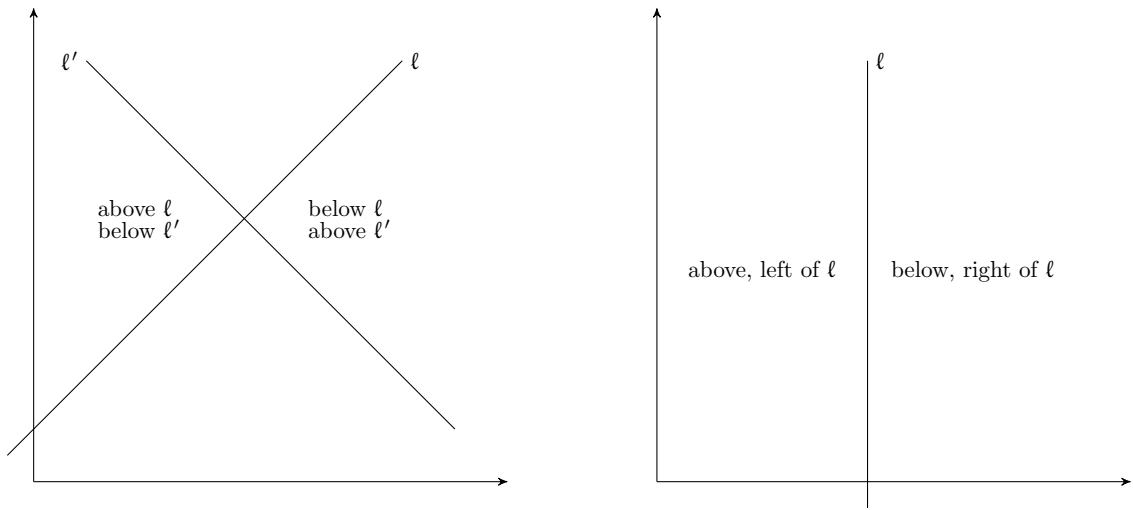

\paragraph{Lines, slopes, axes, points, points above or below lines,
	points on
	the left of or on the right of vertical lines.} (See
	Figure~\ref{fig:above-below}.)
A \emph{vertical line} $\ell$ (in the plane $\Real\times\Real$) is the set
$\ell=\{(x,y)\mid y\in\Real\}$ for a fixed $x\in\Real$; we put
$\slope(\ell)=\infty$. A \emph{non-vertical line} $\ell$ is
the set
$\ell=\{(x,y)\mid y=\gamma\cdot x + c\}$ for some fixed
$\gamma\in\Real$ and $c\in\Real$; here $\slope(\ell)=\gamma$.
A~\emph{line}  is either
a vertical line or a non-vertical line.
By a \emph{$\gamma$-line}, $\gamma\in \Real\cup\{\infty\}$, we mean a
	line whose slope is $\gamma$.
	As expected, by the \emph{horizontal axis}, $\textsc{h-axis}$,
we mean the $0$-line that goes through the \emph{origin} $(0,0)$;
the \emph{vertical axis}, $\textsc{v-axis}$, is the $\infty$-line that  goes through $(0,0)$.

By \emph{points} we mean the pairs $(m,n)\in\Nat\times\Nat$ (hence only the
integer points in the first quadrant of the plane
$\Real\times\Real$), unless explicitly stated that we consider all
integer points (elements of $\Zset\times\Zset$).
Given a non-vertical line  $\ell=\{(x,y)\mid y=\gamma\cdot x + c\}$,
a \emph{point} $(m,n)$ is \emph{above $\ell$} if $n\geq \gamma\cdot m
+ c$,
and it is
 \emph{below $\ell$} if $n\leq \gamma\cdot m + c$.
Given a vertical line  $\ell=\{(x,y)\mid
y\in\Real\}$,
	a \emph{point} $(m,n)$ is \emph{above $\ell$}, or also
	\emph{on the left of $\ell$}, if $m\leq x$, and it is
 \emph{below $\ell$}, or also
	\emph{on the right of $\ell$}, if $m\geq x$.
	A point \textsc{p} is \emph{strictly above} $\ell$ if
	$\textsc{p}$ is above $\ell$ but not on $\ell$; similarly,
	\textsc{p} is \emph{strictly below} $\ell$ if
	$\textsc{p}$ is below $\ell$ but not on $\ell$.

\begin{thm}[Belt theorem]\label{th:qualitative}
	For each $(p,q)\in Q\times Q$, and its respective
	colouring $C=C_{\tu{p,q}}$, there are (a slope) $\gamma\in\langle 0,\infty\rangle $
	that is rational or $\infty$, and
	two (parallel) $\gamma$-lines $\ell_L$ and $\ell_R$
	such that all points above $\ell_L$ are black in $C$ and all
	points below $\ell_R$ are white in $C$. (See
	Figure~\ref{fig:belttheoremclaim}.)
\end{thm}

We recall that we do not discuss how the mapping $C_{\tu{p,q}}$
looks inside the belt
depicted in Figure~\ref{fig:belttheoremclaim};
only in Section~\ref{subsec:polspacealg} we refer to the
fact that it is ultimately periodic there.

Now we introduce further notions useful in the proof of the belt theorem.
\paragraph{Down-black and up-white lines and slopes,
down-limits \texorpdfstring{$\alpha_C$}{alpha_C} and up-limits \texorpdfstring{$\beta_C$}{beta_C}.}
	A~\emph{line} $\ell$, with $\slope(\ell)\in\langle 0,\infty\rangle $, is \emph{down-black in} a colouring
	$C=C_{\tu{p,q}}$ if there are
	infinitely many points below $\ell$ that are black in $C$.
	A (nonnegative) \emph{slope} $\gamma\in \langle 0,\infty\rangle $ is
\emph{down-black in} $C$ if there is a $\gamma$-line
	$\ell$ that is down-black in $C$.

By the \emph{down-limit of} $C$ we mean the value
\begin{center}
$\alpha_C=\inf\,\{\,\gamma\mid \gamma$ is down-black in $C\,\}$
\end{center}
(where $\inf \emptyset =\infty$).
We say that $\alpha$ is a \emph{down-limit} if
	$\alpha=\alpha_C$ for some $C=C_{\tu{p,q}}$.  For a~down-limit $\alpha$, by
	\emph{$\alpha$-colourings} we mean the colourings $C$
	such that $\alpha_{C}=\alpha$.

		A \emph{line} $\ell$ is \emph{up-white in} a colouring
	$C=C_{\tu{p,q}}$ if there are
	infinitely many points above $\ell$ that are white in $C$.
	A (nonnegative) \emph{slope} $\gamma\in \langle 0,\infty\rangle $ is
\emph{up-white in} $C$ if there is a $\gamma$-line
	$\ell$ that is up-white in $C$.
By the \emph{up-limit of} $C$ we mean the value
\begin{center}
$\beta_C=\sup\,\{\,\gamma\mid \gamma$ is up-white in $C\,\}$
\end{center}
(where $\sup \emptyset =0$).

We highlight the following trivial fact.

\begin{prop}\label{prop:trivialdownup}\hfill
  \begin{itemize}
  \item	If $\gamma$ is down-black in $C$,
    then each $\gamma'\in\langle \gamma,\infty\rangle $ is down-black in
    $C$; moreover, if $\gamma'>\gamma$, then each $\gamma'$-line
    is down-black in $C$.
  \item
    If $\gamma$ is up-white in $C$,
    then each $\gamma'\in\langle 0,\gamma\rangle $ is up-white in
    $C$; moreover, if $\gamma'<\gamma$, then each $\gamma'$-line
    is up-white in $C$.
  \end{itemize}
\end{prop}

\noindent
We note that we cannot a priori exclude that some $\alpha_C$ is irrational
and/or not down-black;
 similarly we cannot exclude
that $\beta_C$ differs from $\alpha_C$, and that it is irrational
and/or not up-white.
But we immediately note a simple fact:
\begin{prop}\label{prop:betagreateralpha}
	For each $(p,q)\in Q\times Q$,	$\beta_{C_{\tu{p,q}}}\geq \alpha_{C_{\tu{p,q}}}$.
\end{prop}
\begin{proof}
Let $\alpha_C>\beta_C$ for $C=C_{\tu{p,q}}$, and let
$\alpha_C>\gamma>\beta_C$; hence
$\gamma$ is not down-black and not up-white.
Thus for each $\gamma$-line $\ell$ we have that almost all points
	(i.e., all but finitely many)
below
 $\ell$ are white and
almost
all points above $\ell$ are black; by an obvious shift of
	$\ell$ we deduce that $\gamma$ is up-white and/or
down-black after all, which is a contradiction.
\end{proof}

We now state a lemma that trivially entails
Theorem~\ref{th:qualitative}: the slope $\gamma$ claimed for $C$ in
Theorem~\ref{th:qualitative} is equal to $\alpha_C$ (which is claimed
to be equal to $\beta_C$ by the lemma).

\begin{lem}\label{lem:qualitative}
  For each $(p,q)\in Q\times Q$ and the respective
  colouring $C=C_{\tu{p,q}}$ the following conditions hold:
  \begin{enumerate}
  \item	 $\alpha_C$ is rational or
    $\infty$,
  \item
    $\alpha_C=\beta_C$, and
  \item
    there are $\alpha_C$-lines $\ell_L$ and $\ell_R$ such that
    all points above $\ell_L$ are black in $C$, and all
    points below $\ell_R$ are white in $C$.
  \end{enumerate}
\end{lem}

\noindent
Before proving the lemma
we introduce
a main technical ingredient of the proof,
namely Proposition~\ref{prop:leftline} and Corollary~\ref{cor:leftline},
preceded by
 the needed notions and by a simple, yet very useful, observation
 (Proposition~\ref{prop:neighbour}).

 We can remark that these technical ingredients can hardly be
 ``intuitively anticipated''; they have been ``distilled'' from the
 overall nontrivial proof, as useful technical claims.
 Some intuition about a possible proof
 strategy might be perhaps got by
 looking at the simulation games described in the papers~\cite{DBLP:conf/concur/AbdullaC98}
and~\cite{DBLP:journals/corr/HofmanLMT16}, but our proof is, in fact, not so
tightly related to the simulation problem and could be given in a
more general framework of specific tiling problems (as we also recall
in Section~\ref{sec:addrem}).

\begin{figure}
  \centering
  \begin{subfigure}[t]{0.495\textwidth}
    \centering
    \begin{tikzpicture}[scale=0.7,every node/.style={scale=0.8}]

  
    \draw[->] (0,0) -- (9,0);
    \draw[->] (0,0) -- (0,9);

    \coordinate (start) at (1,1);
    \coordinate (end) at (3,7);

    \draw[-{Stealth[scale=1.3]},thick] (start) -- (end);

    \coordinate (starth) at (1.3,0.5);
    \coordinate (endh)   at (4.3,9.5);

    \coordinate (s1s)    at (1.6,0.2);
    \coordinate (s2s)    at (4.6,9.2);

    \coordinate (i1) at ($(starth)!(start)!(endh)$);
    \coordinate (i2) at ($(starth)!(end)!(endh)$);
    \coordinate (i1p) at ($(s1s)!(start)!(s2s)$);
    \coordinate (i2p) at ($(s1s)!(end)!(s2s)$);

    \draw[<->,dotted] (i1) -- (i2)
    node[midway,anchor=north west,align=center]{d-size \\[3pt] (+)};

    \draw[dotted] (start) -- (i1p);
    \draw[dotted] (end) -- (i2p);
        
    \coordinate (start2) at (8.5,6.5);
    \coordinate (end2)   at (5.5,3.5);

    \coordinate (end2h)   at (5.8,3.2);
    \coordinate (start2h) at (8.8,6.2);

    \coordinate (s3s)     at (6,3);
    \coordinate (s4s)     at (9,6);

    \draw[<->,dotted]
    (start2h) -- (end2h)
    node[midway,anchor=north west,align=center]{d-size\\[3pt](-)};
    
    \draw[dotted] (end2) -- (s3s);
    \draw[dotted] (start2) -- (s4s);

    \draw[-{Stealth[scale=1.4]},thick] (start2) -- (end2);

    

    




    

    
  \end{tikzpicture}
  
    \caption{The d-size of a vector.}
    \label{fig:vector-sizes1}
  \end{subfigure}
    \begin{subfigure}[t]{0.495\textwidth}
      \centering
    \begin{tikzpicture}[scale=0.7,every node/.style={scale=0.8}]

  
    \draw[->] (0,0) -- (9,0);
    \draw[->] (0,0) -- (0,9);

    \begin{scope}[xshift=-3cm,yshift=1.75cm]
    
    \coordinate (start) at (4,1);
    \coordinate (end) at (6,6.5);
   
    \coordinate (ells) at (3.5,0.5);
    \coordinate (elle) at (8.5,5.5);

    \coordinate (ellprimes) at (5.5,7);
    \coordinate (ellprimee) at (8.5,4);

    \coordinate (s1s) at (6.25,6.75);
    \coordinate (s1e) at (8,5);
    \coordinate (s1x) at (6.5,7);

    \coordinate (s2s) at (4.25,0.75);
    \coordinate (s2e) at (8,4.5);
    \coordinate (s2x) at (4.5,0.5);


    \draw[dotted] (end) -- (s1x);
    \draw[<->,dotted]
    (s1s) -- (s1e)
    node[midway, anchor=south west,align=center]{co-$\gamma$-d-size \\(+)};

    \draw[dotted] (start) -- (s2x);
    \draw[<->,dotted]
    (s2s) -- (s2e)
    node[pos=.7, anchor=north west,align=center]{$\gamma$-d-size \\(+)};
    
    \draw[color=gray!80!white,name path=ELL, very thin] (ells) -- (elle);
    \draw[color=gray!80!white,name path=ELLPRIME, very thin] (ellprimes) -- (ellprimee);

    \path[name intersections={of=ELL and ELLPRIME,by=I1}];

    \draw[-{Stealth[scale=1.3]},thick] (start) -- (end);
    \end{scope}
    

    \begin{scope}[xshift=3cm,yshift=-4cm]
    
    \coordinate (start2) at (4.5,7.5);
    \coordinate (end2)   at (1,6);

    \coordinate (ell2s) at (6,9);
    \coordinate (ell2e)  at (1.5,4.5);

    \node[anchor=south] at (ell2s) {$\gamma$-line};

    \coordinate (ellprime2e) at (0.5,6.5);
    \coordinate (ellprime2s) at (2.5,4.5);

    \draw[color=gray!80!white,very thin] (ell2s) -- (ell2e);
    \draw[color=gray!80!white,very thin] (ellprime2s) -- (ellprime2e);

    \coordinate (s3x) at (5,7);
    \coordinate (s3s) at (2.25,4.75);
    \coordinate (s3e) at (4.75,7.25);
    
    \coordinate (s4s) at (1.75, 4.75);
    \coordinate (s4e) at (0.75,5.75);
    \coordinate (s4x) at (0.5,5.5);

    \draw[dotted] (start2) -- (s3x);
    \draw[<->,dotted]
    (s3s) -- (s3e) node[midway,anchor=north west,align=center]
	    {$\gamma$-d-size \\(-)};

    \draw[dotted] (end2) -- (s4x);
    \draw[<->,dotted]
    (s4s) -- (s4e)
    node[pos=0.7,anchor=north east, align=center,shift={(0.2,-0.2)}]
	    {co-$\gamma$-d-size \\(+)};

    \draw[-{Stealth[scale=1.3]},thick] (start2) -- (end2);

    \end{scope}
    
  \end{tikzpicture}


   

    




    
    
    \caption{The d-sizes related to $\gamma\in\langle 0,\infty\rangle$.}
    \label{fig:vector-sizes2}
  \end{subfigure}
	\caption{}
  \label{fig:vector-sizes}
\end{figure}

\paragraph{Vectors, their slopes and d-sizes.} (See
Figure~\ref{fig:vector-sizes1}.)
We view a \emph{vector} $v$ as a pair $(\spt(v),\ept(v))$ where
$\spt(v)\in\Nat\times\Nat$ is the \emph{start-point} of $v$ and
$\ept(v)\in\Nat\times\Nat$ is the
\emph{end-point} of $v$. The \emph{slope of a vector} $v$, $\slope(v)$, is defined when
$\spt(v)\neq\ept(v)$, in which case it is the slope of the line going
through both  $\spt(v)$ and $\ept(v)$.
The \emph{d-size of} $v$ (``\emph{d}'' stands for \emph{direction}) is the Euclidean distance of
the points $\spt(v)$ and $\ept(v)$ with the positive sign ($+$) or the
negative sign ($-$)\,: if $\spt(v)\neq\ept(v)$, then we consider
 the line $\ell$ that is
perpendicular to $v$ and goes through $\spt(v)$; if
$\slope(\ell)\neq \infty$ and $\ept(v)$ is above $\ell$, or
$\slope(\ell)=\infty$ and $\ept(v)$ is to the right of $\ell$, then
the d-size of $v$ is positive, and otherwise  the d-size of $v$ is
negative.

\paragraph{Black-white vectors.}

A \emph{vector} $v$ is \emph{black-white in a colouring} $C$ if $\spt(v)$
is black and $\ept(v)$ is white in $C$.

\paragraph{Neighbour points and vectors.}
A \emph{point} $(m',n')$ is a \emph{neighbour of a point} $(m,n)$ if
$|m'-m|\leq 1$ and $|n'-n|\leq 1$ (which includes the case $m'=m$,
$n'=n$).
A vector $v'$ is a \emph{neighbour vector of} a vector $v$
if $\spt(v')$ is a neighbour
of $\spt(v)$ and $v,v'$ have the same slopes and
sizes (hence $v'$ is a ``small shift'' of $v$).

\medskip

The next proposition is the announced simple observation. It states that
for any vector $v$ where
	$\spt(v)$ is not on the vertical axis,
$\ept(v)$ is not on
	the horizontal axis, and $v$ is black-white in some
 $C_{\tu{p,q}}$ there is its neighbour vector $v'$ that is
black-white in some $C_{\tu{p',q'}}$ and the rank of the white
end-point of $v'$ in  $C_{\tu{p',q'}}$
is smaller than the rank of the white end-point of $v$
in $C_{\tu{p,q}}$.
(Recall the vectors $v_0,v_1$ in Figure~\ref{fig:neighbourvector}a in
Introduction, and the discussion of ranks $\rank(p(m),q(n))$
between the definition of $C_{\tu{p,q}}(m,n)$ and
Proposition~\ref{prop:monotcolour}.)

\begin{prop}[Neighbour black-white vector with smaller
	rank]\label{prop:neighbour}
Let
	\begin{center}
	$C_{\tu{p,q}}(m,n)=\mathrm{black}$,
$C_{\tu{p,q}}(m',n')=\mathrm{white}$
	\end{center}
		where $m>0$ and $n'>0$.
Then there are $p',q'\in Q$ and $i,j\in\{-1,0,1\}$ such that
\begin{center}
$C_{\tu{p',q'}}(m{+}i,n{+}j)=\mathrm{black}$,
$C_{\tu{p',q'}}(m'{+}i,n'{+}j)=\mathrm{white}$,
\end{center}
and  \ $\rank(p'(m'{+}i),q'(n'{+}j))<\rank(p(m'),q(n'))$.
\end{prop}
\begin{proof}
  Let the assumptions hold; we put $\rank(p(m'),q(n'))=r\in\Nat$.
We can thus fix a transition  $p(m')\gt{a}p'(m'{+}i)$
(related to a rule $p\gt{a,i}p'$)
such that for each
 $q(n')\gt{a}q'(n'{+}j)$ we have
 $\rank(p'(m'{+}i),q'(n'{+}j))<r$.
Since $m>0$, we have $p(m)\gt{a}p'(m{+}i)$, and since
$p(m)\simul q(n)$, there is a transition $q(n)\gt{a}q'(n{+}j)$ such
that $p'(m{+}i)\simul q'(n{+}j)$; since $n'>0$, we also have
$q(n')\gt{a}q'(n'{+}j)$.
Hence $C_{\tu{p',q'}}(m{+}i,n{+}j)=\mathrm{black}$,
and $C_{\tu{p',q'}}(m'{+}i,n'{+}j)=\mathrm{white}$, and
	$\rank(p'(m'{+}i),q'(n'{+}j))<r$.
\end{proof}

Now we aim to formulate a crucial ingredient of the proof of
Lemma~\ref{lem:qualitative}, namely Proposition~\ref{prop:leftline}
and its corollary,
for which we need further notions.

\paragraph{Point-to-line distance, perpendicular lines \texorpdfstring{$\ell^{\bot}_b$}{l^bot_b}.}
For an integer point $\textsc{p}$ (in $\Zset\times\Zset$) and a line $\ell$,
by $\dist(\textsc{p},\ell)$ we mean the standard (Euclidean) distance
of $\textsc{p}$ to $\ell$.
Given
a line $\ell$ where $\slope(\ell)\in\langle 0,\infty\rangle $
and a \emph{level}
(which might be also called a \emph{bottom-level})
$b\in\Real_{\geq 0}$,
by $\ell^{\bot}_b$ we denote
the line that is perpendicular to $\ell$, intersects the horizontal
and/or vertical axis in the first quadrant, and
	$\dist((0,0),\ell^{\bot}_b)=b$. (Hence
	$(\ell')^{\bot}_b=\ell^{\bot}_b$ for any $\ell'$ that is
	parallel with $\ell$.
See
	Figure~\ref{fig:perplin-area}a.)

\paragraph{Areas,
	border and interior points of areas.}
By an \emph{area} we mean just a set $A\subseteq \Nat\times\Nat$ of
	points.
	A point $\textsc{p}\in A$ is a \emph{border point of} $A$ if
	it has a neighbour point outside $A$ (hence in
	$(\Nat\times\Nat)\smallsetminus A$);
	we put $\border(A)=\{\textsc{p}\mid \textsc{p}$ is a border
	point of $A\}$.
	Each point $\textsc{p}\in A\smallsetminus\border(A)$
is an \emph{interior point of} $A$.
(E.g., $(0,0)$ is an interior point of $A=\{(0,0), (0,1), (1,0),
(1,1)\}$.)

\begin{figure}
  \centering
  \begin{subfigure}[t]{0.495\textwidth}
    \centering
\begin{tikzpicture}[scale=0.7,every node/.style={scale=0.8}]


  \draw[->] (0,0) -- (9,0);
  \draw[->] (0,0) -- (0,9);
  
  \coordinate (a) at (5,-1);
  \coordinate (b) at (15,9);
  \coordinate (c) at (1,7);
  \coordinate (d) at ($(a)!(1,7)!(b)$);
  \coordinate (e) at (7,8);
  \coordinate (f) at (1,2);
  
  \draw (c) -- (d);
  \draw (f) -- (e);

  \draw[<->,dotted] (0,0) -- ($(c)!(0,0)!(d)$) node[midway,anchor=west,xshift=2mm] {$b$};

  \node[anchor=west] at (d) {$\ell^{\bot}_b$};
  \node[anchor=west] at (e) {$\ell$};
  
\end{tikzpicture}
    \caption{The perpendicular line $\ell^{\bot}_b$.}
    \label{fig:perplin-area1}
    \end{subfigure}
    \begin{subfigure}[t]{0.495\textwidth}
      \centering

\begin{tikzpicture}[scale=0.7,every node/.style={scale=0.8}]
  
  \draw[->] (0,0) -- (9,0);
  \draw[->] (0,0) -- (0,9);
  
  \coordinate (a) at (5,-1);
  \coordinate (b) at (15,9);
  \coordinate (c) at (1,6);
  \coordinate (d) at ($(a)!(c)!(b)$);
  \coordinate (e) at (7,8);
  \coordinate (f) at (1,2);

  \coordinate (g) at (3,2);
  \coordinate (h) at (8,4);

  \draw[name path=P1] (c) -- (d);
  \draw[name path=P2] (f) -- (e);
 
  \draw[<->,dotted] (0,0) -- ($(c)!(0,0)!(d)$) node[midway,anchor=west,xshift=2mm] {$b$};

  \node[anchor=west] at (d) {$\ell^{\bot}_b$};
  \node[anchor=west] at (e) {$\ell$};

  \draw[name path=P3] (g) -- (h);
  \node[anchor=west] at (h) {$\ell'$};
  \path [name intersections={of=P1 and P2,by=I1}];
  \path [name intersections={of=P1 and P3,by=I2}];

  \fill[color=gray!20!white] (I1) -- (I2) -- (h) -- (e) -- (I1);

  \draw[name path=P3] (g) -- (h);
  \draw[name path=P1] (c) -- (d);
  \draw[name path=P2] (f) -- (e);

  \draw[dotted] ($ (I1) + (0.5, 0) $)  -- ($ (e)  + (0, -0.5)$);
  \draw[dotted] ($ (I1) + (0.5, 0) $)  -- ($ (I2) + (0.1,0.4)$);
  \draw[dotted] ($ (I2) + (0.1,0.4) $) -- ($ (h)  + (0,0.32)$);

\end{tikzpicture}
    \caption{\textnormal{$\area((\ell,b),\ell')$}.}
    \label{fig:perplin-area2}
  \end{subfigure}
  \caption{}
  \label{fig:perplin-area}
\end{figure}

\paragraph{Line-level-line areas
\texorpdfstring{\textnormal{$\area((\ell,b),\ell')$}}{Area((l,b),l')} and
\texorpdfstring{\textnormal{$\area(\ell,(b,\ell'))$}}{Area(l,(b,l'))}.} (See
Figure~\ref{fig:perplin-area}b.)
Given two lines $\ell,\ell'$ where
$\infty\geq\slope(\ell)\geq\slope(\ell')\geq 0$,
and
a bottom-level $b\in\Real_{\geq 0}$,
we define
	$\area((\ell,b),\ell')$
as the set of points (in $\Nat\times\Nat$)
that are below $\ell$, above $\ell'$, and also above
	the line $\ell^{\bot}_b$; an exception is the case with
	$\slope(\ell)=0$ (hence $\slope(\ell')=0$ as well)
	where the points
	in $\area((\ell,b),\ell')$ are to the right of
	(the vertical line) $\ell^{\bot}_b$.

By $\area(\ell,(b,\ell'))$ we mean the set of points
that are below $\ell$, above $\ell'$, and also above
the line $(\ell')^{\bot}_b$; again, an exception is the case
with $\slope(\ell')=0$, where the points in $\area(\ell,(b,\ell'))$ are to the right of
(the vertical line) $(\ell')^{\bot}_b$.

In fact, we will not encounter the ``pathological'' case
 where $\slope(\ell)=\slope(\ell')$ and $\ell'$ is
strictly above (to the left of) $\ell$,
in which case the sets $\area((\ell,b),\ell')$ and $\area(\ell,(b,\ell'))$ are
empty. In the special case where $\slope(\ell)=\infty$ and
$\slope(\ell')=0$, we only consider
the sets $\area((\ell,b),\ell')$ where  $(\ell)^{\bot}_b$ coincides with
 $\ell'$, and the sets $\area(\ell,(b,\ell'))$ where  $(\ell')^{\bot}_b$ coincides with
 $\ell$.

\paragraph{Special cases of line-level-line areas:
\texorpdfstring{\textnormal{$\area(\leftofline{\ell})$}}{Area((<-l))} and
\texorpdfstring{\textnormal{$\area(\rightofline{\ell})$}}{Area(l->)}.}
Given a line $\ell$, with $\slope(\ell)\in\langle 0,\infty\rangle $, we put
	$\area(\leftofline{\ell})=\{\,\textsc{p}\in\Nat\times\Nat\mid\textsc{p}$
	is above $\ell\, \}$, and
		$\area(\rightofline{\ell})=\{\,\textsc{p}\in\Nat\times\Nat\mid\textsc{p}$
	is below $\ell\, \}$.
(Hence $\area(\leftofline{\ell})=\area((\textsc{v-axis},0),\ell)$
and $\area(\rightofline{\ell})=\area(\ell,(0,\textsc{h-axis}))$.)

\paragraph{Border points of \texorpdfstring{\textnormal{$\area((\ell,b),\ell')$}}{area((l,b),l')} along \texorpdfstring{$\ell$}{l},
along \texorpdfstring{$\ell'$}{l'}, and along \texorpdfstring{$\ell^{\bot}_b$}{l^bot_b}.}
(See the dotted boundaries in Figure~\ref{fig:perplin-area}b.)
Given  a~point $\textsc{p}\in \border(\area((\ell,b),\ell'))$,
we say that $\textsc{p}$
\emph{lies along}
$\ell$ if $\textsc{p}\in \border(\area(\rightofline{\ell}))$, hence if
$\textsc{p}$ has a neighbour point (integer point in the first quadrant)
strictly above $\ell$; $\textsc{p}$ \emph{lies along}
$\ell'$ if $\textsc{p}\in \border(\area(\leftofline{\ell'}))$.
If $\textsc{p}\in \border(\area((\ell,b),\ell'))$ does not lie along
$\ell$ nor along $\ell'$, then we say that $\textsc{p}$ \emph{lies
along} $\ell^{\bot}_b$.

\paragraph{D-sizes of vectors related to \texorpdfstring{$\gamma\in\langle 0,\infty\rangle $}{gamma € <0,infty>}, \texorpdfstring{$\gamma$}{gamma}-d-size and
co-\texorpdfstring{$\gamma$}{gamma}-d-size.} (See Figure~\ref{fig:vector-sizes2}.)
For
$\gamma\in\langle 0,\infty\rangle $,
the \emph{$\gamma$-d-size of} $v$ and
the \emph{co-$\gamma$-d-size of} $v$ are the following real numbers.
If $\spt(v)=\ept(v)$, then both $\gamma$-d-size and
co-$\gamma$-d-size of $v$ are $0$.
If $\spt(v)\neq\ept(v)$, then
we consider the $\gamma$-line $\ell$ going through $\spt(v)$, and the
line $\ell'$ that is perpendicular to $\ell$ and goes through
$\ept(v)$.
The $\gamma$-d-size of $v$ is $\dist(\spt(v),\ell')$
(hence nonnegative)
if $\slope(\ell')\neq \infty$ (hence $\gamma=\slope(\ell)>0$) and $\spt(v)$ is
below $\ell'$, or if $\slope(\ell')=\infty$ and $\spt(v)$ is
to the left of $\ell'$; the $\gamma$-d-size of $v$ is
$-\dist(\spt(v),\ell')$ (hence nonpositive)
otherwise.
The co-$\gamma$-d-size of $v$
is $\dist(\ept(v),\ell)$ (hence nonnegative)
if $\slope(\ell)\neq\infty$ (hence $0\leq\gamma=\slope(\ell)<\infty$)
and
$\ept(v)$ is above $\ell$, or if
 $\slope(\ell)=\infty$ and $\ept(v)$ is
to the left of $\ell$;
 the co-$\gamma$-d-size of $v$ is
$-\dist(\ept(v),\ell)$ (hence nonpositive)
otherwise.
(If $\gamma=\slope(v)$, then the  $\gamma$-d-size of
$v$ coincides with
the previously defined d-size of $v$.)

\begin{figure}
  \begin{subfigure}[t]{0.495\textwidth}
    \begin{tikzpicture}[scale=0.7,every node/.style={scale=0.8}]

  
  \draw[->] (0,0) -- (9,0);
  \draw[->] (0,0) -- (0,10);
  
  \coordinate (a) at (2.7,-1);
  \coordinate (b) at (12.7,9);
  \coordinate (c) at (0.5,4);
  \coordinate (d) at ($(a)!(c)!(b)$);
  \coordinate (e) at (7,8);
  \coordinate (f) at (1,2);

  \coordinate (g) at (3,1);
  \coordinate (h) at (8,3);
  
  \draw[name path=P1] (c) -- (d); 
  \draw[name path=P2] (f) -- (e); 
  
  \node[anchor=west] at (d) {$\ell^{\bot}_b$};
  \node[anchor=west] at (e) {$\ell$ (slope $\rho$)};

  \draw[name path=P3] (g) -- (h);
  \node[anchor=west] at (h) {$\ell'$};

  \coordinate (h1) at (7,7.5);
  \coordinate (h2) at (2.25,2.75);
  \coordinate (h3) at (3.4,1.6);
  \coordinate (h4) at (7.6,3.2);

  \coordinate (pc) at (4.5,3);
  \drawpoint{(pc)}{black};
  \node[anchor=north west] at (pc) {\small $\textsc{p}_C$};

  
  \draw[dotted] (h1) -- (h2) -- (h3) -- (h4);

   \foreach \x in {2.25,2.5,...,4.5} {
     \drawpoint{(\x,3)}{black};
   }
   
   \foreach \x in {2.5,2.75,...,4.5} {
     \drawpoint{(\x,3.25)}{black};
   }
   
   \foreach \x in {2.75,3,...,4.5} {
     \drawpoint{(\x,3.5)}{black};
   }

   \foreach \x in {3,3.25,...,4.5} {
     \drawpoint{(\x,3.75)}{black};
   }

   \foreach \x in {3.25,3.5,...,4.5} {
     \drawpoint{(\x,4)}{black};
   }

   \foreach \x in {3.5,3.75,...,4.5} {
     \drawpoint{(\x,4.25)}{black};
   }

   \foreach \x in {3.75,4,...,4.5} {
     \drawpoint{(\x,4.5)}{black};
   }

   \foreach \x in {4,4.25,...,4.5} {
      \drawpoint{(\x,4.75)}{black};
   }

   \foreach \x in {4.25,4.5,...,4.5} {
      \drawpoint{(\x,5)}{black};
   }

   \drawpoint{(4.5,5.25)}{black};

   \drawpoint{(2,2.75)}{black};

   \foreach \y in {2.5,2.75} {
     \drawpoint{(2.25,\y)}{black};
   }
   

   \foreach \y in {2.25,2.5,2.75} {
     \drawpoint{(2.5,\y)}{black};
   }

   \drawpoint{(2.75,2)}{white};
   \drawpoint{(3,1.75)}{white};
   \drawpoint{(3.25,1.5)}{white};
   \drawpoint{(3.5,1.5)}{white};
   \drawpoint{(3.75,1.5)}{white};
   \drawpoint{(4,1.5)}{white};
   \drawpoint{(4.25,1.75)}{white};
   \drawpoint{(4.5,1.75)}{white};
   \drawpoint{(4.75,2)}{white};
   \drawpoint{(5,2)}{white};
   \drawpoint{(5.25,2)}{white};
   \drawpoint{(5.5, 2.25)}{white};
   \drawpoint{(5.75,2.25)}{white};
   \drawpoint{(6,2.5)}{white};
   \drawpoint{(6.25,2.5)}{white};
   \drawpoint{(6.5,2.5)}{white};
   \drawpoint{(6.75,2.75)}{white};
   \drawpoint{(7,2.75)}{white};

   \draw[dotted] (2.25,0.75) -- (8.25,6.75);
   
  






  


  

\end{tikzpicture}
    \caption{Assumptions of Proposition~\ref{prop:leftline}.}
    \label{fig:prop-first1}
  \end{subfigure}
    \begin{subfigure}[t]{0.495\textwidth}
      %
%
\begin{tikzpicture}[scale=0.7,every node/.style={scale=0.8}]

  \draw[->] (0,0) -- (9,0);
  \draw[->] (0,0) -- (0,10);
  
  \coordinate (a) at (2.7,-1);
  \coordinate (b) at (12.7,9);
  \coordinate (c) at (0.5,4);
  \coordinate (d) at ($(a)!(c)!(b)$);
  \coordinate (e) at (7,8);
  \coordinate (f) at (1,2);

  \coordinate (g) at (3,1);
  \coordinate (h) at (8,3);
  
  \draw[name path=P1] (c) -- (d); 
  \draw[name path=P2] (f) -- (e); 
  
  \node[anchor=west] at (d) {$\ell^{\bot}_b$};
  \node[anchor=west] at (e) {$\ell$ (slope $\rho$)};

  \draw[name path=P3] (g) -- (h);
  \node[anchor=west] at (h) {$\ell'$};



  \coordinate (bp1) at (2.7,2.7);
  \coordinate (hlp1) at (10,10);
  \coordinate (bp2) at (2.2,8);
  \coordinate (bp3) at (1.7,9.5);
  \coordinate (bp4) at (2.7,9.3);
  \coordinate (bp5) at (4.25,6.9);

  \coordinate (i1) at ($(bp1)!(bp2)!(hlp1)$);
  \coordinate (i2) at ($(bp1)!(bp3)!(hlp1)$);
  \coordinate (i3) at ($(bp1)!(bp4)!(hlp1)$);

  \draw[dotted]
  (bp1) -- (i3)
  (i1) -- (bp2)
  (i2) -- (bp3)
  (i3) -- (bp4);
  
  \drawpoint{(bp1)}{black};
  \drawpoint{(bp2)}{black};
  \drawpoint{(bp3)}{black};
  \drawpoint{(bp4)}{black};

  \drawvec{(bp1)}{black}{(bp5)}{white}{solid}
  
  \coordinate (lstart) at (-0.5,4.75);
  \coordinate (lend)   at (4,9.25);
  \draw (lstart) -- (lend);

  \coordinate (foo) at (4.8,8);
  \coordinate (bi1) at ($(lstart)!(foo)!(lend)$);
  \coordinate (bi2) at ($(e)!(foo)!(f)$);

  \draw[{Latex[scale=1.1]}-{Latex[scale=1.1]}, dotted] (bi1) -- (bi2) node[midway, anchor=south west,yshift=-0.1cm,xshift=-0.1cm] {$>B_0$};
  
  \node[anchor=east,xshift=-0.15cm] at (lend) {$\bar{\ell}$};
  \node at (3.3,5.3) {$v$};

\end{tikzpicture}
      \caption{$B_0$ in the proof yields $\bar{\ell}$.}
      \label{fig:prop-first2}
  \end{subfigure}
	\caption{}
  \label{fig:prop-first}
\end{figure}

\medskip

	The following crucial technical proposition is illustrated in
	Figure~\ref{fig:prop-first}. It will be used several times
	later to bound the up-limits under the described
	circumstances.

	\begin{prop}[Bounding the up-limits $\beta_C$ for a class $\calC$ of
		colourings]\label{prop:leftline}\ \\
	We assume a~nonempty set $\calC$ of colourings (from the set
	$\{C_{\tu{p,q}}\mid p,q\in Q\}$),
	two lines $\ell,\ell'$ and a level $b\in\Real_{\geq 0}$
		such that the following conditions are satisfied
		(cf.~Figure~\ref{fig:prop-first1}):
	\begin{enumerate}
		\item
			$\infty\geq\slope(\ell)\geq\slope(\ell')\geq
			0$,
			and $\area((\ell,b),\ell')$ has no point on
			the vertical axis;
		\item
			each colouring $C\not\in\calC$ is
			monochromatic (all-black or all-white) inside
			$\area((\ell,b),\ell')$;
\item
for each  $C\in\calC$:
			\begin{enumerate}
				\item the points in
					$\border(\area((\ell,b),\ell'))$ that
					lie along $\ell'$ are white in $C$,
				\item
we can fix a point $P_C$
such that $C(P_C)=\mathrm{black}$,
					$P_C$ is an interior point of $\area((\ell,b),\ell')$,
and $\dist(\textsc{p}_C,\ell)\geq \textsc{bd}_C$ where we put
\[
\textsc{bd}_C=\sup\,\{\,\dist(\textsc{p},\ell)\mid
\textsc{p}\in\border(\area((\ell,b),\ell')), C(\textsc{p})=\mathrm{black}\}.
\]
					(Hence there is no point
					in
					$\border(\area((\ell,b),\ell'))$
					that is black in $C$ and has
				a~larger distance from $\ell$
					than $P_C$.)
\end{enumerate}
	\end{enumerate}
	Then there is a line $\bar{\ell}$
	with $\slope(\bar{\ell})=\slope(\ell)$
	such that all points above
	$\bar{\ell}$ are black in each $C\in\calC$.
\end{prop}
Before proving the proposition, we
 note its trivial corollary:
\begin{cor}\label{cor:leftlinetriv}
Under the conditions in Proposition~\ref{prop:leftline},
 $\slope(\ell)\geq \beta_C$ for each $C\in\calC$.
\end{cor}
\begin{proof} To prove Proposition~\ref{prop:leftline}, we let
its assumptions hold, and we put $\rho=\slope(\ell)$.
	If $\rho=\infty$, then the claim is trivial: there are no
	points above, i.e.\ left of, any vertical line $\bar{\ell}$
	that does not intersect the first quadrant; such  $\bar{\ell}$
	trivially satisfies the claim.
Hence we further
	assume
	\[\infty>\rho=\slope(\ell)\geq\slope(\ell')\geq 0.\]

We also note that for almost all $m\in\Nat$ there is $n$ such that
	$(m,n)\in\area((\ell,b),\ell')$; this is obvious if
	$\slope(\ell)>\slope(\ell')$, and in the case
	$\rho=\slope(\ell)=\slope(\ell')$ it is guaranteed by the
	existence of interior
	points of $\area((\ell,b),\ell')$ (they exist due to the
	points $P_C$): if $\textsc{p}=(m,n)$ is
	interior in $\area((\ell,b),\ell')$, then
	we have $(m,n{+}1)\in\area((\ell,b),\ell')$
and
	for any $m'\geq m$
	there is $n'$ such that $(m',n')$ lies below
 $\ell$ and above the
	parallel $\rho$-line going through $\textsc{p}$.

	We call a~\emph{vector} $v$
	\emph{eligible} if $\spt(v)\in\area((\ell,b),\ell')$,
$v$ is black-white in some colouring,
and both the $\rho$-d-size of $v$ and
the co-$\rho$-d-size of $v$ are nonnegative.
(We note that $\ept(v)$ can be outside
 $\area((\ell,b),\ell')$, but then it necessarily lies
	strictly
	above $\ell$.)
E.g., $v$ in Figure~\ref{fig:prop-first2}
	is eligible.

	To finish the proof, it suffices to show that the  co-$\rho$-d-sizes
	of eligible vectors are bounded by some	$B_0\in\Real_{\geq
	0}$; the claimed line $\bar{\ell}$ then surely exists
	(as is illustrated in Figure~\ref{fig:prop-first2}):
	we can take $\bar{\ell}$ above $\ell$ in the distance (to
	$\ell$) greater
	than $B_0$ so that, moreover, $\bar{\ell}$ intersects
	\textsc{v-axis} above the vertical coordinates of all $P_C$,
	$C\in\calC$.

For the sake of contradiction, we assume  that the  co-$\rho$-d-sizes
	of eligible vectors are not bounded; hence
for every $B\in\Real_{\geq 0}$
	the set
	\begin{center}
	$E_B=\{\,v\mid v$ is
	eligible and the co-$\rho$-d-size of $v$ is greater than $B\,\}$
	\end{center}
is nonempty.
	We first note that there is some $B'\in\Real_{\geq 0}$ such that for each
$B\geq B'$ and each $v\in E_B$ we have that
	$v$ is not black-white in any $C\not\in\calC$
	(and thus must be black-white in some $C\in\calC$).

We verify this claim by considering a fixed colouring
	$C\not\in\calC$, which is
 monochromatic in $\area((\ell,b),\ell')$, and by noting that for
	each $B\in\Real_{\geq 0}$ and each $v\in E_B$ we have:
	\begin{itemize}
		\item
If
	$C$ is all-white in $\area((\ell,b),\ell')$, then
			$C(\spt(v))=\mathrm{white}$ (and thus $v$ is not
			black-white in $C$).
\item
If $C$ is  all-black in
			$\area((\ell,b),\ell')$ (and thus
			$C(\spt(v))=\mathrm{black}$), and
			$C(\ept(v))=\mathrm{white}$,
			then $\ept(v)$ is outside  $\area((\ell,b),\ell')$,
			hence strictly above $\ell$; let us consider
			this case, where
			$\ept(v)=(m,n)$. Then
			$C(m,n')=\mathrm{white}$ for all $n'\leq n$
			 (by
			Proposition~\ref{prop:monotcolour}). This
			entails that for this $m$ there is no $n'$
			such that $(m,n')\in \area((\ell,b),\ell')$.
			As discussed above, this can be the case only
			for finitely many $m$, which entails that
			there are only finitely many eligible vectors
			that are black-white in some $C\not\in\calC$.
			Hence the existence of $B'$ is clear.
	\end{itemize}

Now for each $B\geq B'$ we fix a vector $v_B\in E_B$
	with the least possible rank of its white end;
thus $v_B$ is black-white in some $C_{\tu{p,q}}\in\calC$,
	$\ept(v_B)=(m,n)$,
	and
$\rank(p(m),q(n))$ is the least possible, when considering all $v\in E_B$ and all
	$C\in\calC$.
	Since  $\spt(v_B)$ is in $\area((\ell,b),\ell')$, it is
	not on the vertical axis, and thus
Proposition~\ref{prop:neighbour} entails that
	$\spt(v_B)\in\border(\area((\ell,b),\ell'))$ (otherwise $v_B$
	has a~neighbour
	vector that is also in $E_B$ and its white end has a lesser rank).

        \begin{figure}
          \centering
           \begin{subfigure}[t]{0.495\textwidth}
             \centering
		  %
%
\begin{tikzpicture}[scale=0.7,every node/.style={scale=0.8}]

  
  \draw[->] (0,0) -- (0,9);
  \draw[->] (0,0) -- (9,0);

  \coordinate (els) at (0.5,3);
  \coordinate (ele) at (6,8.5);

  \coordinate (elbs) at (0.5,4);
  \coordinate (elbe) at (4,0.5);

  \coordinate (elprimes) at (2.5,1);
  \coordinate (elprimee) at (8.5,3);

  \draw (els) -- (ele);
  \node[anchor=west] at (ele) {$\ell$ (slope $\rho$)};

  \draw (elbs) -- (elbe);
  \node[anchor=west] at (elbe) {$\ell^\bot_b$};

  \draw (elprimes) -- (elprimee);
  \node[anchor=west] at (elprimee) {$\ell'$};

  \coordinate (bp1) at (2,2.75);
  \coordinate (wp)  at (4.5,8);
  \coordinate (bp2) at (4.5,3);


  \coordinate (i2) at (5.75,6.75);
  \coordinate (i3) at (7,5.5);
  
  \draw[dotted] (bp1) -- (i2) -- (wp);
  \draw[dotted] (bp2) -- (i3) -- (wp);

  \drawvec{(bp1)}{black}{(wp)}{white}{solid}
  \drawvec{(bp2)}{black}{(wp)}{white}{solid}
  

  \node[anchor=west,xshift=0.1cm] at (bp2) {$\textsc{p}_C$};
  \node[anchor=west] at (1.75,4) {$v_B$};

\end{tikzpicture}
          \caption{$\rho$-d-size of $(\textsc{p}_C,\ept(v_B))$ is nonnegative.}
          \label{fig:prop-second1}
        \end{subfigure}
          \begin{subfigure}[t]{0.495\textwidth}
            \centering
            \begin{tikzpicture}[scale=0.7,every node/.style={scale=0.8}]


  \draw[->] (0,0) -- (0,9);
  \draw[->] (0,0) -- (9,0);

  \coordinate (els) at (0.5,3);
  \coordinate (ele) at (6,8.5);

  \coordinate (elbs) at (0.5,4);
  \coordinate (elbe) at (4,0.5);

  \coordinate (elprimes) at (2.5,1);
  \coordinate (elprimee) at (8.5,3);

  \draw (els) -- (ele);
  \node[anchor=west] at (ele) {$\ell$ (slope $\rho$)};

  \draw (elbs) -- (elbe);
  \node[anchor=west] at (elbe) {$\ell^\bot_b$};

  \draw (elprimes) -- (elprimee);
  \node[anchor=west] at (elprimee) {$\ell'$};

  \coordinate (bp1) at (2,3);
  \coordinate (wp)  at (3,6);
  \coordinate (bp2) at (8,7);

  \coordinate (i1) at (2,7);
  \coordinate (i2) at (4,5);
  \coordinate (i3) at (5,4);
  

  \foreach \x in {2,2.25,...,7.75}
  {
    \drawpoint{(\x,7)}{black};
  }

  \foreach \y in {3.25,3.5,...,6.75}
  {
    \drawpoint{(2,\y)}{black};
  }

  \draw[dotted] (bp1) -- (i2) -- (wp);
  \draw[dotted] (bp2) -- (i3) -- (wp);

  \drawvec{(bp1)}{black}{(wp)}{white}{solid}
  \drawvec{(bp2)}{black}{(wp)}{white}{solid}
  

  \node[anchor=west] at (bp2) {$\textsc{p}_C$};
  \node[anchor=west] at (2.5,4.5) {$v_B$};

\end{tikzpicture}
            \caption{$\rho$-d-size of $(\textsc{p}_C,\ept(v_B))$ is negative.}
            \label{fig:prop-second2}
        \end{subfigure}
        \caption{}
        \label{fig:prop-second}
        \end{figure}

For any fixed $B\geq B'$, by $3(b)$ we thus have
	$\dist(\spt(v_B),\ell)\leq\dist(\textsc{p}_C,\ell)$ for the
	respective colouring $C=C_{\tu{p,q}}\in\calC$ (in which $\ept(v_B)$
	has the least possible rank);
the vector $(\textsc{p}_C,\ept(v_B))$, which is also black-white in
	$C$, thus has the co-$\rho$-d-size also greater than $B$ (since
	it is not smaller than the co-$\rho$-d-size of $v_B$).
If the vector $(\textsc{p}_C,\ept(v_B))$ were eligible
(as is depicted in Figure~\ref{fig:prop-second1}), thus
belonging to $E_B$,
	Proposition~\ref{prop:neighbour} would yield a contradiction
	(since $\textsc{p}_C$ is an interior point of
	$\area((\ell,b),\ell')$, the vector $(\textsc{p}_C,\ept(v_B))$
 would have a neighbour vector
	that is also in $E_B$
	and its white end has a lesser rank).
Hence the $\rho$-d-size of
	$(\textsc{p}_C,\ept(v_B))$ is negative
	(as depicted in Figure~\ref{fig:prop-second2}); this entails that
$\rho=\slope(\ell)>0$ (since in the case $\rho=0$ the fact
$C(\textsc{p}_C)=\mathrm{black}$ would entail $C(\ept(v_B))=\mathrm{black}$, by
Proposition~\ref{prop:monotcolour}, which contradicts the
assumption $C(\ept(v_B))=\mathrm{white}$).

The fact that $\rho=\slope(\ell)>0$ and the $\rho$-d-size of
	$(\textsc{p}_C,\ept(v_B))$ is negative entails that
$\ept(v_B)$ is strictly below the line
that is	perpendicular to
	$\ell$ and goes through $\textsc{p}_C$; moreover,
$\ept(v_B)$ is below the horizontal line going through $\textsc{p}_C$
since $C(\textsc{p}_C)=\mathrm{black}$ and  $C(\ept(v_B))=\mathrm{white}$
(see Figure~\ref{fig:prop-second2}).
There are thus only finitely many points that can be
$\ept(v_B)$, independently of the chosen  $B\geq B'$.
In other words, the set $\{\ept(v_B)\mid B\geq B'\}$ is
finite, which also entails that the set
 $\{v_B\mid B\geq B'\}$ is finite (recall that $v_B$ is
 eligible, and thus the
 $\rho$-d-size of $v_B$ is nonnegative).
 This contradicts our choice of $v_B$ that entails
 that the co-$\rho$-d-size of $v_B$ is greater than $B$, for each
 $B\in\Real_{\geq 0}$.

 The  co-$\rho$-d-sizes
	of eligible vectors are thus indeed bounded by some
	$B_0\in\Real_{\geq 0}$.
\end{proof}

Besides Corollary~\ref{cor:leftlinetriv}, which is trivial,
we also derive the next corollary, showing that if the assumptions of 	Proposition~\ref{prop:leftline} are slightly
strengthened, then we get a strict upper bound on the up-limits
$\beta_C$.

\begin{cor}[Bounding the up-limits $\beta_C$
	strictly]\label{cor:leftline}\hfill\\
	If $\calC,\ell,\ell',b$ satisfy the conditions of
	Proposition~\ref{prop:leftline}
	and, moreover, $\slope(\ell)>\slope(\ell')$ and
	for each $C\in\calC$ we can choose
	$\textsc{p}_C$ (which is an interior point of
	$\area((\ell,b),\ell')$) so that it has no neighbour point on $\ell$ and
	the inequality  in $3(b)$ is strict
	($\dist(\textsc{p}_C,\ell)>\textsc{bd}_C$),
	then $\slope(\ell)>\beta_C$ for each $C\in\calC$.
\end{cor}
\begin{proof}
	Let the described conditions be satisfied (in fact, such a
	situation is
	depicted already in Figure~\ref{fig:prop-first1};
	Figure~\ref{fig:rotation-1} makes clear that all neighbour points
	of $\textsc{p}_C$ are strictly below $\ell$).

	Since $\rho=\slope(\ell)>\slope(\ell')\geq 0$,
	we can very slightly rotate $\ell$ to the
	right around the intersection-point of $\ell$ and
	$\ell^{\bot}_b$ (see Figure~\ref{fig:rotation-2}), by which we get
	$\ell''$ with a slope $\rho''$,
	$\rho>\rho''\geq\slope(\ell')$,
	so that the resulting $\area((\ell'',b''),\ell')$
	($\ell^{\bot}_b$ has rotated to ${(\ell'')}^{\bot}_{b''}$)
	is a subset of $\area((\ell,b),\ell')$
	(in other words, the rotation of $\ell^{\bot}_b$ to ${(\ell'')}^{\bot}_{b''}$
	is so small that there is no [integer] point in
	$\area((\ell'',b''),\ell')\smallsetminus
	\area((\ell,b),\ell')$, i.e.\ in the grey triangle in Figure~\ref{fig:rotation-2})
	 and so that the original points $\textsc{p}_C$
	are interior points also in $\area((\ell'',b''),\ell')$
while the
	conditions of Proposition~\ref{prop:leftline}
	(in particular $\dist(\textsc{p}_C,\ell)\geq \textsc{bd}_C$)
	are kept. (To be very precise, in the case where
	$\slope(\ell)=\infty$ and $\slope(\ell')=0$ we recall our
	stipulation that
	$\ell^{\bot}_b$ coincides with $\ell'$; the discussed
	slight rotation then trivially has the property
that there is no point in $\area((\ell'',b''),\ell')\smallsetminus
	\area((\ell,b),\ell')$.)
	Hence
 Proposition~\ref{prop:leftline} entails that
	$\rho''\geq\beta_C$, and thus
	$\rho>\beta_C$ (since $\rho>\rho''$).
\end{proof}

\begin{figure}
  \begin{subfigure}[t]{0.49\textwidth}
    \centering
  %
%
\begin{tikzpicture}[scale=0.7,every node/.style={scale=0.8}]

  
  \draw[->] (0,0) -- (10,0);
  \draw[->] (0,0) -- (0,9);

  \coordinate (i) at (3,3);
  \coordinate (l1s) at (1,1);
  \coordinate (l1e) at (8,8);
  \coordinate (l2s) at (1,2);
  \coordinate (l2e) at (9,6);
  \coordinate (l1ps) at (1,5);
  \coordinate (l1pe) at (5,1);
  \coordinate (l2ps) at (1.8, 5.4);
  \coordinate (l2pe) at (4.2, 0.6);

  \coordinate (bp1) at (4,2.5);
  \coordinate (bp2) at (7, 3.5);
  \coordinate (bp2boxtl) at (6.5, 4); 

  \coordinate (i1) at ($(l1s)!(bp1)!(l1e)$);
  \coordinate (i2) at ($(l1s)!(bp2)!(l1e)$);

  \coordinate (rls) at (1,0.9);
  \coordinate (rle) at (9,2);

  \path[name path=P1] (l1ps) -- (l1pe);
  \path[name path=P2] (l2ps) -- (l2pe);
  \path[name path=LPRIME]  (rls) -- (rle);

  \draw (l1s) -- (l1e);
  \draw (l1ps) -- (l1pe);
  \node[anchor=south] at (l1e) {$\ell$};
  \node[anchor=west] at (l1pe) {$\ell^{\bot}_b$};
  
  \draw[dotted] (i1) -- (bp1);
  \draw[dotted] (i2) -- (bp2);

  \draw[dashed]
      (bp2) rectangle (bp2boxtl);
  
  \drawpoint{(bp1)}{black};

  \drawpoint{(bp2)}{black};
  \node[anchor=west] at (bp2) {$\textsc{p}_C$};

  \draw (rls) -- (rle);
  \node[anchor=west] at (rle) {$\ell'$};

\end{tikzpicture}
  \caption{Assumptions of Corollary \ref{cor:leftline} (the unit square shows that $\textsc{p}_C$ has no neighbour on $\ell$).}
  \label{fig:rotation-1}
\end{subfigure}\hfill %
\begin{subfigure}[t]{0.49\textwidth}
  \centering
  %
%
\begin{tikzpicture}[scale=0.7,every node/.style={scale=0.8}]

  \draw[->] (0,0) -- (10,0);
  \draw[->] (0,0) -- (0,9);
  

  
  



  \coordinate (i) at (3,3);
  \coordinate (l1s) at (1,1);
  \coordinate (l1e) at (8,8);
  \coordinate (l2s) at (1,2);
  \coordinate (l2e) at (9,6);
  \coordinate (l1ps) at (1,5);
  \coordinate (l1pe) at (5,1);
  \coordinate (l2ps) at (1.8, 5.4);
  \coordinate (l2pe) at (4.2, 0.6);

  \coordinate (bp1) at (4,2.5);
  \coordinate (bp2) at (7, 3.5);

  \coordinate (i1) at ($(l2s)!(bp1)!(l2e)$);
  \coordinate (i2) at ($(l2s)!(bp2)!(l2e)$);

  \coordinate (rls) at (1,0.9);
  \coordinate (rle) at (9,2);

  \path[name path=P1] (l1ps) -- (l1pe);
  \path[name path=P2] (l2ps) -- (l2pe);
  \path[name path=LPRIME]  (rls) -- (rle);

  \path[name intersections={of=P1 and LPRIME,by=i3}];
  \path[name intersections={of=P2 and LPRIME,by=i4}];
  
  \fill[color=gray!20!white] (i) -- (i3) -- (i4) -- (i);
  
  \draw (l1s) -- (l1e);
  \draw (l1ps) -- (l1pe);
  \node[anchor=south] at (l1e) {$\ell$};
  \node[anchor=west] at (l1pe) {$\ell^{\bot}_b$};

  \draw[dashed] (l2s) -- (l2e);
  \draw[dashed] (l2ps) -- (l2pe);
  \node[anchor=west] at (l2e) {$\ell''$};
  \node[anchor=east] at (l2pe) {$(\ell'')^{\bot}_{b''}$};
  
  \draw[dotted] (i1) -- (bp1);
  \draw[dotted] (i2) -- (bp2);
  
  \drawpoint{(bp1)}{black};
  \drawpoint{(bp2)}{black};
 \node[anchor=west] at (bp2) {$\textsc{p}_C$};
  
  \draw (rls) -- (rle);
  \node[anchor=west] at (rle) {$\ell'$};

  \draw[->] (8.5,8) .. controls (8.9,8) and (9.3,7.5) .. (9.5,6.5);
  
\end{tikzpicture}
  \caption{Rotation is so slight that the grey area
    contains no integer points, except possibly on
    $\ell^{\bot}_b$.}
  \label{fig:rotation-2}
\end{subfigure}
\caption{}
\label{fig:rotation}
\end{figure}

\paragraph{Proof of Lemma~\ref{lem:qualitative}.}

We denote the down-limits, i.e.\ the elements of the set
$\{\alpha_C\mid C=C_{\tu{p,q}}, (p,q)\in Q\times Q\}$)
by $\alpha_1,\alpha_2,\cdots,\alpha_k$  ($k\leq |Q|^2$)
so that
\begin{equation}\label{eq:downlimits}
0\leq\alpha_1<\alpha_2\cdots <\alpha_k\leq\infty.
\end{equation}
We now assume that the statement of Lemma~\ref{lem:qualitative}
holds for the
colourings related to $\alpha_1,\alpha_2,\dots,\alpha_{i-1}$
($i\in\{1,2,\dots,k\}$), and
prove the statement for
the $\alpha_i$-colourings.
The proof is given by Claims~\ref{cl:betaalpha} -- \ref{cl:ellelb}.

\begin{clm}\label{cl:betaalpha}
	For each $\alpha_i$-colouring $C$ we have $\beta_C=\alpha_i$
	(hence the down-limit $\alpha_C=\alpha_i$ and the up-limit $\beta_C$
	coincide).
\end{clm}
\begin{proof}
If $\alpha_i=\infty$, then the claim is trivial
	(since $\infty\geq\beta_C\geq\alpha_C$, by
	Proposition~\ref{prop:betagreateralpha});
	hence we assume
	$\infty>\alpha_i$.
	We fix some slope $\rho$ where $\infty>\rho>\alpha_i$ and
	$\alpha_{i+1}>\rho$ if $i<k$.
	If $\alpha_i=0$, then we put $\rho'=0$,
	and if $\alpha_i>0$, then we fix $\rho'$ so that $\alpha_i>\rho'>0$ and
 $\rho'>\alpha_{i-1}$ if $i>1$.
We fix
a $\rho$-line $\ell$ and a $\rho'$-line $\ell'$,
	both going through the origin $(0,0)$. (We can recall $\ell$
	and $\ell'$ from
	Figure~\ref{fig:prop-first1}, and imagine that they go through
	$(0,0)$.)
	By $\calC$ we denote the set of all $\alpha_i$-colourings, and
	note:
	\begin{itemize}
		\item
for each $C\in\calC$ (where $\alpha_C=\alpha_i$), there are
	infinitely many black points
			in $\area((\ell,0),\ell')$, since
the $\rho$-line $\ell$ must be down-black in $C$ due to the fact
			$\rho>\alpha_i=\alpha_C$ (recall
			Proposition~\ref{prop:trivialdownup}),
			but almost all
			border points of $\area((\ell,0),\ell')$ along
			$\ell'$ (i.e. those having neighbours strictly
			below $\ell'$)
			are white:
if $\alpha_i=\alpha_C>\rho'$,
			then no $\rho'$-line can be down-black (in
			particular, no shift of $\ell'$ to the left can
			be down-black), and if
			$\alpha_i=\rho'=0$, then
$\ell'$ is the horizontal axis
and almost all points on $\ell'$ are
			interior, having no neighbours (i.e.
			neighbour integer points in the first quadrant)
			outside $\area((\ell,0),\ell')$, since
			$\rho=\slope(\ell)>0$;
\item
for each $C\not\in\calC$ (where $\alpha_C\neq\alpha_i$)
			 we have:
			\begin{itemize}
				\item
	if $\alpha_C>\alpha_i$, hence $\alpha_C=\alpha_j$ for some
					$j>i$,
					then only finitely many points of
					$\area((\ell,0),\ell')$ are
					black in $C$
 (since	$\alpha_C>\rho=\slope(\ell)$, and thus $\ell$ cannot be
					down-black in $C$);
				\item
					if $\alpha_C<\alpha_i$,
hence  $\alpha_C=\alpha_j$ for some
					$j<i$,
					then only finitely many points of
					$\area((\ell,0),\ell')$ are
					white in $C$
					(here
					$\alpha_i>\rho'=\slope(\ell')>\alpha_C=\alpha_j$,
					and we use that
					$\beta_C=\alpha_C=\alpha_j$ by
					the induction hypothesis of
	our proof of Lemma~\ref{lem:qualitative},
					which entails that $\ell'$ is
					not up-white in $C$).
			\end{itemize}
	\end{itemize}
Hence we can choose a sufficiently large level $b\in\Real_{\geq 0}$ so that the set
$\calC$ (of $\alpha_i$-colourings) and $\ell,\ell',b$
satisfy the conditions $1$, $2$, and $3(a)$ of
Proposition~\ref{prop:leftline};
in particular, the value
\[
\textsc{bd}_C=\sup\,\{\,\dist(\textsc{p},\ell)\mid
\textsc{p}\in\border(\area((\ell,b),\ell')), C(\textsc{p})=\mathrm{black}\}
\]
is
finite for each $C\in\calC$.

Since $\slope(\ell)=\rho>\alpha_i$,
we can fix some
$\rho''$ satisfying
$\rho>\rho''>\alpha_i$ and also
fix the~$\rho''$-line $\ell''$ going through $(0,0)$.
The line $\ell''$ is down-black
	in all $C\in \calC$ (by Proposition~\ref{prop:trivialdownup})
	and the fact
	$\slope(\ell)>\slope(\ell'')>\slope(\ell')$ entails
that for each $C\in \calC$
there are infinitely many points in $\area((\ell,b),\ell')$
that are below $\ell''$ and black
	in $C$.
Hence for
each $C\in\calC$
	the value $\sup\,\{\,\dist(\textsc{p},\ell)\mid
	\textsc{p}\in\area((\ell,b),\ell'), C(\textsc{p})=\mathrm{black}\}$ is
	infinite
(in each $C\in\calC$, the distance of
	the black points in $\area((\ell,b),\ell')$
	to $\ell$ increases above any bound when the level
	increases).

Thus the condition $3(b)$ of
Proposition~\ref{prop:leftline} is also satisfied
(for each $C\in\calC$, the point $\textsc{p}_C$ can be chosen in a bigger distance from $\ell$ than
the finite value $\textsc{bd}_C$,
as is also depicted in Figure~\ref{fig:prop-first1},
so even the assumptions of Corollary~\ref{cor:leftline} are satisfied);
Proposition~\ref{prop:leftline} thus
entails that
	$\rho\geq\beta_C$ for each $C\in\calC$.
In fact,  $\rho>\alpha_i$ could be chosen arbitrarily close to
$\alpha_i$;
we have thus shown  $\rho\geq\beta_C$ for each $\rho>\alpha_i$.
By recalling that $\beta_C\geq \alpha_C$
(Proposition~\ref{prop:betagreateralpha}), we derive that
$\beta_C=\alpha_C=\alpha_i$ for each $\alpha_i$-colouring $C$.
\end{proof}

\begin{clm}\label{cl:eller}
There is an $\alpha_i$-line $\ell_R$ such that all points below $\ell_R$
are white
	in all $\alpha_i$-colourings.
\end{clm}
\begin{proof}
	If $\alpha_i=0$ (which can happen for $i=1$), then the claim
	is trivial (we can consider a~horizontal line $\ell_R$ strictly
	below $\textsc{h-axis}$, below which there are no points
	[i.e. integer points in the first quadrant]).

	So we assume $\alpha_i>0$ and
we fix an $\alpha_i$-line $\ell$ such that all points below $\ell$
	are white in the maximum number  of $\alpha_i$-colourings.
Let $\calC$ be the set of remaining $\alpha_i$-colourings; hence
	each $\alpha_i$-line is down-black in each
	$C\in\calC$ (which has not been so far excluded though
	$\alpha_C=\alpha_i$ is the down-limit of $C$).
	We assume
	$\calC\neq\emptyset$ (otherwise we are done by putting
	$\ell_R=\ell$), and show a
	contradiction.

Each $C\in\calC$ has black points below $\ell$ in unbounded
	distances from $\ell$ (since each $\alpha_i$-line is down-black in each
	$C\in\calC$).
	We fix some $\ell'$ such that
	$\alpha_i=\slope(\ell)>\slope(\ell')\geq 0$, and
	$\slope(\ell')>\alpha_{i-1}$ if $i>1$. (We can again look at
	Figure~\ref{fig:prop-first1}.)
It is now a routine to verify that
there is
	some $b\in\Real_{\geq 0}$ such that $\calC$ and $\ell,\ell',b$
	satisfy the assumptions of Proposition~\ref{prop:leftline},
	and even the assumptions of Corollary~\ref{cor:leftline}.
	This entails that
	$\slope(\ell)=\alpha_i>\beta_C$, for each $C\in\calC$.
But since $\calC$ is a set of $\alpha_i$-colourings (hence
	$\alpha_C=\alpha_i$ for each $C\in\calC$), we must have
	 $\beta_C\geq\alpha_i$, for each $C\in\calC$ (by
	 Proposition~\ref{prop:betagreateralpha}).
	 Hence $\calC=\emptyset$ after all.
\end{proof}

We note that $\ell_R$ from Claim~\ref{cl:eller} can be
shifted to the right, which entails a useful corollary
(for which we recall that
$\area(\leftofline{\ell})=\area((\textsc{v-axis},0),\ell)$ and that
all border points of $\area(\leftofline{\ell})$ lie along $\ell$):

\begin{cor}
\label{cor:eller}
There is an $\alpha_i$-line $\ell_R$ such that all border points of
	$\area(\leftofline{\ell_R})$ are white
	in all $\alpha_i$-colourings.
\end{cor}
We note that the claim also comprises the case $\alpha_i=0$ with $\ell_R$ being below
$\textsc{h-axis}$, in which case $\area(\leftofline{\ell_R})$
has no border points at all.

Now we aim to show that there is an $\alpha_i$-line $\ell_L$
such that all points above $\ell_L$
are black
	in all $\alpha_i$-colourings, and that $\alpha_i$ is rational or
	$\infty$. (By this the proof of Lemma~\ref{lem:qualitative} will be
	finished.)
Since the case $\alpha_i=\infty$ is trivial, we further assume that
$\infty>\alpha_i$.

We will use the notion of the
rightmost down-black
$\alpha_i$-lines $\ell_C$ for $\alpha_i$-colourings $C$, after which we
again apply Proposition~\ref{prop:leftline}.
But, in fact, we can not a priori assume that the rightmost down-black
$\alpha_i$-lines exist and that $\alpha_i$ is rational,
so we partition and analyse the class $\calC$ of
$\alpha_i$-colourings without any such assumptions. We define the line $\ell_C$
as the ``right-hand limit'' for the respective down-black lines, if some
$\alpha_i$-lines that are down-black in $C$ exist, which does not
exclude
the case that $\ell_C$ itself is not down-black.

\paragraph{Partition of \texorpdfstring{$\alpha_i$}{alpha_i}-colourings by rightmost down-black
lines \texorpdfstring{$\ell_C$}{l_C}.}
Let $\calC$ be the set of all $\alpha_i$-colourings.
For technical convenience we use a fixed $\alpha_i$-line $\ell_R$
guaranteed by
Claim~\ref{cl:eller} (or Corollary~\ref{cor:eller}), to define the partition of $\calC$ into the
following classes:
	\begin{itemize}
		\item	$\calC_1$
	contains each $C\in\calC$ for which there is an
	$\alpha_i$-line that is down-black in $C$;
\\
			for each $C\in\calC_1$ we define the \emph{rightmost
			down-black line} $\ell_C$ as the $\alpha_i$-line
	whose distance from $\ell_R$ is the infimum of distances of
	the $\alpha_i$-lines that are down-black in $C$.
We further partition $\calC_1$ as follows:
			\begin{itemize}
				\item $\calC_{11}$ contains all
					$C\in\calC_1$ for
	which $\ell_C$ is down-black;
\item
$\calC_{12}$
contains all $C\in\calC_1$ for which $\ell_C$ is not down-black.
			\end{itemize}
\item	 $\calC_2=\calC\smallsetminus \calC_1$; hence for each
	$C\in\calC_2$ and each 	$\alpha_i$-line $\ell$
	only finitely many points below $\ell$ are black in $C$.
\end{itemize}

We now show that there is a desired ``left-bound-line'' for all $C\in\calC_1$, by
Claim~\ref{cl:ellela} that is preceded by a simple auxiliary fact in
Claim~\ref{cl:auxmu}; it turns out that $\calC_{12}$ is empty, while
we still do not exclude that  $\calC_{11}$ is empty as well.
Then we show that $\calC_2$ is empty (by
Claim~\ref{cl:ellelb}); hence the (nonempty) set $\calC$ of $\alpha_i$-colourings
is equal to $\calC_{11}$, and Claim~\ref{cl:ellela} thus  finishes
the proof of Lemma~\ref{lem:qualitative} (including the fact that
$\alpha_i$ is rational or $\infty$).

\begin{clm}[auxiliary]\label{cl:auxmu}\hfill
	\begin{enumerate}
		\item
			If $\gamma\in\Real_{\geq 0}$ is rational,
			then for any $\gamma$-line $\ell$ we have:
\begin{enumerate}
				\item
$\ell$ either contains infinitely many
		integer points (in $\Zset\times\Zset$), or no
		integer point;
	\item
there exists the least value $\mu\in\Real_{>0}$ such that
		$\mu=\dist(\textsc{p},\ell)$ for some integer point
		$\textsc{p}$ that is not on $\ell$.
			\end{enumerate}
		\item
If $\gamma\in\Real_{> 0}$ is irrational, then for each
	$B\in\Real_{> 0}$ there
	is a (tiny) real number $\mu>0$ such that for any two $\gamma$-lines
        $\ell,\ell'$ whose (Euclidean) distance is $\mu$ we have that
	the distance of any two different integer points lying
	between the lines $\ell$ and $\ell'$ is larger than $B$.
	\end{enumerate}
\end{clm}
\begin{proof}\par
	1.
	For $\gamma=0$ both parts of the claim are obvious; so we assume
that $\gamma=\frac{\Delta_y}{\Delta_x}$ for $\Delta_x,
	\Delta_y\in\Nat_{>0}$, and consider a $\gamma$-line $\ell$.
We note that
	if an integer point $(z_1,z_2)\in\Zset\times\Zset$ is
	on $\ell$, then the integer points
	$(z_1+i\cdot\Delta_x,z_2+i\cdot\Delta_y)$,  for all
	$i\in\Zset$, are on $\ell$ as well.
	Hence 1(a) is clear.
	Now we note that for any integer point  $\textsc{p}=(z_1,z_2)\in\Zset\times\Zset$
outside $\ell$ we have that
	$\dist(\textsc{p},\ell)=\dist(\textsc{p}',\ell)$ for all
	(integer) points
	$\textsc{p}'=(z_1+i\cdot\Delta_x,z_2+i\cdot\Delta_y)$ where
	$i\in\Zset$. Hence $\mu$ claimed in 1(b) is the least value in
	the set
\begin{equation*}
	\left\{\,\dist(\textsc{p},\ell)\mid
	\textsc{p}=(z_1,z_2)\in\Zset\times\Zset,
	\textsc{p}\not\in\ell, 0\leq z_2\leq \Delta_y\,\right\},
\end{equation*}
which
	obviously exists (since it exists for each fixed
	$z_2\in\{0,1,\dots,\Delta_y\}$).

	2.
We fix an irrational $\gamma\in\Real_{> 0}$
and some $B\in\Real_{>0}$.
	We note that for each vector $v=(\textsc{p},\textsc{p}')$
where $\textsc{p},\textsc{p}'$ are two different integer points and
	$\dist(\textsc{p},\textsc{p}')\leq B$
	(by $\dist(\textsc{p},\textsc{p}')$ we refer to the Euclidean
	distance of points)
we have that the absolute value of the co-$\gamma$-d-size of $v$ is
	greater than zero (since $\slope(v)$ is rational and thus differs from
	$\gamma$). Moreover, the set
	\begin{equation*}
    M_{\gamma,B} = \left\{\mu'~\middle|~\begin{array}{l}\mu' \text{ is the absolute value of the co-$\gamma$-d-size of some $(\textsc{p},\textsc{p}')$} \\ \text{where $\textsc{p},\textsc{p}'$ are two different integer points and $\dist(\textsc{p},\textsc{p}')\leq B$}\end{array}\right\}
	\end{equation*}
is obviously finite. We can thus choose $\mu$ satisfying $0<\mu<\min
	M_{\gamma,B}$; such $\mu$ satisfies the claim.
	Indeed:
if there were two $\gamma$-lines
$\ell,\ell'$
			whose (Euclidean) distance
is $\mu$ and two different integer points $\textsc{p},\textsc{p}'$
lying 	between $\ell$ and $\ell'$ and satisfying
$\dist(\textsc{p},\textsc{p}')\leq B$, then the absolute value of the
	co-$\gamma$-d-size of the vector $(\textsc{p},\textsc{p}')$
	would belong to  $M_{\gamma,B}$ and would be not bigger than
	$\mu$ --- a contradiction.
\end{proof}

\begin{clm}\label{cl:ellela}
If $\calC_1\neq\emptyset$, then there is an $\alpha_i$-line $\ell'_L$ such
	that all points above $\ell'_L$
are black
	in all colourings from $\calC_1$; moreover,
	$\alpha_i$ is rational and $\calC_{12}$ is empty. (We recall that
	$\infty>\alpha_i$, as stipulated after
	Corollary~\ref{cor:eller}.)
\end{clm}
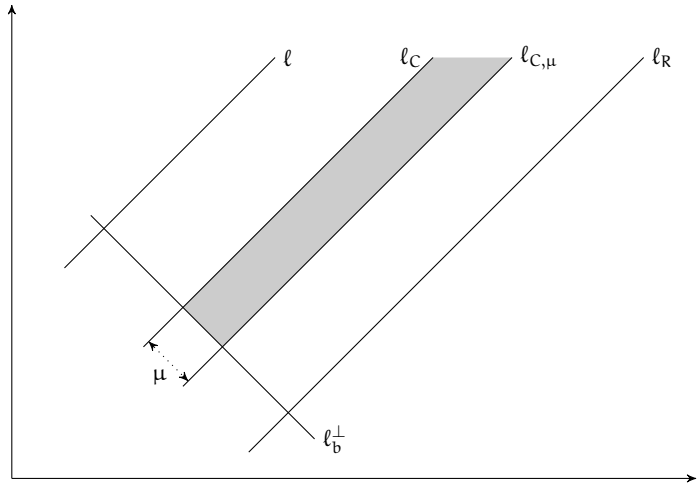
\begin{figure}[ht]
\centering
  %
\begin{tikzpicture}[scale=0.7,every node/.style={scale=0.8}]

  
  \draw[->] (0,0) -- (13,0);
  \draw[->] (0,0) -- (0,9);

  \coordinate (els) at (1,4);
  \coordinate (ele) at (5,8);


  \coordinate (elcs) at (2.5,2.5);
  \coordinate (elce) at (8,8);

  \coordinate (elrs) at (4.5,0.5);
  \coordinate (elre) at (12,8);

  \coordinate (elps) at (1.5,5);
  \coordinate (elpe) at ($(6,1)!(elps)!(13,8)$);


  \coordinate (ms) at (3.25,1.75);
  \coordinate (me) at (9.5,8);

  \path[name path=p1] (elps) -- (elpe);
  \path[name path=p2] (elcs) -- (elce);
  \path[name path=p3] (elrs) -- (elre);
  \path[name path=pm] (ms) -- (me);

  \path[name intersections={of=p1 and p2,by=i1}];
  \path[name intersections={of=p1 and pm,by=im}];
  \path[name intersections={of=p1 and p3,by=i2}];

  \fill[color=gray!40!white]
  (i1) -- (im) -- (me) -- (elce) -- (i1);

  \draw (els) -- (ele);
  \node[anchor=west] at (ele) {$\ell$};

  \draw (elcs) -- (elce);
  \node[anchor=east,inner sep=5pt] at (elce) {$\ell_C$};
  
  \draw (elrs) -- (elre);
  \node[anchor=west] at (elre) {$\ell_R$};

  \draw (elps) -- (elpe);
  \node[anchor=west] at (elpe) {$\ell_b^{\bot}$};

  \draw (ms) -- (me);
  \node[anchor=west] at (me) {$\ell_{C,\mu}$};

  \coordinate (foo) at (2.8,2.4);
  \draw[<->,dotted]
      ($(ms)!(foo)!(me)$) -- ($(elcs)!(foo)!(elce)$)
      node[midway,below,xshift=-1.5mm] {$\mu$};
  
\end{tikzpicture}
\caption{
 The situation considered in the proof of Claim \ref{cl:ellela}.
}\label{fig:ellelone}
\end{figure}
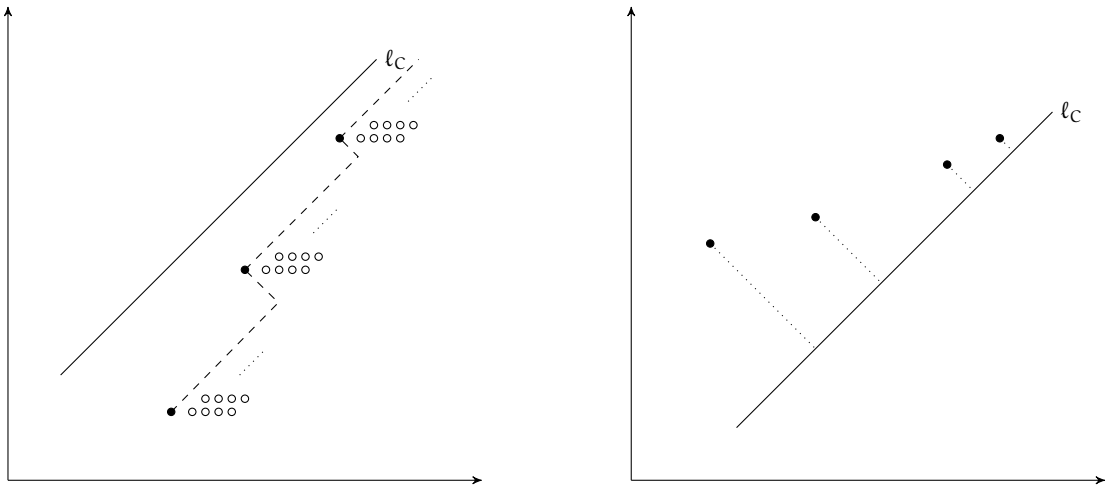
\begin{figure}
  \centering
  \begin{subfigure}[t]{0.495\textwidth}
    \centering
\newcommand{\dhs}[4]{
  \foreach \x in {#1,...,#2}
   \drawpoint{(\x,#3)}{#4};
}

\begin{tikzpicture}[scale=0.7,every node/.style={scale=0.8}]

  \draw[->] (0,0) -- (9,0);
  \draw[->] (0,0) -- (0,9);
  
  \coordinate (hlps) at (5,-1);
  \coordinate (hlpe) at (15,9);
  \coordinate (lbs) at (1,6);
  \coordinate (lbe) at ($(hlps)!(lbs)!(hlpe)$);
  \coordinate (ele) at (7,8);
  \coordinate (els) at (1,2);

  \coordinate (g) at (3,2);
  \coordinate (h) at (8,4);

  \draw (ele) -- (els);

  \node[anchor=west] at (ele) {$\ell_C$};

  \coordinate (bp1) at (3.1,1.3);
  \coordinate (bp2) at (4.5,4);
  \coordinate (bp3) at (6.3,6.5);

  \coordinate (bp1he) at (8, 6.3);
  \coordinate (bp1e)  at ($(bp1)!(bp2)!(bp1he)$);
  
  \coordinate (bp2he) at (8, 7.5);
  \coordinate (bp2e)  at ($(bp2)!(bp3)!(bp2he)$);

  \coordinate (bp3e) at (7.8,8);


  
  \draw[dashed] (bp1) -- (bp1e) -- (bp2);
  \draw[dashed] (bp2) -- (bp2e) -- (bp3);
  \draw[dashed] (bp3) -- (bp3e);
  
  \drawpoint{(bp1)}{black};
  \drawpoint{(bp2)}{black};
  \drawpoint{(bp3)}{black};

  \dhs{3.5,3.75}{4.25}{1.3}{white};
  \begin{scope}[shift={(0.25,0.25)}]
    \dhs{3.5,3.75}{4.25}{1.3}{white};
  \end{scope}
  \dhs{4.9,5.15}{5.65}{4}{white};
  \begin{scope}[shift={(0.25,0.25)}]
    \dhs{4.9,5.15}{5.65}{4}{white};
  \end{scope}
  \dhs{6.7,6.95}{7.45}{6.5}{white};
  \begin{scope}[shift={(0.25,0.25)}]
    \dhs{6.7,6.95}{7.45}{6.5}{white};
  \end{scope}

  \draw[dotted] (4.4,2) -- (4.9,2.5);
  \draw[dotted] (5.8,4.7) -- (6.3,5.2);
  \draw[dotted] (7.6,7.2) -- (8.1,7.7);
\end{tikzpicture}
   \caption{$C\in\calC_{11}$.}
   \label{fig:defc11c12-1}
 \end{subfigure}
\begin{subfigure}[t]{0.495\textwidth}
  \centering
  \begin{tikzpicture}[scale=0.7,every node/.style={scale=0.8}]

  \draw[->] (0,0) -- (9,0);
  \draw[->] (0,0) -- (0,9);
  
  \coordinate (hlps) at (5,-1);
  \coordinate (hlpe) at (15,9);
  \coordinate (lbs) at (1,6);
  \coordinate (lbe) at ($(hlps)!(lbs)!(hlpe)$);
  \coordinate (ele) at (8,7);
  \coordinate (els) at (2,1);

  \coordinate (g) at (3,2);
  \coordinate (h) at (8,4);

  \draw (ele) -- (els);

  \node[anchor=west] at (ele) {$\ell_C$};

  \coordinate (bp1) at (1.5,4.5);
  \coordinate (bp2) at (3.5,5);
  \coordinate (bp3) at (6,6);
  \coordinate (bp4) at (7,6.5);
  
  \coordinate (ip1) at ($(els)!(bp1)!(ele)$);
  \coordinate (ip2) at ($(els)!(bp2)!(ele)$);
  \coordinate (ip3) at ($(els)!(bp3)!(ele)$);
  \coordinate (ip4) at ($(els)!(bp4)!(ele)$);
  
  \draw[dotted] (bp1) -- (ip1);
  \draw[dotted] (bp2) -- (ip2);
  \draw[dotted] (bp3) -- (ip3);
  \draw[dotted] (bp4) -- (ip4);

  \drawpoint{(bp1)}{black};
  \drawpoint{(bp2)}{black};
  \drawpoint{(bp3)}{black};
  \drawpoint{(bp4)}{black};

\end{tikzpicture}
  \caption{$C\in\calC_{12}$.}
  \label{fig:defc11c12-2}
\end{subfigure}
\caption{A closer look at $\ell_C$ when $\alpha_C$ is irrational.}
\label{fig:defc11c12}
\end{figure}
\begin{proof}
We fix an $\alpha_i$-line $\ell_R$ so that
all points in
	$\border\,(\,\area(\leftofline{\ell_R})\,)$ are white in each
	$C\in\calC=\calC_1\cup\calC_2$
	(such $\ell_R$ exists by Corollary~\ref{cor:eller}).
We assume that
	$\calC_1\neq\emptyset$, and we fix an $\alpha_i$-line
	$\ell$ sufficiently above all rightmost down-black lines
	$\ell_C$, $C\in\calC_1$ (defined after Corollary~\ref{cor:eller}),
	so that $\area((\ell,0),\ell_C)$ has infinitely many interior
	points for each $C\in\calC_1$.	(See
	Figure~\ref{fig:ellelone}.)
	Finally we choose (a sufficiently large) $b\in\Real_{\geq 0}$ so
	that $\area((\ell,b),\ell_R)$ does not intersect with the
	vertical axis and is monochromatic in each
	colouring
	$C\not\in\calC_1$; we can easily verify that such $b$ exists
	since we have:
	\begin{itemize}
		\item
			for each ($\alpha_i$-colouring) $C\in\calC_2$
and for each colouring $C$ for which
 $\alpha_C>\alpha_i$
			there are 	only finitely many points
			below the $\alpha_i$-line $\ell$ that are black in $C$ (and we
			choose $b$
			so that these colourings are all-white in
			$\area((\ell,b),\ell_R)$);
		\item
			for each colouring $C$ for which
			$\alpha_C<\alpha_i$,
			hence $\alpha_C=\alpha_j$
			for some $j\in\{0,1,\dots,i{-}1\}$,
			we have $\alpha_C=\beta_C$  by the
	induction hypothesis of our proof of
			Lemma~\ref{lem:qualitative}, and there are
			thus only finitely many points above the
			$\alpha_i$-line $\ell_R$ that are white in $C$ (and we
			choose $b$
			so that these colourings are all-black in
			$\area((\ell,b),\ell_R)$).
	\end{itemize}
	Figure~\ref{fig:ellelone} depicts $\area((\ell,b),\ell_R)$,
and $\ell_C$ just for one $C\in\calC_1$; let us ignore the additional dark-grey
	belt and the line $\ell_{C,\mu}$ for this moment.

	Let us now \emph{assume that $\alpha_i$} (the slope of $\ell$, $\ell_C$,
	$\ell_R$) \emph{is rational}. By using Claim~\ref{cl:auxmu}(1)
we easily deduce that for each $C\in\calC_1$ the rightmost down-black line
$\ell_C$ must be down-black; moreover, it contains
infinitely many  points that are
	black in $C$, while only finitely many points strictly below
	$\ell_C$ can be black in $C$.
	(If $\ell_C$ was not down-black then also a line arising by
a tiny shift of $\ell_C$ to the left, by less
	than $\mu$ in Claim~\ref{cl:auxmu}(1b), would be not
	down-black; this would contradict the definition of
	$\ell_C$. On the other hand, if there were infinitely many points strictly below
	$\ell_C$ that are black in $C$, then a tiny shift of $\ell_C$
	to the right would give us a down-black line, which also
	contradicts the definition of $\ell_C$.)

Hence if $\alpha_i$ is rational, then $\calC_{12}$ is empty, and thus
$\calC_1=\calC_{11}$. Moreover,
we can simply increase
$b$ so that for each $C\in\calC_1$ all the infinitely many points
in $\area((\ell_C,b),\ell_{R})$ that are black in $C$ are on
the line $\ell_C$.
We can thus apply
Proposition~\ref{prop:leftline} (depicted in
Figure~\ref{fig:prop-first}) to
	 $\calC_1,\ell,\ell_R,b$; for each $C\in\calC_1$, the point
	 $\textsc{p}_C$ can be chosen as one of the infinitely many points
	 on $\ell_C$ that are black in
	 $C$ and interior in $\area((\ell,b),\ell_R)$ (hence not
	 close to $\ell^\bot_b$). Proposition~\ref{prop:leftline}
	 gives us
 the claimed $\ell'_L$.

To finish the proof, we now \emph{assume that
	$\alpha_i$ is irrational} (which entails that any $\alpha_i$-line
	contains at most one integer
	point), and we will lead this assumption to a contradiction
	with the assumption $\calC_1\neq\emptyset$.
Recalling Figure~\ref{fig:ellelone}, we aim to increase $b$ so that
not only Proposition~\ref{prop:leftline}, but even
Corollary~\ref{cor:leftline}, could be applied; this will entail that
$\alpha_C>\beta_C$ for some $C\in\calC_1$, which is the desired
contradiction.

Now we cannot assume that  $\calC_{12}=\emptyset$, but
the colourings $C\in\calC_{12}$ (where $\ell_C$ are not down-black)
create no problem for our aim:
We can increase $b$ so that for each  $C\in\calC_{12}$
there are no
points in $\area((\ell_C,b),\ell_{R})$ that are black in $C$.
By definition of $\ell_C$, now
for each point $\textsc{p}$ in  $\area((\ell,b),\ell_{R})$ that is black in some
$C\in\calC_{12}$ (hence $\textsc{p}$ is strictly above $\ell_C$) there
are infinitely many interior points $\textsc{p}'$ in $\area((\ell,b),\ell_{R})$ that
are black in $C$ and for which
$\dist(\textsc{p}',\ell_C)<\dist(\textsc{p},\ell_C)$, hence
$\dist(\textsc{p},\ell)<\dist(\textsc{p}',\ell)$.
(Figure~\ref{fig:defc11c12-2} aims to illustrate this, by depicting black points
strictly above $\ell_C$
that
``converge'' to $\ell_C$.) Hence for each $C\in\calC_{12}$ and each
sufficiently large $b$ there surely is some $\textsc{p}_C$,
$C(\textsc{p}_C)=\mathrm{black}$, such that
$\dist(\textsc{p},\ell)<\dist(\textsc{p}_C,\ell)$ for all points
$\textsc{p}\in\border(\area((\ell,b),\ell_R))$ that are black in $C$.

Now we look at the set $\calC_{11}$, considering a fixed colouring $C\in\calC_{11}$.
By definition, the line $\ell_C$ (with the irrational slope
$\alpha_i$) is down-black in $C$; hence infinitely many points in
$\area((\ell_C,b),\ell_R)$
(recall Figure~\ref{fig:ellelone}) are black in $C$, while there is at
most one (integer) point on $\ell_C$.
By definition of $\ell_C$, for any (however tiny) $\mu\in\Real_{>0}$ there are
only finitely many points in $\area((\ell_{C,\mu},b),\ell_R)$
that are black in $C$,
where
$\ell_{C,\mu}$ arises by shifting $\ell_C$ by $\mu$ to the right
(see Figure~\ref{fig:ellelone}); Figure~\ref{fig:defc11c12-1}
 aims to illustrate this ``converging'' of black points to $\ell_C$ from below.
Hence for any chosen $\mu\in\Real_{>0}$ we can surely increase $b$ so
that all the infinitely many points in  $\area((\ell_{C},b),\ell_R)$
that are black in $C$ are in  $\area((\ell_{C},b),\ell_{C,\mu})$
(i.e., in the grey belt between $\ell_C$ and $\ell_{C,\mu}$
in Figure~\ref{fig:ellelone}).

If we manage to choose the above discussed $\mu$ and $b$ so that
for each $C\in\calC_{11}$ there is no border point of
$\area((\ell,b),\ell_R)$ that is below $\ell_C$ and black in $C$
(such a point could only be in the grey belt, i.e.\ in
$\area((\ell_{C},b),\ell_{C,\mu})$, close to $\ell^\bot_b$), then
we can
surely choose the points
$\textsc{p}_C$, required by Corollary~\ref{cor:leftline}
for $\calC_1,\ell,\ell_R,b$, not only for all $C\in\calC_{12}$ but
also for all
$C\in\calC_{11}$ (for each $C\in\calC_{11}$ we let $\textsc{p}_C$ be
one of the infinitely points in the grey belt that are black in~$C$).
But this yields a contradiction
 since
Corollary~\ref{cor:leftline} in this case
entails that $\slope(\ell)=\alpha_i=\alpha_C>\beta_C$ for all $C\in\calC_1$
(which contradicts $\beta_C\geq\alpha_C$ established by
Proposition~\ref{prop:betagreateralpha}).

Now we show that we can indeed choose $\mu$ and $b$ that yield
the described contradiction. We can choose
 $\mu\in\Real_{>0}$ sufficiently small
so that any two different integer points in
$\area((\ell_{C},b),\ell_{C,\mu})$ (i.e.\ in the grey belt)
have very large distances, for any $C\in\calC_{11}$.
(This follows from Claim~\ref{cl:auxmu}(2).)
We can thus surely choose $b$ so that there is no (integer) point
that is a border point of $\area((\ell,b),\ell_R)$ and lies in a grey
belt (i.e.\ in $\area((\ell_{C},b),\ell_{C,\mu})$ for some
$C\in\calC_{11}$).

The assumptions that $\alpha_i$ is irrational and $\calC_1\neq\emptyset$
have thus yielded a contradiction.
Hence the assumption $\calC_1\neq\emptyset$ entails
that $\alpha_i$ is rational, $\calC_{12}=\emptyset$ (hence
$\calC_{1}=\calC_{11}$),
and there is the claimed line
$\ell'_L$.
\end{proof}

\begin{clm}\label{cl:ellelb}
$\calC_2=\emptyset$, and thus Claim~\ref{cl:ellela} finishes the proof
	of Lemma~\ref{lem:qualitative}.
\end{clm}

\begin{figure}[ht]
  \centering
\begin{subfigure}[t]{0.495\textwidth}
  \centering
    \begin{tikzpicture}[scale=0.7,every node/.style={scale=0.8}]

  
  \draw[->] (0,0) -- (9,0);
  \draw[->] (0,0) -- (0,9);

  \begin{scope}[xshift=-1cm]
  \coordinate (elprimes) at (1.5,4);
  \coordinate (elprimee) at (3,9);
   
  \coordinate (els) at (4,0);
  \coordinate (ele) at (9,5);

  \coordinate (elps) at (1.5,5);
  \coordinate (elpe) at ($(6,1)!(elps)!(13,8)$);
  
  \draw (elps) -- (elpe);
  \node[anchor=west] at (elpe) {$\ell_b^{\bot}$};

  \draw (els) -- (ele);
  \node[anchor=south,xshift=0.2cm] at (ele) {$\ell$ (slope $\alpha_i$)};

  \draw (elprimes) -- (elprimee);
  \node[anchor=west] at (elprimee) {$\ell'$};

  \coordinate (ellpps) at (3.5,2);
  \coordinate (ellppe) at (8.16,9);


  \draw[dotted] (4.75,1.75) -- (7.75,4.75);

  \drawpoint{(5.25,1.75)}{white};
  \drawpoint{(5.5,2)}{white};
  \drawpoint{(5.75,2.25)}{white};
  \drawpoint{(6,2.5)}{white};
  \drawpoint{(6.25,2.75)}{white};

  \draw[loosely dotted] (6.75,3.25) -- (7.75,4.25);

  \drawpoint{(2,5)}{black};
  \drawpoint{(2.15,5.25)}{black};
  \drawpoint{(2.25,5.5)}{black};
  \drawpoint{(2.30,5.75)}{black};
  \drawpoint{(2.4,6)}{black};
  \drawpoint{(2.5,6.25)}{black};

  \draw[dotted] (2.25,4.25) -- (3.65,9);
  \draw[loosely dotted] (2.75, 7) -- (3.15,8.5);
  \end{scope}

  
\end{tikzpicture}
    \caption{$\area(\ell',(b,\ell))$ for proving Claim~\ref{cl:ellelb}.}\label{fig:elleltwo-1}
  \end{subfigure}
  \begin{subfigure}[t]{0.495\textwidth}
    \centering
    %
%
\begin{tikzpicture}[scale=0.7,every node/.style={scale=0.8}]

  
  \draw[->] (0,0) -- (9,0);
  \draw[->] (0,0) -- (0,9);

  \begin{scope}[xshift=-1cm]
  \coordinate (elprimes) at (1.5,4);
  \coordinate (elprimee) at (3,9);
   
  \coordinate (els) at (4,0);
  \coordinate (ele) at (9,5);

  \coordinate (elps) at (1.5,5);
  \coordinate (elpe) at ($(6,1)!(elps)!(13,8)$);

  \coordinate (wp1) at (3.5, 3.5);
  \coordinate (wp2) at (4.5, 5);
  \coordinate (wp3) at (5.5, 6.5);
  \coordinate (wp4) at (6.5, 8);
 
  \draw (elps) -- (elpe);
  \node[anchor=west] at (elpe) {$\ell_b^{\bot}$};

  \draw (els) -- (ele);
  \node[anchor=south,xshift=0.2cm] at (ele) {$\ell$ (slope $\alpha_i$)};

  \draw (elprimes) -- (elprimee);
  \node[anchor=west] at (elprimee) {$\ell'$};

  \coordinate (wp1hlp) at (10,10);
  \coordinate (i1)     at ($(wp1)!(wp2)!(wp1hlp)$);

  \draw[dotted] (wp1) -- (i1) -- (wp2);

  \coordinate (wp2hlp) at (10,10.5);
  \coordinate (i2)     at ($(wp2)!(wp3)!(wp2hlp)$);

  \draw[dotted] (wp2) -- (i2) -- (wp3);
  
  \coordinate (wp3hlp) at (10,11);
  \coordinate (i3)     at ($(wp3)!(wp4)!(wp3hlp)$);

  \draw[dotted] (wp3) -- (i3) -- (wp4);

  \drawvec{(wp1)}{white}{(wp2)}{white}{solid};
  \drawvec{(wp2)}{white}{(wp3)}{white}{solid};
  \drawvec{(wp3)}{white}{(wp4)}{white}{solid};

  \node[anchor=south,xshift=0.05cm] at (6.8,7.75) {\small $\mu_0$};
  \node  at (6.5,7) {\small $B_0$};

  \coordinate (ellpps) at (4,1.5);
  \coordinate (ellppe) at (8.56,8.5);

  \node[anchor=east,xshift=4mm,yshift=9mm] at (wp1) {$v_1$};
  \node[anchor=east,xshift=4mm,yshift=9mm] at (wp2) {$v_2$};
  \node[anchor=east,xshift=4mm,yshift=9mm] at (wp3) {$v_3$};

  
  \draw[dotted] (4.75,1.75) -- (7.75,4.75);

  \drawpoint{(5.25,1.75)}{white};
  \drawpoint{(5.5,2)}{white};
  \drawpoint{(5.75,2.25)}{white};
  \drawpoint{(6,2.5)}{white};
  \drawpoint{(6.25,2.75)}{white};

  \draw[loosely dotted] (6.75,3.25) -- (7.75,4.25);

  \drawpoint{(2,5)}{black};
  \drawpoint{(2.15,5.25)}{black};
  \drawpoint{(2.25,5.5)}{black};
  \drawpoint{(2.30,5.75)}{black};
  \drawpoint{(2.4,6)}{black};
  \drawpoint{(2.5,6.25)}{black};

  \draw[dotted] (2.25,4.25) -- (3.65,9);
  \draw[loosely dotted] (2.75, 7) -- (3.15,8.5);

  \end{scope}
  
\end{tikzpicture}
    \caption{An up-white slope larger than $\alpha_i$.}\label{fig:elleltwo-2}
  \end{subfigure}
  \caption{}
  \label{fig:elleltwo}
\end{figure}

\begin{proof}
For the sake of contradiction, we assume
	$\calC_2\neq\emptyset$, and recall that
for each $C\in\calC_2$ there is no $\alpha_i$-line that is down-black
	in $C$ (hence there are only finitely many points below any given
	$\alpha_i$-line that are black in $C$).
We also recall our assumption $\infty>\alpha_i$ (stipulated  after
	Corollary~\ref{cor:eller}).

	Now we fix an $\alpha_i$-line $\ell$ so that all points above
	$\ell$ are black in all $C\in\calC_1$; the existence of such
	$\ell$ is trivial if $\calC_1=\emptyset$, and otherwise it is
	guaranteed by
	Claim~\ref{cl:ellela}.
We also fix a~line $\ell'$ where $\infty>
\slope(\ell')>\slope(\ell)=\alpha_i$, and
$\alpha_{i+1}>\slope(\ell')$ if $i<k$ (where $k$ is the number of
	all down-limits, recall~(\ref{eq:downlimits}));
finally we fix some
	$b\in\Real_{\geq 0}$ so that the following conditions
are satisfied (see
	Figure~\ref{fig:elleltwo-1}):
	\begin{enumerate}
		\item
	$\area(\ell',(b,\ell))$ is monochromatic for all
$C\not\in\calC_2$;
\item
 all border points of $\area(\ell',(b,\ell))$
	along $\ell'$ (having neighbours strictly above $\ell'$)
	are black in all $C\in\calC_2$;
\item
 all border points of $\area(\ell',(b,\ell))$
	along $\ell$ (having neighbours strictly below $\ell$)
	are white in all $C\in\calC_2$.
	\end{enumerate}
Any sufficiently large $b$ will satisfy these conditions since we have:
	\begin{itemize}
		\item
for each colouring $C$ for which
			$\alpha_C>\alpha_i$, in which case
			$\alpha_C>\slope(\ell')$,
			there are 	only finitely many points
			below $\ell'$ that are black in $C$
			(and we
			choose $b$
			so that these colourings are all-white in
			$\area(\ell',(b,\ell))$);
		\item
			for each ($\alpha_i$-colouring) $C\in\calC_1$
all points in  $\area(\ell',(b,\ell))$ are black in $C$ by our
			choice of $\ell$;
		\item
			for each colouring $C$ for which
			$\alpha_C<\alpha_i$,
 in which case
			$\slope(\ell)>\alpha_C$ and $\alpha_C=\beta_C$  by the
	induction hypothesis of our proof of
			Lemma~\ref{lem:qualitative},
			there are only finitely many points above
			 $\ell$ that are white in $C$
			 (and we
			choose $b$
			so that these colourings are all-black in
			$\area(\ell',(b,\ell))$);
		\item
for each ($\alpha_i$-colouring) $C\in\calC_2$ we have
			$\alpha_i=\alpha_C=\beta_C$ (by
			Claim~\ref{cl:betaalpha}); hence above any
			line that is parallel with $\ell'$ (and whose
			slope is thus greater than $\beta_C$) there
			are only finitely many points that are white
			in $C$ (and we can thus
			choose $b$
			so that the border points of $\area(\ell',(b,\ell))$
			along $\ell'$ are all black in all
			$C\in\calC_2$);
		\item
by definition of $\calC_2$, for any $C\in\calC_2$ and any $\alpha_i$-line
			$\bar{\ell}$ (which is thus parallel with
			$\ell$)
			there are only finitely many points below
	$\bar{\ell}$ that are black in $C$ (and we can thus
			choose $b$
			so that the border points of $\area(\ell',(b,\ell))$
			along $\ell$ are all white in all
			$C\in\calC_2$).
	\end{itemize}
Let us now call a vector $v$ to be
a \emph{$(B,\mu)$-vector},
for $B\in\Real_{\geq 0}$ and $\mu\in\Real_{>0}$, if $B$ is the
$\alpha_i$-d-size of $v$
and $\mu$ is the co-$\alpha_i$-d-size of $v$.
We can see (looking at Figure~\ref{fig:elleltwo} and recalling
that each $\alpha_i$-line is not down-black in any $C\in\calC_2$)
that we can fix some $\mu_0>0$ such that for any $B\in\Real_{>0}$
there is $B_0\geq B$ and
a  $(B_0,\mu_0)$-vector $v$
where $\spt(v)\in\border(\area(\ell',(b,\ell)))$
and $C(\spt(v))=C(\ept(v))=\mathrm{white}$ for some $C\in\calC_2$
(which entails that $\ept(v)$ is in $\area(\ell',(b,\ell))$).
Moreover, we can even choose a  sufficiently large $B_0$
so that
\begin{itemize}
	\item
there is
a  $(B_0,\mu_0)$-vector $v$
where $\spt(v)\in\border(\area(\ell',(b,\ell)))$
and $C(\spt(v))=C(\ept(v))=\mathrm{white}$ for some $C\in\calC_2$;
\item
	the slope of 	 $(B_0,\mu_0)$-vectors is strictly between
		$\slope(\ell')$ and $\slope(\ell)$;
	\item
for each $(B_0,\mu_0)$-vector $v$
where $\spt(v)\in\border(\area(\ell',(b,\ell)))$, and for each colouring
$C$ we have
	\begin{equation}\label{eq:whitewhite}
\textnormal{if $C(\spt(v))=\mathrm{white}$,
		then 	$C(\ept(v))=\mathrm{white}$.}
	\end{equation}
\end{itemize}
	But then each $(B_0,\mu_0)$-vector $v$ with $\spt(v)\in\area(\ell',(b,\ell))$
	(not only with $\spt(v)\in\border(\area(\ell',(b,\ell)))$)
	satisfies~(\ref{eq:whitewhite}).
	Indeed, a violating $(B_0,\mu_0)$-vector $v$ with the least
	rank of its white start-point $\spt(v)$, where $\spt(v)$ is necessarily an interior
	point of $\area(\ell',(b,\ell))$, leads to a
	contradiction by Proposition~\ref{prop:neighbour} applied to
	the opposite black-white vector (which would yield a
violating $(B_0,\mu_0)$-vector with a smaller rank of its start-point).

	This entails that the slope of $(B_0,\mu_0)$-vectors, which is larger
than $\alpha_i$, is up-white in each $C\in\calC_2$ (consider the line
containing white-white vectors $v_1,v_2,v_3,\dots$
in	Figure~\ref{fig:elleltwo-2}); this is a contradiction
since $\beta_C=\alpha_i$ for all $C\in\calC_2$ (by Claim~\ref{cl:betaalpha}).
Hence $\calC_2=\emptyset$ after all.
\end{proof}

\subsection{Quantitative belt theorem}\label{sec:quantitbelttheorem}

We have proven the belt theorem (Theorem~\ref{th:qualitative}),
but we have not derived anything specific about the slopes, widths,
and positions of the belts (one of which is depicted in
Figure~\ref{fig:belttheoremclaim}).
Now we show that these parameters can be presented by
polynomially bounded integers (in the size of the underlying
one-counter net $\calN=(Q,Act,\delta)$ that we have fixed).

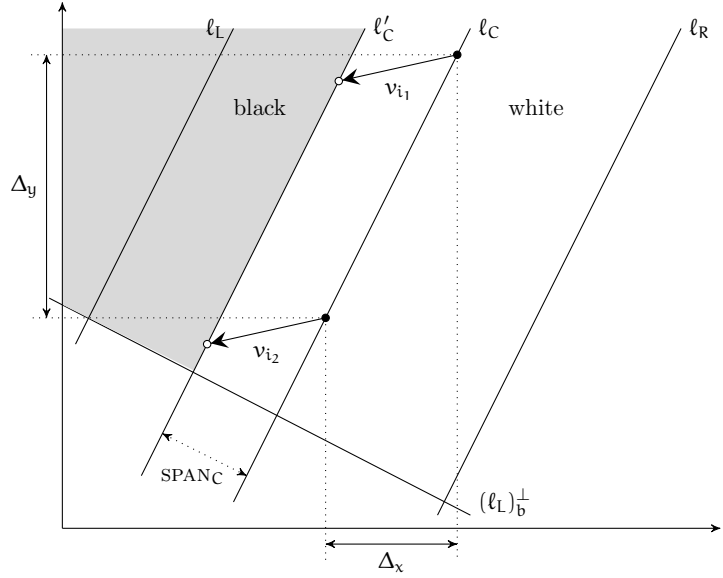
\begin{figure}
  \centering
\begin{tikzpicture}[scale=0.7,every node/.style={scale=0.8}]

  
  \draw (-1.25,3.375) -- (6.75,-0.75);
  
  \fill[color=gray!30!white]
     (1.5,2) -- (-1,3.25) -- (-1,8.5) -- (4.75,8.5) -- (1.5,2); 

  \node at (2.75,7) {black};
  \node at (8,7) {white};
   
  \draw[->] (-1,-1) -- (11.5,-1);
  \draw[->] (-1,-1) -- (-1,9);
 
  \coordinate (els) at (0.5,0);
  \coordinate (ele) at (4.75,8.5);

  \coordinate (elLs) at (-0.75,2.5);
  \coordinate (elLe) at (2.25,8.5);
  
  \coordinate (elps) at (2.25,-0.5);
  \coordinate (elpe) at (6.75,8.5);

  \coordinate (elRs) at (6.125,-0.75);
  \coordinate (elRe) at (10.75,8.5);
  
  \draw (els) -- (ele);
  \node[anchor=west] at (ele) {$\ell_C'$};

  \draw (elps) -- (elpe);
  \node[anchor=west] at (elpe) {$\ell_C$};

  \draw (elLs) -- (elLe);
  \node[anchor=east] at (elLe) {$\ell_L$};

  \draw (elRs) -- (elRe);
  \node[anchor=west] at (elRe) {$\ell_R$};

  \coordinate (foo) at (2.5,0);
  \coordinate (i1)  at ($(els)!(foo)!(ele)$);

  \draw[<->,dotted]
  (i1) -- (foo)
  node[anchor=north,midway,yshift=-0.1cm,xshift=-0.2cm] {$\spand_C$};

  \coordinate (bp1) at (6.5, 8);
  \coordinate (wp1) at (4.25,7.5);

  
  \coordinate (bp2) at (4,3);
  \coordinate (wp2) at (1.75,2.5);

  \draw[dotted]
  (bp1) -- (6.5,-1.6)
  (bp1) -- (-1.6,8)
  (bp2) -- (4,-1.6)
  (bp2) -- (-1.6,3);

  \draw[<->]
  (6.5,-1.3) -- (4,-1.3)
  node[midway,anchor=north] {$\Delta_x$};

  \draw[<->]
  (-1.3,8) -- (-1.3,3)
  node[midway,anchor=east] {$\Delta_y$};
  
  \drawvec{(bp1)}{black}{(wp1)}{white}{solid}
  \drawvec{(bp2)}{black}{(wp2)}{white}{solid}

  \node at (5.4,7.3) {$v_{i_1}$};
  \node at  (2.9,2.3) {$v_{i_2}$};

  \node[anchor=west] at (6.75,-0.5) {$(\ell_L)^\bot_b$};

\end{tikzpicture}

  

  



  
  \caption{The vectors $v_{i_1}$ and $v_{i_2}$ imply
	$\alpha = \frac{\Delta_y}{\Delta_x}$ (where
	$\alpha=\slope(\ell_L)$).}\label{fig:belt-slope}
\end{figure}

We start with the \emph{belt-slopes}, which is another
name for the down-limits $\alpha_C$ (that coincide with the up-limits
$\beta_C$ by Lemma~\ref{lem:qualitative});
we recall that we have ordered the set $\{\alpha_C\mid C=C_{\tu{p,q}},
(p,q)\in Q\times Q\}$, denoting its elements
(in~(\ref{eq:downlimits})) as
\begin{equation}\label{eq:beltslopesaredownlimits}
0\leq\alpha_1<\alpha_2\cdots <\alpha_k\leq\infty.
\end{equation}
Hence when we refer to a belt-slope $\alpha$, we understand that
$\alpha=\alpha_i$ for some $i\in[1,k]$.
We can also recall that by an $\alpha$-colouring we mean a colouring
$C$ ($C=C_{\tu{p,q}}$ for some $(p,q)\in Q\times Q$) for which
$\alpha_C=\alpha$.
Before clarifying that the belt-slopes are fractions of small integers
(or $\infty$), which is shown by
Proposition~\ref{prop:beltslopes},  we formulate a useful
corollary of Proposition~\ref{prop:neighbour}, which captures
the black-white vector travel discussed around Figure~\ref{fig:neighbourvector} in
Introduction. (The black-white vectors in
Figure~\ref{fig:neighbourvector} are directed upwards but the
corollary also handles the vectors directed downwards, like $v_{i_1}$
in Figure~\ref{fig:belt-slope}.)

\begin{cor}[of Proposition~\ref{prop:neighbour}]\label{cor:neighbour}
Any
	vector $v_0$ that is black-white in some $C_0$ gives rise to a sequence
	$v_0,v_1,\dots,v_n$ where $\spt(v_n)\in\textsc{v-axis}$ or
	$\ept(v_n)\in\textsc{h-axis}$, and
 for $i=0,1,2,\dots,n{-}1$ we have that $v_{i+1}$ is a neighbour vector of
	$v_i$ that is black-white in some $C_{i+1}$ and the rank of
	the white end of  $v_{i+1}$
	in $C_{i+1}$ is smaller than the rank
	of the white end of $v_i$ in
	$C_i$.
\end{cor}

\begin{prop}[Belt slopes]\label{prop:beltslopes}
	If $\alpha$, $\infty>\alpha>0$, is a belt-slope, then
$\alpha=\frac{\Delta_y}{\Delta_x}$
for some integers $\Delta_x$, $\Delta_y$ from $\{1,2,\cdots,|\calC|\}$
	where $\calC$ is the set of
	$\alpha$-colourings.
\end{prop}
\begin{proof}
Let $\calC$ be the set of $\alpha$-colourings, for some fixed
	belt-slope $\alpha$ where
$\infty>\alpha>0$, and
let $\ell_L,\ell_R$ be some $\alpha$-lines such that
all points above $\ell_L$ are black in all $C\in\calC$ and
all points below $\ell_R$ are white in all $C\in\calC$;
the belt theorem (Theorem~\ref{th:qualitative}) guarantees that
	such lines exist, and that $\alpha$ is rational.

	By reasoning as in the proof of Claim~\ref{cl:ellela} (when assuming that
	$\alpha_i$ is rational) we derive that
for each $C\in\calC$ there is the rightmost $\alpha$-line $\ell_C$
	that is down-black in $C$; we also recall that there are infinitely
	many points on $\ell_C$ that are black in $C$ (but only
	finitely many points strictly below $\ell_C$ are black in $C$).
Analogously we derive that there is the leftmost $\alpha$-line $\ell'_C$
	that is up-white in $C$;  there are infinitely
	many points on $\ell'_C$ that are white in $C$.
	(The colourings depicted in Figure~\ref{fig:solvedsimul}, related
	to the simple net
in Figure~\ref{fig:exsocn}, might suggest that
$\ell'_C$ is below $\ell_C$, but
Figure~\ref{fig:belt-slope} depicts a more typical situation.)
By $\spand_C$ we denote the distance of $\ell_C$ and
	$\ell'_C$, with the negative sign if $\ell'_C$ is below
 $\ell_C$.
	We put
	\begin{center}
	$s=\max\,\{\,\spand_C\mid C\in\calC\,\}$, and
	$\calC_\textsc{ms}=\{\,C\in\calC\mid \spand_C=s\}$
	\end{center}
	(where $\textsc{ms}$ refers to ``Maximal Span'').

Now we choose $v_0$ so that $v_0$ is black-white in some
	$C_0\in\calC_\textsc{ms}$, $\spt(v_0)\in\ell_{C_0}$, and
	$\ept(v_0)\in\ell'_{C_0}$; moreover, we choose $v_0$ sufficiently
	high in the respective $\alpha$-belt so that each $v_i$ in the prefix
	$v_0,v_1,v_2,\dots,v_{|\calC_{\textsc{ms}}|}$ of the sequence
	guaranteed by Corollary~\ref{cor:neighbour} must necessarily
	satisfy that $\spt(v_i)\in\ell_{C_i}$ and
	$\ept(v_i)\in\ell'_{C_i}$ for some $C_i\in\calC_{\textsc{ms}}$.
Figure~\ref{fig:belt-slope} depicts this ``sufficiently high'', where
$b$ is chosen so that
	\begin{itemize}
		\item
			for each  $C\in\calC$
	the points strictly above $\ell'_C$ in
	$\area(\textsc{v-axis},(b,\ell'_C))$
	are black in $C$
and the points strictly below $\ell_C$
in $\area((\ell_C,b),\textsc{h-axis})$
	are white in $C$,
\item
		and each colouring $C\not\in\calC$ (for which
$\alpha_C\neq\alpha$) is
	monochromatic in
$\area((\ell_L,b),\ell_R)$.
	\end{itemize}

By the pigeonhole principle,
in the above discussed sequence $v_0,v_1,v_2,\dots,v_{|\calC_{\textsc{ms}}|}$
	we have some $i_1,i_2$ such that $0\leq i_1<i_2\leq
	|\calC_{\textsc{ms}}|$ and both $\spt(v_{i_1})$ and
	$\spt(v_{i_2})$ are on $\ell_C$ for the same
	$C\in\calC_{\textsc{ms}}$
	(and $\ept(v_{i_1})$ and
	$\ept(v_{i_2})$ are on $\ell'_C$).
Since $\slope(\ell_C)=\alpha$, and $\spt(v_{i_1})$,
	$\spt(v_{i_2})$ are two different points on $\ell_C$,
we deduce that $\alpha$ is the slope of the vector
	$(\spt(v_{i_1}),\spt(v_{i_2}))$.
        This entails
	that  $\alpha=\frac{\Delta_y}{\Delta_x}$
	where $\Delta_x, \Delta_y\in\{1,2,\cdots,i_2{-}i_1\}$;
$\Delta_x$ is the absolute value of the $0$-d-size [horizontal
	d-size] of the vector $(\spt(v_{i_1}),\spt(v_{i_2}))$,
	and $\Delta_y$ is the absolute value of its
	co-$0$-d-size [vertical d-size] (see Figure
	\ref{fig:belt-slope}).
\end{proof}

To deal with the belt-widths and the belt positions, it is useful to
define two quasi-orders on the set of colourings $\{C\mid
C=C_{\tu{p,q}}, (p,q)\in Q\times Q\}$,
based on the belt-slopes and the lines  $\ell^C_L$, $\ell^C_R$
defined below (which can differ from the lines
$\ell_L$, $\ell_R$ in Theorem~\ref{th:qualitative} and
$\ell'_C$, $\ell_C$ used in the proof of
Proposition~\ref{prop:beltslopes}).
For technical reasons we also extend the notion
of border points to \emph{almost-border points} of particular areas.

\paragraph{Lines  \texorpdfstring{$\ell^C_L$}{l^C_L} (leftmost with a white point) and
\texorpdfstring{$\ell^C_R$}{l^C_R} (rightmost with a black point).}
For each colouring $C$ we define $\ell^C_L$ as
the leftmost $\alpha_C$-line that contains a point that is white in
$C$, and $\ell^C_R$ as
the rightmost $\alpha_C$-line that contains a point that is black in
$C$; we put $\ell^C_L=\ell^C_R=\textsc{h-axis}$ if $C$ is all-black
(in which case $\alpha_C=0$), and
$\ell^C_L=\ell^C_R=\textsc{v-axis}$ if $C$ is all-white (in which case
$\alpha_C=\infty$).
We note that Theorem~\ref{th:qualitative}
(depicted in Figure~\ref{fig:belttheoremclaim}), which includes the
fact
that the belt-slopes $\alpha_C$ are rational or $\infty$,
entails that the lines
$\ell^C_L$
and
$\ell^C_R$
are well-defined.

We again note that the simulation preorder depicted
in~Figure~\ref{fig:solvedsimul},
related to our simple example in Figure~\ref{fig:exsocn}, can give an
impression that $\ell^C_L$ is typically below (to the right of)
$\ell^C_R$; in this case
(whenever $\ell^C_L$ is below $\ell^C_R$), the distance between
$\ell^C_L$ and
$\ell^C_R$ is clearly at most $1$ (and the distance $1$ is achieved
in the colouring $C_{\tu{p_2,p_1}}$ in Figure~\ref{fig:solvedsimul}).
In more complicated examples, $\ell^C_L$ can be surely above (to
the left of) $\ell^C_R$; in Figure~\ref{fig:belt-slope}, $\ell^C_L$
would be between $\ell_L$ and $\ell'_C$, and
 $\ell^C_R$
would be between $\ell_C$ and $\ell_R$. In the case when $\ell^C_L$
is above $\ell^C_R$
 the distance between
$\ell^C_L$ and $\ell^C_R$ can be surely larger than $1$; nevertheless
it is polynomially bounded (in the size of $Q$ of our fixed OCN
$\calN=(Q,Act,\delta)$), as we discuss below.

\paragraph{A counterclockwise quasi-order \texorpdfstring{$\qocl$}{<=_L} and a clockwise
quasi-order
\texorpdfstring{$\qocr$}{<=_R} on the set of colourings.}
The quasi-orders $\qocl$ and $\qocr$ on the set $\{\,C\mid
C=C_{\tu{p,q}}, (p,q)\in Q\times Q\,\}$
 are defined as follows:
\begin{center}
$C\qocl C'$ if either $\alpha_C<\alpha_{C'}$, or
$\alpha_C=\alpha_{C'}$ and $\ell^C_L$ is below (to the right of)
$\ell^{C'}_L$;
\end{center}
\begin{center}
$C\qocr C'$ if either $\alpha_C>\alpha_{C'}$, or
$\alpha_C=\alpha_{C'}$ and $\ell^C_R$ is above (to the left of)
$\ell^{C'}_R$.
\end{center}
By $C\sqocl C'$ we denote that $C\qocl C'$ and  $C'\not\qocl C$;
the case $C\sqocr C'$ is analogous.

We can recall the order $\alpha_1<\alpha_2\cdots <\alpha_k$
of belt-slopes (\ref{eq:beltslopesaredownlimits}).
Hence $C\qocl C'$ iff $\alpha_C=\alpha_i$ and $\alpha_{C'}=\alpha_j$
where $i<j$, or $i=j$ and $\ell^C_L$ is below (to the right of)
$\ell^{C'}_L$; analogously we can express  $C\qocr C'$.

\paragraph{Almost-border points of
\texorpdfstring{\textnormal{$\area(\leftofline{\ell})$}}{Area(<-l)}
and \texorpdfstring{\textnormal{$\area(\rightofline{\ell})$}}{Area(l->)}.}

We say that $\textsc{p}\in \area(\leftofline{\ell})$ is an
\emph{almost-border point of} $\area(\leftofline{\ell})$ if $\textsc{p}$ has a neighbour point
in $\area(\rightofline{\ell})$ (hence $\textsc{p}$ is a border point
of $\area(\leftofline{\ell})$ or has a neighbour point on $\ell$).
Similarly,  $\textsc{p}\in \area(\rightofline{\ell})$ is an
\emph{almost-border point of} $\area(\rightofline{\ell})$ if $\textsc{p}$ has a neighbour point
in $\area(\leftofline{\ell})$.

\medskip

The next proposition (Proposition~\ref{prop:polydistlines}) and its
corollary (Corollary~\ref{cor:polydistlines})
show that the distances of lines
$\ell^C_L$ and  $\ell^C_R$ to the origin  $(0,0)$ are ``small'',
bounded by a polynomial in the number of colourings (hence in $|Q\times Q|$ for
the underlying OCN  $\calN=(Q,Act,\delta)$); this also yields the
announced polynomial bound on the distance between $\ell^C_L$ and
$\ell^C_R$ (for any colouring $C=C_{\tu{p,q}}$).

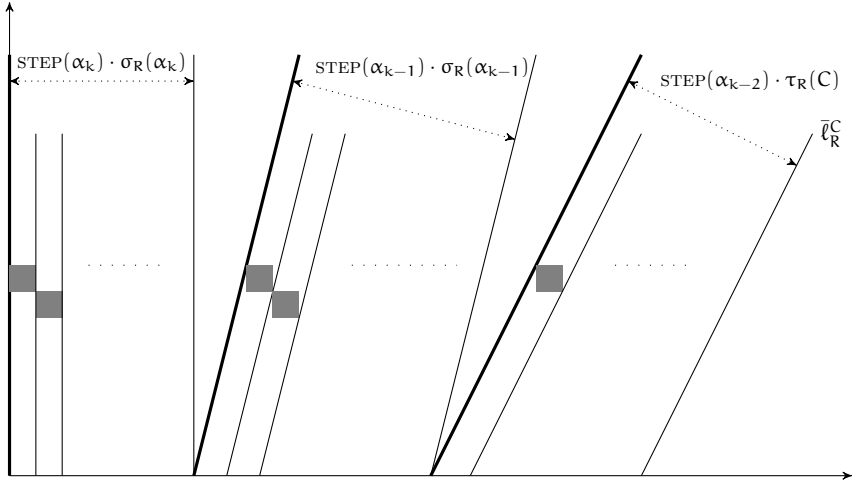
\begin{figure}
  \centering
 \begin{tikzpicture}[scale=0.7,every node/.style={scale=0.8}]

  
    \draw[->] (0,0) -- (16,0);
    \draw[->] (0,0) -- (0,9);

    \draw[very thick] (0,0) -- (0,8);
    \fill[color=gray] (0,4) rectangle (0.5,3.5);
    \fill[color=gray] (0.5,3.5) rectangle (1,3);
    
    \draw[thin] (0.5,0) -- (0.5,6.5);
    \draw[thin] (1,0) -- (1,6.5);
    \draw[loosely dotted] (1.5,4) -- (3,4);
    \draw (3.5,0) -- (3.5,8);
    \draw[<->,dotted]
    (0,7.5) -- (3.5,7.5)
    node[midway,anchor=south,align=center] {\small
	$\step(\alpha_k)\cdot\sigma_R(\alpha_k)$};

    \draw[very thick] (3.5,0) -- (5.5,8);
    \draw[thin] (8,0) -- (10,8);
    \draw[<->,dotted] (5.375,7.5) -- ($(8,0)!(5.375,7.45)!(10,8)$)
    node[midway,anchor=south
	west,align=center,xshift=-1.6cm,yshift=0.4cm]
	{\small $\step(\alpha_{k-1})\cdot\sigma_R(\alpha_{k-1})$};
    \fill[color=gray] (4.5,4) rectangle (5,3.5);
    \draw[thin] (4.125,0) -- (5.750,6.5);
    \fill[color=gray] (5,3.5) rectangle (5.5,3);
    \draw[thin] (4.750,0) -- (6.375,6.5);
    \draw[loosely dotted] (6.5,4) -- (8.5,4);
    
    \draw[very thick] (8,0) -- (12,8);
    
    \draw[thin] (12,0) -- (15.25,6.5);
	\node[anchor=west] at (15.25,6.5) {$\bar{\ell}^C_R$};
    \draw[<->,dotted] (11.75,7.5) -- ($(12,0)!(11.75,7.5)!(16,8)$)
    node[midway,anchor=south west,xshift=-1cm,yshift=0.4cm]
	{\small$\step(\alpha_{k-2})\cdot\tau_R(C)$};
    \fill[color=gray] (10,4) rectangle (10.5,3.5);
    \draw[thin] (8.75,0) -- (12,6.5);
    \draw[loosely dotted] (11.5,4) -- (13,4);




    
    



    
\end{tikzpicture}
\caption{
	The actual $\ell^C_R$ is above (to the left of) the depicted  $\bar{\ell}^C_R$;
the grey squares have the size $1\times 1$.}\label{fig:width}
\end{figure}

Figure~\ref{fig:width} (the notation on which will be explained below)
aims to illustrate that $\ell^C_R$ cannot intersect $\textsc{h-axis}$ in a
large distance to the right from the origin $(0,0)$.
This follows by Proposition~\ref{prop:polydistlines}(1) that entails
that if the intersection of some $\ell^C_R$ with $\textsc{h-axis}$ is
to the right of $(0,0)$, then the distance of this intersection
to $(0,0)$ is bounded by the product of a small number with the
cardinality of the set $\{C'\mid C'\qocr C \}$ (as follows by a
repeated use of Proposition~\ref{prop:polydistlines}(1)).

\begin{prop}\label{prop:polydistlines}
For each colouring $C$, and its respective lines  $\ell^C_L$ and
	$\ell^C_R$, we have:
	\begin{enumerate}
		\item
If $(0,0)$ is strictly above $\ell^C_R$,
			then there is $C'\sqocr C$
such that $\area(\leftofline{\ell^{C'}_R})$ contains a point that is
			an almost-border point of
			$\area(\leftofline{\ell^C_R})$.
		\item
If $(0,0)$ is strictly below $\ell^C_L$,
			then there is $C'\sqocl C$
such that $\area(\rightofline{\ell^{C'}_L})$ contains a point that is
			an almost-border point of
			$\area(\rightofline{\ell^C_L})$.
	\end{enumerate}
	\end{prop}
\begin{proof}
1.
Let us fix some $C_0$
	such that $(0,0)$ is strictly above $\ell^{C_0}_R$~(see
	Figure~\ref{fig:gapeller}); we will show that there is
$C'_0\sqocr C_0$
such that $\area(\leftofline{\ell^{C'_0}_R})$ contains a point that is
			an almost-border point of
			$\area(\leftofline{\ell^{C_0}_R})$.

We put $\alpha=\alpha_{C_0}$, and
	$\calC=\{C\mid
	\alpha_{C}=\alpha$ and $C_0\qocr C\}$.
	Let $\ell$ be the leftmost $\alpha$-line that is up-white in some
	$C\in\calC$; we fix such $\bar{C}\in\calC$.
	Hence $\ell$ contains infinitely many points
that are white in $\bar{C}$ but only
	finitely many points strictly above $\ell$ can be white
	in any colouring $C$ such that $C_0\qocr C$ (which includes all
	$C\in\calC$).
	We can thus choose $b\in\Real_{\geq 0}$ so that
 all points in
	$\area(\textsc{v-axis},(b,\ell))$
	(i.e., in the grey area in Figure~\ref{fig:gapeller})
	that are strictly above
	$\ell$ are black in all $C$ such that $C_0\qocr C$
	(hence those points can be white only in some $C\sqocr C_0$).

	In $\bar{C}$ we
	consider a black-white vector $v_0$ such that
	$\spt(v_0)\in\area(\rightofline{\ell^{C_0}_R})$ (there must be a respective point that is  black in $\bar{C}$
	since $C_0\qocr\bar{C}$) and
	$\ept(v_0)\in\ell$; we have infinitely many possibilities on
	$\ell$ for
	such $\ept(v_0)$ (see Figure~\ref{fig:gapeller}).

We fix a sequence $v_0,v_1,\dots,v_n$ guaranteed by
	Corollary~\ref{cor:neighbour}. There must be the least
	$i\in[1,n]$
	such that $\spt(v_i)$ is strictly above $\ell^{C_0}_R$
	(since $\spt(v_n)$ is on \textsc{v-axis} and thus strictly
	above $\ell^{C_0}_R$);
hence $\spt(v_i)$
	is an almost-border point of
	$\area(\leftofline{\ell^{C_0}_R})$
	(see again Figure~\ref{fig:gapeller}).
We could choose $v_0$ with
	a sufficiently large $\alpha$-d-size,
	so that $\ept(v_i)$ is necessarily in $\area(\textsc{v-axis},(b,\ell))$
	(in the grey area in Figure~\ref{fig:gapeller}); moreover,
$\ept(v_i)$ is strictly above $\ell$. Hence $v_i$ is black-white in
	some $C'_0\sqocr C_0$, which entails that $\spt(v_i)$ (which is
	an almost-border point of
	$\area(\leftofline{\ell^{C_0}_R})$) is in
	$\area(\leftofline{\ell^{C'_0}_R})$.

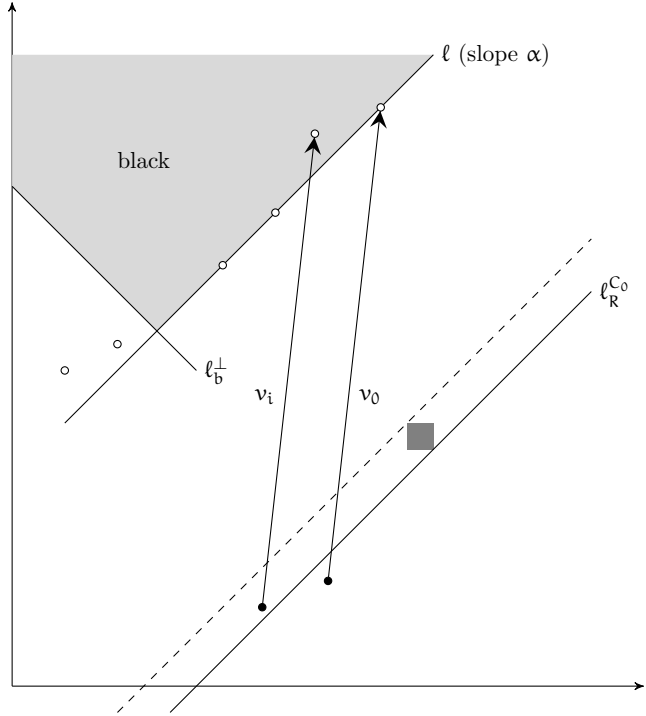
\begin{figure}
	\begin{center}
\begin{tikzpicture}[scale=0.7,every node/.style={scale=0.8}]

  \draw[->] (0,0) -- (12,0);
  \draw[->] (0,0) -- (0,13);

  \coordinate (elrcs) at (3,-0.5);
  \coordinate (elrce) at (11,7.5);

  \coordinate (els) at (1,5);
  \coordinate (ele) at (8,12);

  \coordinate (pers) at (0,9.5);
  \coordinate (pere) at (3.5,6);

  \coordinate (pars) at (2,-0.5);
  \coordinate (pare) at (11,8.5);

  \fill[color=gray] (7.5,5) rectangle (8,4.5);

  \fill[color=gray!30!white]
  (2.75, 6.75) -- (pers) -- (0,12) -- (ele) -- (2.75, 6.75);
  
  \draw (elrcs) -- (elrce);
  \node[anchor=west] at (elrce) {$\ell_R^{C_0}$};

  \draw (els) -- (ele);
  \node[anchor=west] at (ele) {$\ell$ (slope $\alpha$)};
 
  \draw (pers) -- (pere);
	  \node[anchor=west] at (pere) {$\ell^{\bot}_b$};
  \draw[dashed] (pars) -- (pare);

  \coordinate (bp1) at (6,2);
  \coordinate (wp1) at (7,11);

  \coordinate (bp2) at (4.75, 1.5);
  \coordinate (wp2) at (5.75, 10.5);
  
  \drawvec{(bp1)}{black}{(wp1)}{white}{solid}
  \drawvec{(bp2)}{black}{(wp2)}{white}{solid}

  \node at (6.8,5.5) {$v_0$};
  \node at (4.8,5.5) {$v_i$};

  \drawpoint{(1,6)}{white};
  \drawpoint{(5,9)}{white};
  \drawpoint{(2,6.5)}{white};
  \drawpoint{(4,8)}{white};

  \node at (2.5,10) {black};

\end{tikzpicture}
	\end{center}
	\caption{Vector $v_0$ that is black-white in $\bar{C}$ gives rise
	to $v_i$ that is black-white in $C'_0\sqocr C_0$.}\label{fig:gapeller}
\end{figure}

2. This condition is shown analogously as $1$.
Here we assume that
 $(0,0)$ is strictly below $\ell^{C_0}_L$, and put
$\alpha=\alpha_{C_0}$ and
	$\calC=\{C\mid
	\alpha_{C}=\alpha$ and $C_0\qocl C\}$.
Now $\ell$ is the rightmost $\alpha$-line that is down-black in some
	$\bar{C}\in\calC$.
We choose $v_0$, with a sufficiently large absolute value of $\alpha$-d-size,
so that	$\ept(v_0)$ is in $\area(\leftofline{\ell^{C_0}_L})$ and
	white in $\bar{C}$, and $\spt(v_0)$ is on $\ell$ and black in
	$\bar{C}$.
\end{proof}

To formulate the announced corollary,
we introduce further technical notions (that also appear in
Figure~\ref{fig:width}).

\paragraph{Horizontal and vertical distances,
values \texorpdfstring{\sc $\step(\alpha)$}{Step(alpha)}, \texorpdfstring{\sc $\hstep(\alpha)$}{H-Step(alpha)}, \texorpdfstring{\sc $\vstep(\alpha)$}{V-Step(alpha)},
and \texorpdfstring{$\sigma_R(\alpha)$}{sigma_R(alpha)},
\texorpdfstring{$\sigma_L(\alpha)$}{sigma_L(alpha)},
\texorpdfstring{$\tau_R(C)$}{tau_R(C)}, \texorpdfstring{$\tau_L(C)$}{tau_L(C)}.}
\begin{itemize}
	\item
		For a slope (a belt-slope in particular) $\alpha$, $0<\alpha<\infty$, let $\ell$ and $\ell'$
be two $\alpha$-lines such that a unit square fits between them, with
		its top-left corner on $\ell$ and its bottom-right
		corner on $\ell'$ (see a grey square in Figure~\ref{fig:width}).
By $\step(\alpha)$ we mean the (Euclidean) distance
of the lines $\ell$ and $\ell'$, by $\hstep(\alpha)$ their horizontal
		distance (i.e., the distance of their intersections
		with $\textsc{h-axis}$), and by
 $\vstep(\alpha)$ their vertical distance.
		We put $\step(\infty)=\hstep(\infty)=1$,
		and  $\step(0)=\vstep(0)=1$.
	\item
		By $\hd((0,0),\ell)$ ($\vd((0,0),\ell)$)
		we mean the distance of
		$(0,0)$ to the intersection of $\ell$ with
		$\textsc{h-axis}$ (with $\textsc{v-axis}$).
	\item
For a belt-slope $\alpha$,
by $\sigma_R(\alpha)$ we denote the number of equivalence classes of
the equivalence $\equiv_R\mathop{=}\qocr \cap \sqsupseteq_R$ in the
set of $\alpha$-colourings.
The value $\sigma_L(\alpha)$ is the number of classes of
$\equiv_L\mathop{=}\qocl \cap \sqsupseteq_L$ in the set of
		$\alpha$-colourings.
\item
For an $\alpha$-colouring $C$, we define $\tau_R(C)$ to be the number of
equivalence classes of $\equiv_R$ in the set
$\{ C' \mid C' \mbox{ is an $\alpha$-colouring and } C' \qocr C\}$.
The value $\tau_L(C)$ is the number of classes of
		$\equiv_L$ in the set
$\{ C' \mid C' \mbox{ is an $\alpha$-colouring and } C' \qocl C\}$.
		\end{itemize}
We note that $\step(\alpha)\leq \sqrt{2}$ (with the equality for $\alpha = 1$).
If $\alpha$ is a belt-slope and $\calC$ is the set of
$\alpha$-colourings,
then Proposition~\ref{prop:beltslopes}
entails that
\begin{equation}\label{eq:polboundstep}
	\textnormal{$\hstep(\alpha)\leq 1+|\calC|$ if $\alpha>0$ and
	$\vstep(\alpha)\leq 1+|\calC|$ if $\alpha<\infty$.}
\end{equation}
Now we state the announced corollary of
Proposition~\ref{prop:polydistlines};
we recall the order
 $\alpha_1<\alpha_2\cdots <\alpha_k$
of belt-slopes (\ref{eq:beltslopesaredownlimits}) and look
at Figure~\ref{fig:width}.

\begin{cor}\label{cor:polydistlines}
	For each $\alpha_i$-colouring $C$ (where $i\in[1,k]$) we have:
  \begin{enumerate}
    \item
	    if $(0,0)$ is above $\ell^C_R$ and $\alpha_i>0$, then
\[\hd((0,0),\ell^C_R) \;\leq\;
\left(\sum_{j=i+1}^{k}\hstep(\alpha_j)\cdot\sigma_R(\alpha_j)\right)+
	\hstep(\alpha_i)\cdot\tau_R(C);\]
		  \item if $(0,0)$ is below $\ell_L^C$ and
			  $\alpha_i<\infty$, then
\[\vd((0,0),\ell^C_L) \;\leq\;
\left(\sum_{j=1}^{i-1}\vstep(\alpha_j)\cdot\sigma_L(\alpha_j)\right)+
	\vstep(\alpha_i)\cdot\tau_L(C).\]
 \end{enumerate}
\end{cor}

\noindent
By Proposition~\ref{prop:beltslopes},
Corollary~\ref{cor:polydistlines}, and the
fact~(\ref{eq:polboundstep}) we get:

\begin{thm}[Quantitative belt theorem]\label{th:quantitative}
	The slopes, widths, and positions of belts from
	Theorem~\ref{th:qualitative} can be presented by integers that
	are polynomial in $|Q|$, referring to the underlying OCN
	$\calN=(Q,Act,\delta)$.
\end{thm}

We could describe the respective polynomials more specifically,
but this is not necessary for deriving the \PSPACE-membership for the
simulation problem on one-counter nets, which is sketched in the next
subsection.

\subsection{Polynomial-space algorithm}\label{subsec:polspacealg}

Based on the basic belt theorem, a straightforward algorithm that
decides
whether $p(m)\simul q(n)$ for a given OCN $\calN$,
or more generally constructs a description of $\simul$ on the LTS
$\calL_{\calN}$, was given
in~\cite{DBLP:conf/sofsem/JancarMS99,DBLP:conf/stacs/JancarKM00};
taking the ``polynomial'' slopes and widths of belts into account, we
get polynomial-space algorithms~\cite{DBLP:journals/corr/HofmanLMT16}.
In principle, we can use ``brute force'' and simply construct an initial rectangle of the
planes $\Nat\times\Nat$, first assuming that all points are black in
all colourings, and
stepwise recolouring to white when finding points with ranks
$1,2,3,\dots$. It is easy to note that the colouring inside the belts is
(ultimately) periodic, i.e., after a possible initial segment another segment
repeats forever.
The belt periods can be (and sometimes are)
exponential, hence the above ``brute-force'' algorithm would also use
exponential space to discover the repeated belt-segments and provide
a complete description of the relation $\simul$ (on the LTS
$\calL_{\calN}$).
Nevertheless, it is a technical routine to modify
the algorithm so that it uses only
a polynomial work-space.

%
%
\section{Remark on Double-exponential Periods}\label{sec:doubleexpperiod}

We have mentioned the periodic colouring of the
simulation-belts for OCNs, where the periods can be exponential.
It is not so surprising then that these periods
for SOCNs can be double-exponential, which we now discuss in more
detail.
In fact, by our constructions this is a corollary of the fact that
the periods of sequences $\calSD$ given by sequence descriptions
$\calD=(\Delta,D,m)$ can be double-exponential w.r.t. the size
of $\calD$ (where $m$ is presented in binary); this also entails that
the period in the sequence $W(0),W(1),W(2),\dots$ for the countdown
games (recall Section~\ref{subsec:ECGinEXPS}) can be double-exponential.
Hence we concentrate just on the sequences $\calSD$.

Given $\calD=(\Delta,D,m)$ (where $D:\Delta^3\to\Delta$), we recall
that $\calSD:\Nat\to\Delta$ satisfies
$\calSD(0)=\hash$, $\calSD(1)=\calSD(2)=\cdots=\calSD(m)=\blank$,
and $\calSD(i)=D(\calSD(i{-}m{-}1),\calSD(i{-}m),\calSD(i{-}m{+}1))$
for $i>m$.
It is thus obvious that there are  $i_0,d \leq
\abs{\Delta}^{m{+}1}$, where $d\geq 1$,
such that $\calSD(i{+}d)=\calSD(i)$ for each $i\geq i_0$
(in $\calSD$, after an initial segment of length $i_0$, a segment of the
period-length $d$ is repeated forever). We now show that the least
such $d$ can be double-exponential.

\begin{lem}
\label{lem:sd_periodic}
For each $n\in\Nat$ there is a~sequence description~$\calD$ of size polynomial
in~$n$ such that for some $d\geq 2^{2^n}$ we have $\calSD(i)=\hash$
iff $i$ is a~multiple of~$d$; moreover,
	$\calSD(i{+}d)=\calSD(i)$ for each $i\geq 0$.
\end{lem}
\begin{proof}
We fix a~Turing machine
$\calM=(Q,\Sigma,\Gamma,\delta,q_0,\{\qacc\})$,
	with $\Sigma=\{\TT{0}\}$ and $\Gamma\supseteq
	\{\TT{0},\TT{1},\blank,\$\}$  that behaves as follows.
	Starting from the configuration $q_0\TT{0}^n$, $\calM$
	writes the word $\$\TT{0}^m\$$ on the tape, where  $m=2^n$,
	while visiting only the cells $0,1,\dots,m{+}1$.
	Now, inside this space (the cells $0,1,\dots,m{+}1$), all
	$w\in\{\TT{0},\TT{1}\}^m$ are stepwise generated;
the tape content, always of length $m{+}2$,
	will stepwise become $\$\TT{0}\cdots\TT{0}\TT{0}\TT{0}\$$,
	$\$\TT{0}\cdots\TT{0}\TT{0}\TT{1}\$$,
	$\$\TT{0}\cdots\TT{0}\TT{1}\TT{0}\$$,
	$\$\TT{0}\cdots\TT{0}\TT{1}\TT{1}\$$, $\cdots\cdots$,
	 $\$\TT{1}\cdots\TT{1}\$$.
Finally $\calM$ rewrites all cells
with~$\blank$, and halts at the cell~$0$ in~$\qacc$.
Hence $\calM$ uses only cells $0,1,\dots, m{+}1$ and performs
 $t$ steps where $t>2^m=2^{2^n}$.

Now for each $n\in\Nat$ we consider
a~Turing machine~$\calM_n$ that, when starting in $q'_0$ on the empty
tape, first
	 writes $\TT{0}^n$, then invokes the computation of $\calM$
	on $\TT{0}^n$, and when $\calM$ halts (in~$\qacc$), then $\calM_n$
 restores its initial configuration with the empty tape; the initial
	state $q'_0$ will be visited only at the start and at the end
	of the described computation.

Hence the computation $C_0,C_1,C_2,\dots$ of $\calM_n$ from $q'_0$ and the empty tape is
	infinite, visits only the cells $0,1,\dots,m{+}1$ for $m=2^n$,
and there is $d_0\geq 2^{2^n}$ such that
	the control state~$q_0'$ is visited
exactly in the configurations $C_i$ where $i$ is a multiple of $d_0$.
We also note
	that the size of $\calM_n$ is $\bigO(n)$.

	The infinite word $C_0C_1C_2\dots$ (each $C_i$ of length $m{+}2$)
	can be viewed as $\calSD$ for
 a~sequence description~$\calD$ of size polynomial with respect to~$n$
	(recall the proof of
	Proposition~\ref{prop:EGSP_exspace_complete}).
	For 	$d=d_0\cdot(m+2)$ we thus have
	$\calSD(i)=\calSD(i{+}d)$ for all $i\in\Nat$.
Moreover, we identify $\hash$ with the pair $(q_0',\blank)$, which entails
that $\calSD(i)=\hash$ iff $i$ is a multiple of $d$.
\end{proof}

%
%

\section{Additional Remarks}\label{sec:addrem}

One particular application of countdown games was shown by
Kiefer~\cite{DBLP:journals/ipl/Kiefer13} who
modified them
to show \EXPTIME-hardness of
bisimilarity on BPA processes.
Our \EXPSPACE-complete modification does
not seem easily implementable by BPA processes, hence
the \EXPTIME-hardness result
in~\cite{DBLP:journals/ipl/Kiefer13} has not been improved here.
(The known upper bound for bisimilarity on BPA is \doubleEXPTIME.)

We also mention that the involved result
in~\cite{DBLP:conf/icalp/GollerHOW10}
shows that, given any fixed language $L$ in \EXPSPACE, for any word $w$
(in the alphabet of $L$) we can construct
a succinct one-counter automaton that performs a computation which
is accepting iff $w\in L$.
Such a computation needs to access
concrete bits in the (reversed) binary presentation of the counter value.
A straightforward direct access to such bits is destructive (the counter
value is lost after the bit is read) but this can be avoided:
instead of a ``destructive reading'' the computation just ``guesses''
the respective bits, and it is forced to guess correctly by a carefully
constructed CTL formula that is required to be satisfied by the
computation.
This result is surely deeper than
the \EXPSPACE-hardness of existential countdown games, though the
former does not seem to entail the latter immediately.

Finally we note that the theorems and proofs in Section~\ref{sec:structsimulOCN}
could be formulated for specific tiling problems, without a direct
reference to one-counter nets.

\section*{Acknowledgements.}
We thank the anonymous reviewers for their helpful comments.

%
%

\bibliographystyle{alphaurl}
\bibliography{bibliography}

\end{document}